# Theoretical and Observational Aspects of Cosmological Inflation

**Barun Kumar Pal**

**Physics and Applied Mathematics Unit**

**Indian Statistical Institute, Kolkata**

A Thesis Submitted To
**JADAVPUR UNIVERSITY**
For The Degree Of

`Doctor Of Philosophy`

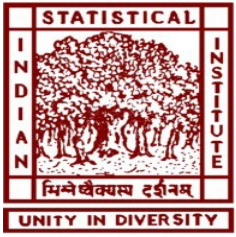

# INDIAN STATISTICAL INSTITUTE
203 B. T. Road, Kolkata – 700108

**Physics and Applied Mathematics Unit**
**Fax: +91 33 2577 3026**
**URL: http://www.isical.ac.in/~pamu**

## CERTIFICATE FROM THE SUPERVISORS

This is to certify that the thesis entitled "**THEORETICAL AND OBSERVATIONAL ASPECTS OF COSMOLOGICAL INFLATION**" submitted by Sri **Barun Kumar Pal** who got his name registered on **18/06/2010 (Index No.: 107/10/Maths./20)** for the award of Ph. D. (Science) degree of Jadavpur University, is absolutely based upon his own work under the supervisions of `Prof. Banasri Basu` and `Dr. Supratik Pal`, and that neither this thesis nor any part of it has been submitted for either any degree / diploma or any other academic award anywhere before.

**Dr. Banasri Basu**                                      **Dr. Supratik Pal**
Professor                                        Assistant Professor
Phone: +91 33 2575 3037                    Phone: +91 33 2575 3027
E-mail: banasri@isical.ac.in             E-mail: supratik@isical.ac.in

*I would like to dedicate this thesis to*

**My Father**

*who has left us two years ago.*

# ACKNOWLEDGEMENTS

*First of all, I want to acknowledge Council of Scientific and Industrial Research, Govt. of India, for financial support through a Research Fellowship (Grant No. 09/093 (0119)/2009). It has allowed me the opportunity to carry on research work in the grand atmosphere of Physics and Applied Mathematics Unit of Indian Statistical Institute, Kolkata.*

*I am very much grateful to my thesis advisors, Prof. Banasri Basu & Dr. Supratik Pal, for their ceaseless encouragements, assistances, suggestions and guidance throughout my research work, all the credit goes to them.*

*I am thankful to my collaborators – Prof. Barnana Roy of our Unit and Ms. Hamsa Padmanabhan of IUCAA. I also want to acknowledge Prof. Tarun Souradeep of IUCAA for helpful suggestions and discussions.*

*I heartily thank all the research scholars, faculty members and office staff of our unit for their charitable cooperation, support whenever and wherever needed.*

*I would like to pay sincere gratitude to my father and grandmother for their never-ending love and blessing. They dreamed of this day, but are no more to see it.*

*I am also very much thankful to all my brothers, uncles, aunties and cousins together with loved ones for their kind and loving support.*

*Last but not the least, I want to specially honor my mom for her carefulness, love, encouragement, .............................., without whom this thesis would have remained incomplete.*

# PUBLICATIONS

The thesis is based on the following research works:

1. MUTATED HILLTOP INFLATION: A NATURAL CHOICE FOR EARLY UNIVERSE
   *Barun Kumar Pal, Supratik Pal and B. Basu*
   **Journal of Cosmology and Astro-particle Physics 01**, 029 (2010).

2. A SEMI-ANALYTICAL APPROACH TO PERTURBATIONS IN MUTATED HILLTOP INFLATION
   *Barun Kumar Pal, Supratik Pal and B. Basu*
   **International Journal of Modern Physics D 21**, 1250017 (2012).

3. CONFRONTING QUASI-EXPONENTIAL INFLATION WITH WMAP SEVEN
   *Barun Kumar Pal, Supratik Pal and B. Basu*
   **Journal of Cosmology and Astro-particle Physics 04**, 009 (2012).

4. THE BERRY PHASE IN INFLATIONARY COSMOLOGY
   *Barun Kumar Pal, Supratik Pal and B. Basu*
   **Classical and Quantum Gravity 30**, 125002 (2013).

5. TOWARDS RECONSTRUCTION OF UNLENSED, INTRINSIC CMB POWER SPECTRA FROM LENSED MAP
   *Barun Kumar Pal, Hamsa Padmanabhan and Supratik Pal*
   **arXiv**: 1309.1827. (**Submitted**)

# Abstract

This thesis piles up the results of five works in the field of *cosmological inflation*, *inflationary cosmological perturbations*, *cosmic microwave background* (CMB) and *weak gravitational lensing of cosmic microwave background*.

A simple single field phenomenological model of inflation – the *mutated hilltop inflation*, has been presented. The presence of infinite number of terms in the *Taylor* series expansion of the inflaton potential makes the mutated hilltop inflation attractive and accurate at the same time. A brand new way to modify the *consistency relation*, by taking into account the direct effect of scalar field evolution in estimating observable parameters at the time of *horizon exit*, has been discussed.

Also, the inflationary cosmological perturbations of quantum mechanical origin have been studied from a different perspective. The complete *wave-functions* for the inflationary cosmological perturbations (both scalar and tensor) have been derived by solving the associated *Schrödinger* equations. The wave-functions have then been used to derive a cosmological analogue of the *Berry phase* which may serve as a complementary probe of inflationary cosmology.

With a phenomenological *Hubble* parameter the *quasi-exponential* models of inflation have been discussed, using *Hamilton-Jacobi* formulation. After constraining the model parameter from the theoretical as well as observational ground, we confront *quasi-exponential inflation* with *WMAP–7* data using publicly available code CAMB (Code for Anisotropies in the Microwave Background). The observable parameters are found to fair extremely well with *WMAP–7* data. The analysis also predicts a ratio of the tensor to scalar amplitudes that may even be detected in near future.

Finally, weak gravitational lensing in the context of CMB has been discussed. A simple method to extract the unlensed, intrinsic CMB temperature and polarization power spectra from the observed (i.e., lensed) data has been proposed. Using a matrix inversion technique, we demonstrate how one can reconstruct the intrinsic CMB spectra directly from the lensed data. The delensed spectra obtained by the technique are calibrated against CAMB using *WMAP–7* best-fit data and applied to *WMAP–9* unbinned data as well. In principle, our methodology may help to subtract the $E$-mode lensing contribution in order to obtain the intrinsic $B$-mode power.

# CONTENTS

# LIST OF FIGURES

# LIST OF TABLES

# CHAPTER 1

# Basics of Cosmological Inflation

## 1.1 Introduction

Over the decades *Big-Bang* theory has come across many successes by explaining the *nucleosynthesis, expanding universe, formation of cosmic microwave background radiation*, but it has some flaws related to the initial conditions which can not be explicated within its limit. One of the basic ideas of modern cosmology is that, there was a brief early epoch in the history of our Universe when the Universe expanded almost exponentially – the so-called *Cosmic Inflation* [1]. During that time our Universe grew very rapidly making it almost *homogeneous* and *spatially flat*. A sufficiently long epoch of primordial inflation leads to a homogeneous and isotropic Universe that is old and flat. In that era a small smooth spatial region of size less than the *Hubble* radius can grow large enough to encompass the present volume of the observable Universe. Although, the inflationary scenario, so far the best resolution for the early Universe, is in vogue for quite some time, but still it is more like a paradigm than a theory as a specific compelling model is yet to be separated out from the spectrum of possible models and alternatives.

The most striking feature of the inflationary scenario is that they supply quantum seeds for the origin of cosmological fluctuations observed in the large scale structure of matter and in *Cosmic Microwave Background* (CMB). The observed anisotropies in CMB [2, 3, 4, 5] are in tune with the inflationary prediction for the primordial perturbations. The finespun observational data [4, 5, 6, 7] are now enforcing the models of inflation to face the challenge of passing through the crucial tests of observations thereby making inflationary model building a sticky task. Apart from that, these precise observational data are now leading us towards the fundamental physics of the early Universe. Attempts in this direction include *standard particle physics* *motivated models* [8, 9] and *allied phenomenological models* [10, 11, 12, 13], *string theory inspired models* [14, 15, 16, 17], inflation from *super-symmetry* [18, 19], *warm inflation* [20, 21], *multi-field inflation* [22] and *brane-world models* [23, 24, 25, 26, 27, 28, 29] among others. Armed with plenty of successes, despite many efforts, inflation has not been superseded by its various challengers [30, 31, 32, 32, 33, 34, 35, 36, 37, 38, 39, 40, 41], and this scenario has gradually become an integral part of modern cosmology.

In this chapter, I shall briefly review the standard *Big-Bang* scenario and the paradigm of *single field slow-roll [10, 12, 42] inflation* together with its impact on CMB temperature anisotropy by analyzing the evolution of cosmological perturbations originated during inflation. The main motivation behind concentrating on the single field *slow-roll* inflation is that, they are the simplest to understand and at the same time consistent with recent observational data coming from the latest observational probes like *Planck* [4], *Wilkinson Microwave Anisotropy Probe* (*WMAP*) [5] etc., which are now very well complemented by the data from ground based telescopes – *Atacama Cosmology Telescope* (ACT) [43, 44] and *South Pole Telescope* (SPT) [45, 46].

## 1.2 COSMOLOGICAL INFLATION

The paradigm of cosmic inflation does not replace the *Big-Bang* theory, but they are complement to each other and when taken together it is the best theory compatible with the observations.

### 1.2.1 THE HOT BIG-BANG COSMOLOGY

The most powerful as well as the most utile assumption about our Universe, so far confirmed by various observations [2, 3, 4, 5], is that on the large scales it is *homogeneous* and *isotropic, i.e.,* invariant under space-time *translation* and *rotation*, which resembles the *cosmological principle*.

Modern cosmology is soundly based on *Einstein*'s theory of *General Relativity* (GR), according to which our Universe is described by a *four-dimensional* geometry satisfying the following *Einstein* equations –

$$G_{\mu\nu} \equiv R_{\mu\nu} - \tfrac{1}{2}R g_{\mu\nu} = 8\pi G T_{\mu\nu}. \tag{1.1}$$

*Einstein* was the first to apply GR in cosmology, but after getting *non-stationary solution* he wrongly interpreted it as a shortcoming of the theory. It was only after *Edwin Hubble*, who established that Universe is not static but expanding, GR truly sets its foot into cosmology.

#### A. FRW Metric and Cosmological Dynamics

The observed homogeneity and isotropy of our Universe is perfectly described by the *Friedmann-Robertson-Walker* (FRW) metric which in the appropriate *co-ordinate* system has the following form

$$ds^2 = -dt^2 + a^2(t)\left[\frac{dr^2}{1-\kappa r^2} + r^2\left(d\theta^2 + sin^2\theta\, d\phi^2\right)\right] \tag{1.2}$$

where $a(t)$ is the scale factor which measures the expansion of our Universe and depending on the curvature of spatial *hyper-surfaces*, $\kappa = 0, \pm 1$. The coordinates $r$, $\theta$ and $\phi$ are referred to as *co-moving coordinates*.

The matter content compatible with the homogeneity and isotropy of the space-time is expressed by the energy momentum tensor, which in the same *co-moving coordinate* reads

$$T^\mu_\nu = \text{Diag}\left[-\rho(t), P(t), P(t), P(t)\right] \tag{1.3}$$

where $\rho$ and $P$ are energy and pressure densities respectively. The *Einstein* Eqn.(1.1), when combined with Eqns.(1.2) and (1.3), brings forth following *Friedmann* equations[1]

$$H^2 \equiv \left(\frac{\dot{a}}{a}\right)^2 = \frac{1}{3M_P^2}\rho - \frac{\kappa}{a^2} \tag{1.4}$$

$$\frac{\ddot{a}}{a} = -\frac{1}{6M_P^2}(\rho + 3P) \tag{1.5}$$

where $M_P \equiv \frac{1}{\sqrt{8\pi G}}$, is the reduced *Planck* mass and $H$ is the *Hubble* parameter. The conservation of the energy momentum tensor, $T^\mu_{\nu;\mu} = 0$, where ';' denotes *covariant derivative*, leads to the following continuity equation

$$\dot{\rho} + 3H(\rho + P) = 0. \tag{1.6}$$

The set of Eqns.(1.4), (1.5) and (1.6) are not independent, as a result the system is closed and unique solution exists whenever initial conditions are specified. The Eqn.(1.5) also points out that for accelerated expansion, $\ddot{a} > 0$, we require the pressure to be negative as energy is always positive. In other words, negative pressure is the root of cosmological acceleration.

To proceed further we now define at any given time, a critical energy density $\rho_C$ and various density parameters as follows –

$$\rho_C = 3H^2/8\pi G, \ \Omega_\kappa \equiv -\kappa/a^2H^2, \ \Omega_B \equiv \rho_B/\rho_C, \ \Omega_{DM} \equiv \rho_{DM}/\rho_C, \ \Omega_{DE} \equiv \rho_{DE}/\rho_C, \ \Omega_{Total} \equiv \rho/\rho_C \tag{1.7}$$

where $\Omega_\kappa$, $\Omega_B$, $\Omega_{DM}$, $\Omega_{DE}$, $\Omega_{Total}$ are the density parameters associated with the *curvature, baryonic matter, dark matter, dark energy* and *total energy density* of the Universe respectively, and $\rho_B$, $\rho_{DM}$, $\rho_{DE}$ are the energy densities for the *baryonic matter, dark matter* and *dark energy* respectively. The current measurements of the *Hubble* parameter, $H_0 = 69.32 \pm 0.80$ km/s/Mpc [47] (recent data from *Planck* favors $H_0 = 67.3 \pm 1.2$ km/s/Mpc [4, 48]), together with $\kappa$ when taken with Eqn.(1.4) provides the present estimate for the total energy density of the Universe. The relative contributions of different forms of energies are found from various cosmological observations. The total energy density at present is very close to the critical one, *i.e.*, $\Omega_{Total} \approx 1.0027^{+0.0039}_{-0.0038}$ and $\Omega_\kappa(t = t_0) \equiv -\kappa/a_0^2 H_0^2 = -0.0027^{+0.0039}_{-0.0038}$ [5, 47], which has been

---

[1]Here an over-dot, "$\cdot$", represents derivative with respect to cosmic time $t$ and unless otherwise stated an over-prime, " $'$ ", denotes derivative with respect to the argument of the function.

calculated from the first peak position of the CMB angular power spectrum [4, 5]. The baryonic matter content of the present Universe is given by $\Omega_B = 0.04628 \pm 0.00093$ [47] which has been estimated from the heights of the CMB acoustic peaks. The existence of *dark matter* was predicted by analyzing the *galactic rotation curves* and recent observational data favors $\Omega_{DM} = 0.2402^{+0.0088}_{-0.0087}$ [47]. The bound set on the total matter content (both baryonic and dark matter) of our Universe by the satellite mission *Planck* is $\Omega_{Mat} \equiv \Omega_M + \Omega_{DM} = 0.315 \pm 0.017$ [4, 48]. The rest is attributed to a mysterious substance, *Dark Energy*, which is the cause of present day acceleration. From the high redshift *supernovae*, large scale structure, CMB power spectra and the age of the Universe, $t_0 = 13.772 \pm 0.059$ Gyr [47], it has been found that $\Omega_{DE} = 0.7135^{+0.0095}_{-0.0096}$ [47]. Whereas the combined data from *Planck* and *WMAP* favors $t_0 = 13.817 \pm 0.048$ Gyr and $\Omega_{DE} = 0.685^{+0.018}_{-0.016}$ [48].

The solution of the *Friedmann* equation is obtained by specifying the *equation of state parameter*, $\omega \equiv P/\rho$. The scale factor for $\kappa = 0$ then turns out to be

$$a(t) \propto t^{\frac{2}{3(1+\omega)}}. \tag{1.8}$$

Therefore, standard *Big-Bang* theory predicts *decelerated expansion* as $a(t) \propto t^p$ with $p < 1$, for ordinary matter ($\omega = 0$) and radiation ($\omega = 1/3$). But, recent observations suggest that our Universe is accelerating at present. This gives rise to some severe problems which can not be answered within the limit of *Big-Bang* theory.

### B. Big-Bang Puzzles

At present the shortcomings of standard cosmology are well appreciated – *the horizon or large-scale smoothness problem; the small-scale inhomogeneity problem (origin of density perturbations); and the flatness or oldness problem*. But, they do not indicate any logical inconsistencies of the theory, rather very special initial data seems to be required for the evolution to a Universe that is qualitatively similar to ours today.

Before discussing those puzzles, we first rewrite *Friedmann* Eqn.(1.4) in the following form

$$\Omega_{Total} - 1 = \frac{\kappa}{a^2 H^2}. \tag{1.9}$$

1. **Homogeneity Problem:** This puzzle can be stated as, why the assumptions of homogeneity and isotropy stand out to be so good? Inhomogeneities grow with time due to gravitational attraction but recent observations suggest that our Universe is almost homogeneous. At the time of last scattering inhomogeneities were $\delta\rho/\rho \sim 10^{-5}$ and nowadays it is almost $\sim 1$ (in clusters of clusters). This means in the remote past the inhomogeneities were much smaller than today. How to explain a smooth Universe in the past?

2. **Flatness Problem:** The latest observations suggest that our Universe is very nearly

spatially flat. But in a decelerated *universe*, a *flat* solution is *unstable* as $|\Omega_{Total} - 1|$ increases with time which is clear from Eqn.(1.9). In order to get the correct value of $|\Omega_{Total} - 1| < 10^{-2}$ at present, the value of $|\Omega_{Total} - 1|$ at very early times has to be fine-tuned to values amazingly close to zero, but without being exactly zero. This is why the flatness problem is also dubbed as *fine-tuning problem*.

3. **Horizon Problem:** The CMB, which is a form of electromagnetic radiation filling the Universe, has temperature $T_0 = 2.73$ K almost same in all directions. This tells us that, before the formation of CMB they were in thermal equilibrium. But, in standard *Big-Bang* scenario there were not enough time for those regions to come into thermal equilibrium. Because, there were almost $10^6$ casually disconnected regions within the volume that now corresponds to our *horizon*. This is the so called horizon problem.

### 1.2.2 INFLATION

A solution to the *Big-Bang puzzles* is provided by the inflationary scenario. The basic prescription is to invert the behavior of the *co-moving Hubble* radius ($1/aH$), *i.e.*, make it decrease sufficiently in the very early Universe. The corresponding requirement is $\ddot{a} > 0$, *i.e.*, we need an early phase of accelerated expansion.

Inflation is defined to be an early epoch of accelerated expansion. So inflation means $\ddot{a} > 0$, which from the second *Friedmann* Eqn.(1.5) can be interpreted as

$$\ddot{a} > 0 \iff \rho + 3p < 0, \tag{1.10}$$

*i.e.*, violation of the *strong energy condition*. A very simple example of such equation of state is the cosmological constant, which corresponds to a cosmological fluid with $\omega = -1$. But a strict cosmological constant will lead to an ever accelerating *universe* which spoils the successes of *Big-Bang* theory and it does not provide any explanation for the origin of observed structures in our Universe either. Another possibility, which we are interested in here is the *scalar field*, known as *inflaton* in the context of inflation.

#### A. Solution to Big-Bang Puzzles

Inflation is not the first attempt to address these shortcomings, over the past three decades cosmologists have pondered this question and proposed alternative solutions. Inflation is a solution based upon well-defined, albeit speculative, early Universe micro-physics describing the *post-Planckian* epoch.

Since the *co-moving Hubble* radius $(aH)^{-1}$ decreases almost exponentially during inflation, the term $\kappa/a^2H^2$ is driven towards zero very rapidly. As a result, at the end of inflation $|\Omega_{Total} - 1|$ is pushed very close to zero. Though during radiation and matter domination the *co-moving Hubble* radius grows but $|\Omega_{Total} - 1|$ is still very small.

Before the inflation, size of the Universe was very small and within that small patch *thermal equilibrium* was established. Inflation then magnifies that homogeneous patch larger than present size of the observable Universe. So different portions of the last scattering surface which seem to have no causal contact were actually in *equilibrium* before the inflation. This is why we see CMB so homogeneous and isotropic.

### B. Inflationary Dynamics

The dynamics of a homogeneous scalar field with potential energy $V(\phi)$, minimally coupled to gravity, is governed by the action

$$S_\phi = -\int d^4x \sqrt{-g} \left[ \tfrac{1}{2} \partial_\mu \phi \partial^\mu \phi + V(\phi) \right]. \qquad (1.11)$$

Corresponding energy momentum tensor is given by

$$T_{\mu\nu} = \partial_\mu \phi \partial_\nu \phi - g_{\mu\nu} \left[ \tfrac{1}{2} \partial^\alpha \phi \partial_\alpha \phi + V(\phi) \right]. \qquad (1.12)$$

The energy and pressure densities of a homogeneous scalar field with potential $V(\phi)$ are

$$\rho = \tfrac{1}{2}\dot\phi^2 + V(\phi), \; P = \tfrac{1}{2}\dot\phi^2 - V(\phi). \qquad (1.13)$$

The *Friedmann* Eqns.(1.4) and (1.5) for the scalar field are now given by

$$H^2 = \tfrac{1}{3\mathrm{M}_P^2}\left[\tfrac{1}{2}\dot\phi^2 + V(\phi)\right] \qquad (1.14)$$

$$\frac{\ddot a}{a} = -\tfrac{1}{3\mathrm{M}_P^2}\left[\dot\phi^2 - V(\phi)\right] \qquad (1.15)$$

and the corresponding energy conservation turns out to be

$$\ddot\phi + 3H\dot\phi + V'(\phi) = 0 \qquad (1.16)$$

which is nothing but the *Klein-Gordon* equation for a scalar field.

The system of equations (1.14), (1.15) and (1.16) does not always possess accelerated expansion, but does so in the *slow-roll* regime [10, 12, 42], where potential energy of the inflaton dominates over its kinetic energy.

### C. Slow-Roll Approximation

To get the insight of inflationary scenario we need to solve Eqns.(1.14), (1.15) and (1.16), which is very difficult with their full generalities. But, the mathematical expressions are necessary to compare inflationary predictions with the observational data. The *slow-roll approximation* provides an elegant way to overcome that difficulty by reducing the complexity to a great extent.

In the *slow-roll approximation*, the kinetic energy term, $\dot\phi^2/2$, in Eqns.(1.14), (1.15) and the acceleration term, $\ddot\phi$, in Eqn.(1.16) are ignored, which lead to the modified *Friedmann* equations governing the dynamics of a scalar field

$$H^2 \approx \tfrac{1}{3\text{M}_P^2} V(\phi) \tag{1.17}$$

$$3H\dot{\phi} + V'(\phi) \approx 0. \tag{1.18}$$

The conditions for neglecting the kinetic energy and acceleration terms can be equivalently written by the smallness of two *slow-roll parameters*, $\epsilon_V \ll 1$ and $|\eta_V| \ll 1$ where [49]

$$\epsilon_V \equiv \tfrac{\text{M}_P^2}{2}\left[\frac{V'(\phi)}{V(\phi)}\right]^2, \quad \eta_V \equiv \text{M}_P^2 \left[\frac{V''(\phi)}{V(\phi)}\right]. \tag{1.19}$$

The *slow-roll approximation* is applicable only when $\epsilon_V \ll 1$, $|\eta_V| \ll 1$ are satisfied, *i.e.*, where the *slope* and the *curvature* of the inflaton potential are very small. These two parameters $\epsilon_V$, $\eta_V$ are generally called *potential slow-roll parameters*. Inflation ends when the *slow-roll* approximation is violated and the corresponding value of the inflaton is computed from the following equation

$$\max_{\phi}\{\epsilon_V, |\eta_V|\} = 1. \tag{1.20}$$

If instead, the *Hubble* parameter is considered as the fundamental quantity then we can define another set of parameters, *Hubble slow-roll parameters,* as follows [42]

$$\epsilon_H \equiv 2\,\text{M}_P^2\left[\frac{H'(\phi)}{H(\phi)}\right]^2, \quad \eta_H \equiv 2\,\text{M}_P^2\left[\frac{H''(\phi)}{H(\phi)}\right]. \tag{1.21}$$

These two *Hubble slow-roll parameters* are extremely useful and *inflation* is very naturally described by them [42, 50, 51], making them superior choice over the *potential slow-roll parameters*. The condition for inflation is then precisely given by

$$\ddot{a} > 0 \iff \epsilon_H < 1. \tag{1.22}$$

The parameters $\epsilon_H$ and $\eta_H$ are not the usual *slow-roll* parameters. But in the *slow-roll* limit $\epsilon_H \to \epsilon_V$ and $\eta_H \to \eta_V - \epsilon_V$ [42]. The exact end point of inflation is provided by the *Hubble slow-roll parameter*, $\epsilon_H = 1$, whereas in the case of *potential slow-roll approximation*, the end of inflation given by Eqn.(1.20) is only a first order result [42]. In Chapter 4, using *Hubble slow-roll approximation* with a phenomenological *Hubble parameter*, we have modeled *quasi-exponential inflation* and confronted with *WMAP–7* data [5].

### D. Amount of Inflation

In order to resolve *Big-Bang* puzzles, we need sufficient amount of accelerated expansion. The amount of inflaton is generally expressed as the *logarithmic* difference between the final and initial values of the scale factor which is called *number of e-foldings*. So the *number of e-foldings*, $N$, is found to be

$$N \equiv \ln \frac{a_E}{a_I} = \int_{a_I}^{a_E} \frac{da}{a} = -\frac{1}{M_P} \int_{\phi_I}^{\phi_E} \frac{d\phi}{\sqrt{2\epsilon_H}} \approx -\frac{1}{M_P} \int_{\phi_I}^{\phi_E} \frac{d\phi}{\sqrt{2\epsilon_V}}, \qquad (1.23)$$

where $a_I$ & $a_E$ and $\phi_I$ & $\phi_E$ are the values of the scale factor and inflaton field at the start & at the end of inflation respectively.

To solve the cosmological puzzles we need about 70 *e-foldings*, so that the scale factor should have at least increased by a factor of $e^{70}$ during inflation. The *co-moving* scale corresponding to the present estimated value of the *Hubble* radius, crossed the horizon when nearly 60 *e-foldings* are still left before the end of inflation and all the other relevant scales following within the next few *e-foldings*.

### 1.2.3 SPECTRUM OF INFLATIONARY MODELS

More than three decades have slipped away since *Guth*'s original paper [1] attracted considerable attention to cosmological community. Thirty years later inflation is still alive in a much stronger position than ever based on highly precise observational data available of late. But still there is no unique model of inflation. The fundamental microscopic origin of the *inflaton* is a mystery too. Inflation is believed to occur at a very high energy scale $\sim 10^5$ GeV – *Planck scale*, depending on the field strength. The form of the inflaton potential is still unknown. As a result, there are plenty of inflationary models available in the literature.

The models of inflation can be broadly classified into two categories – *small field models*, where the variation of the inflaton, *i.e.*, the difference between initial and final values of the inflaton, $\Delta\phi$, is less than the *Planck* mass, *i.e.*, $\Delta\phi < M_P$ and *large field models*, where the variation of the inflaton is of the order of the *Planck* mass. The generic form of single field inflaton potentials is given by

$$V(\phi) = V_0 f\left(\frac{\phi}{\mu}\right) \qquad (1.24)$$

where $V_0$ is the height corresponding to the vacuum energy during inflation and $\mu$ is the width which corresponds to the change in inflaton value, $\Delta\phi$, during inflation. Different models have different form for the function $f$.

#### A. Large Field Models

In large-field models, also known as *chaotic inflation* [52], the inflaton field starts at a large field value and then evolves to a minimum at the origin $\phi = 0$. This type of models are characterized by $V''(\phi) > 0$. Large field models are mostly governed by the simplest type of potentials. A general set of large field polynomial potentials can be written as [53, 54],

$$V(\phi) = V_0 \left(\frac{\phi}{\mu}\right)^p \qquad (1.25)$$

where a particular model is chosen by specifying the parameter $p$. In general $p$ is a positive integer but models with fractional values of $p$ are also available [14, 55]. There is another set

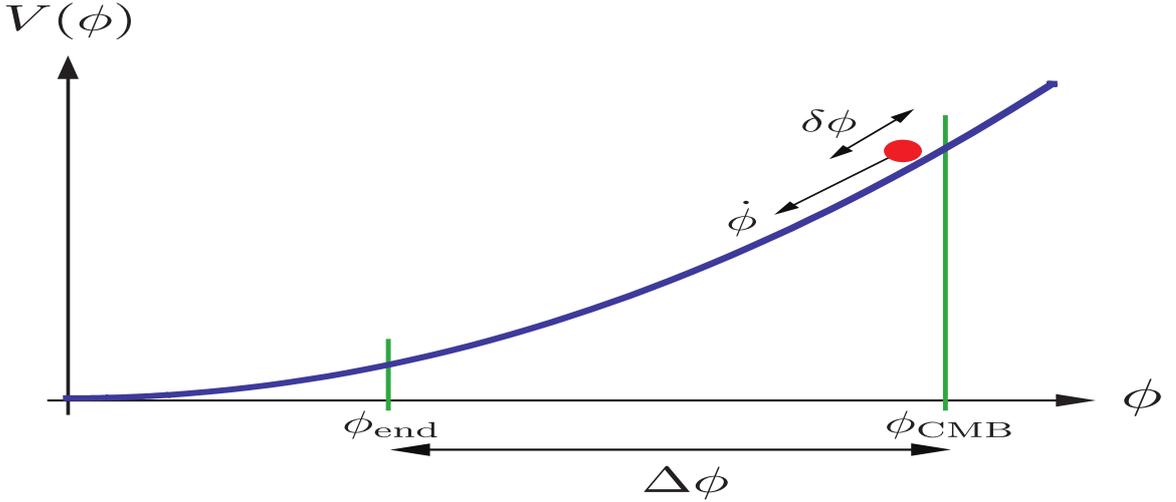

**FIG. 1.1:** *In these models the inflaton field evolves over a super-Planckian range during inflation, $\Delta\phi > M_P$. The $\phi_{CMB}$ represents value of the inflaton when the CMB fluctuations are created. Figure reproduced from [62].*

of large field potentials having the following form [56, 57, 58, 59, 60, 61],

$$V(\phi) = V_0 \exp\left(\frac{\phi}{\mu}\right)^p. \tag{1.26}$$

`Fig.1.1` shows variation of the inflaton potential with the inflaton for large field model.

### B. Small Field Models

In the small field models [10, 12, 63, 64, 65, 66, 67, 68, 69, 70] the inflaton moves over a small distance, $\Delta\phi < M_P$. The potentials that give rise to such small field evolution often arise in mechanism of *spontaneous symmetry breaking*. Small field models are characterized by the negative curvature of the potential, *i.e.*, $V''(\phi) < 0$. A simple example is

$$V(\phi) = V_0 \left[1 - \left(\frac{\phi}{\mu}\right)^p\right], \tag{1.27}$$

which can be visualized as a lowest order *Taylor* series expansion of a potential about the origin. The `Fig.1.2` shows typical behavior of the small field inflationary potential.

Let us now turn our attention to a very compelling class of inflationary models that has made great impact on inflationary model building of late – the *hilltop inflation*.

### C. Hilltop Inflation

*Hilltop inflation* [71] falls into wide class of small field models, which is characterized by the negative curvature of their potentials. For sufficient inflation, the inflaton potential should be flat enough and this condition is satisfied if the potential obeys the *slow-roll* condition. As inflation occurs while the inflaton field rolls down from the maximum of the potential, the *flatness* condition is vacuously true for hilltop inflation.

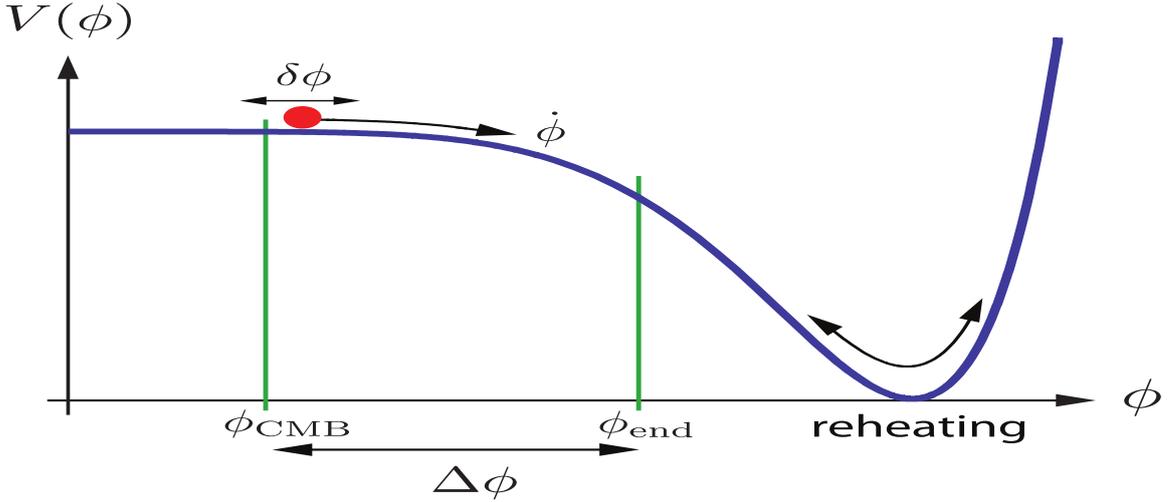

**FIG. 1.2:** *In these models the inflaton field evolves over a sub-Planckian range during inflation, $\Delta\phi < M_P$. This figure has been reproduced from [62].*

The potential describing hilltop inflation can be written as [71]

$$V(\phi) = V_0 - \tfrac{1}{2}m^2\phi^2 + \textit{Higher Order Terms} = V_0\left[1 - \tfrac{1}{2\mathrm{M}_P^2}|\eta_0|\phi^2 + ...\right] \quad (1.28)$$

where $V \approx V_0$ and $\eta_0$ is the value of the *slow-roll* parameter $\eta_V$, at the maximum. Assuming mass term dominates over the higher order terms, the *slow-roll* parameters turn out to be

$$\epsilon_V = \tfrac{1}{2\mathrm{M}_P^2}|\eta_0|^2\phi^2\left(1 - \tfrac{1}{2\mathrm{M}_P^2}|\eta_0|\phi^2\right)^{-2}, \quad \eta_V = -\mathrm{M}_P^2|\eta_0|\left(1 - \tfrac{1}{2\mathrm{M}_P^2}|\eta_0|\phi^2\right)^{-2}. \quad (1.29)$$

Here the expression for the number of *e-foldings* is given by

$$N - N_E = \ln(\phi/\phi_E) - \tfrac{1}{4\mathrm{M}_P^2}\left[\phi^2 - \phi_E^2\right]. \quad (1.30)$$

Most of the observable parameters as calculated from hilltop inflation are in very good agreement with the latest data [4, 5].

Nowadays, hilltop inflation has turned out to be a very prospective model to explain early universe phenomenon in the sense that in these models a flat potential can easily be converted to one with a maximum with the addition of one or two terms from a power series expansion. Consequently, many existing inflationary models can be converted to hilltop by suitably tuning the model parameters. In Chapter 2 we have discussed a variant of hilltop inflation – the *mutated hilltop inflation* [72, 73].

### D.  Multi-field, Hybrid and Modified Gravity Models

So far, from the observational point of view the simplest models of inflation, *i.e.*, models involving single scalar field, are in excellent harmony with the latest available data. But, there are inflationary models [74, 75, 76, 77, 78] involving several scalar fields and naturally they are more complicated. The general action describing the evolution of $N$ scalar fields $\phi^A$ ($A$=1, 2,...,

N) can be written as,

$$\mathcal{A} = -\int d^4x \sqrt{-g}\left[\tfrac{\mathrm{M}_P^2}{2}R + \tfrac{1}{2}G_{AB}(\phi)\partial^\mu\phi^A\partial_\mu\phi^B + V(\phi)\right] \quad (1.31)$$

where $V$ is the scalar potential and $G_{AB}$ is *symmetric positive definite* matrix. The expressions for the energy and pressure in this case turn out to be

$$\rho = \tfrac{1}{2}G_{AB}\dot\phi^A\dot\phi^B + V(\phi), \quad P = \tfrac{1}{2}G_{AB}\dot\phi^A\dot\phi^B - V(\phi). \quad (1.32)$$

For sufficient inflation we need $\tfrac{1}{2}G_{AB}\dot\phi^A\dot\phi^B \ll V(\phi)$, which leads to the *slow-roll* equations, $3H\dot\phi^A = -G^{AB}\frac{\partial V}{\partial \phi^B}$. These *slow-roll* conditions remain good approximations for an appreciable time provided the multi-scalar generalizations of the *slow-roll* parameters are small over a broad enough region.

There are also *hybrid inflation* [79, 80] models, where two scalar fields are involved but only one of them is responsible for inflation and other is held at a specific constant value which fixes the energy scale of inflation. In a typical hybrid inflation model, the scalar field responsible for inflation evolves toward a minimum with nonzero vacuum energy. Inflation ends as a result of instability in the second field when the inflaton field reaches its critical value which may be regarded as a phase transition. Effectively, hybrid inflation scenarios correspond to single field models with a potential characterized by $V''(\phi) > 0$ and $0 < \epsilon_V < \eta_V$ [9, 81].

An entirely different strand of inflationary model building is the *modified gravity inflation* which relies on the modification to the gravity sector. One way of achieving this is simply by adding higher order terms constructed entirely from curvature tensor to the *Einstein-Hilbert* action. The simplest example of such a model is the $R^2$ inflation [82, 83] where the term $R^2/(6M^2)$ is added to the *Einstein-Hilbert* action. There are also *scalar-tensor* theories [84], where modification to gravity is caused by one or more scalar fields and for a single scalar field the action can be written as

$$\mathcal{A} = -\int d^4x \sqrt{-g}\left[\tfrac{f(\phi)}{2}R + \tfrac{1}{2}\partial^\mu\phi\partial_\mu\phi + V(\phi) - \mathcal{L}_{matter}\right] \quad (1.33)$$

where $\mathcal{L}_{Matter}$ does not contain terms involving $\phi$. By the following redefinitions

$$d\phi \longmapsto \mathrm{M}_P\sqrt{\frac{2f + 3f'(\phi)^2}{4f^2}}d\phi, \quad V(\phi) \longmapsto \frac{\mathrm{M}_P^4}{f^2}V(\phi) \quad (1.34)$$

the above action can be reduced to that similar to a single scalar field minimally coupled to gravity and inflation can be studied in the usual way [81]. There are also approaches based on *extra dimensions* [85], but we shall not enter into those.

Though, inflation was originally introduced in order to explain the *Big-Bang puzzles*, by far, the best attraction of inflationary scenario is their ability to produce correct spectra of *cosmolog-* *ical fluctuations* observed in the large scale structure of the matter and in CMB. The quantum fluctuations of the inflaton produce microscopic perturbation seeds and inflation stretched them outside the *horizon*. Long after inflation they re-enter the *horizon* and create observable cosmological fluctuations. Therefore, the study of cosmological perturbations is indispensable in order to compare inflationary predictions with the latest observations and to discriminate among different classes of inflationary models.

## 1.3 Inflationary Cosmological Perturbations

Nowadays, the study of cosmological perturbations has emerged as an essential tool for modern cosmology to confront any theoretical prediction with the observations. The recent data are compatible with the current paradigm of the early Universe cosmology, *i.e.,* the *inflationary universe scenario*.

It is well known that the inhomogeneities grow with time due to the attractive nature of gravity. So inhomogeneities were smaller in the past. As we are interested in fluctuations produced during inflation we can treat them as linear perturbations about their background values. Though the theory of higher order perturbations is also well studied [86, 87, 88], which are important in the context of *non-Gaussianities* [89, 90], but till date the observational data are very much consistent with the first order perturbation theory.

### 1.3.1 Perturbed Geometry

There is a basic distinction between general relativistic perturbations and the perturbations in other field theories where the underlying space-time is fixed. In general relativity, space-time is not a frozen background but must also be perturbed if the matter is perturbed. Mathematically, the problem of describing the growth of small perturbations in the context of general relativity reduces to solving the *Einstein* equations linearized about an expanding background. This seems very straightforward, but it is actually very difficult as well as tedious task in practice due to our *gauge freedom*.

Fixing a *gauge* in GR implies specifying a particular *co-ordinate* system. The choice of *co-ordinates* defines a *threading* of spacetime into lines (corresponding to fixed spatial coordinates $x$) and a *slicing* of spacetime into hyper-surfaces (corresponding to fixed time $t$). Life is more difficult as there is no unique choice for *gauge*. In order to get rid of *gauge* ambiguities, there are two possible way-outs –

  1  *Construct gauge invariant quantities and then calculate* **OR**
  2  *Choose a particular gauge and perform the calculations there.*

Here we shall follow the *gauge invariant* approach first formulated by *Bardeen* [91] and later improved by many others [92, 93]. Before going into details, we first define *conformal time*, which has been found very useful in studying the evolution of cosmological perturbations as,

$$\eta = \int \frac{dt}{a(t)}. \tag{1.35}$$

In a decelerated *universe* $\eta$ takes *positive* values whereas in an accelerating *universe* it has *negative* values. The FRW metric (1.2) for $\kappa = 0$, in terms of the conformal time, now takes the following unanalyzable form

$$ds^2 = a^2(\eta)\left[-d\eta^2 + \delta_{ij}dx^i dx^j\right]. \tag{1.36}$$

The linear evolution of the cosmological perturbations is obtained by perturbing the FRW metric at the first-order,

$$g_{\mu\nu} = \bar{g}_{\mu\nu}(\eta) + \delta g_{\mu\nu}(\eta, x); \; \delta g_{\mu\nu}(\eta, x) \ll \bar{g}_{\mu\nu}(\eta) \tag{1.37}$$

where $\bar{g}_{\mu\nu}(\eta)$ and $\delta g_{\mu\nu}(\eta, x)$ are the background homogeneous metric and metric fluctuations respectively. The most useful way to examine the metric fluctuations is to classify them according to their spin under spatial rotations and there are *scalar*, *vector* and second rank *tensor* perturbations, which are *spin 0*, *spin 1* and *spin 2* objects respectively. In linear theory, there is no coupling between the different fluctuation modes, hence they evolve independently resulting in autonomous study of each kind with simplified algebra. The most general form of the perturbed spatially flat FRW metric can be written as

$$ds^2 = a^2\left[-(1+2A)d\eta^2 + 2B_i dx^i d\eta + (\delta_{ij} + h_{ij})dx^i dx^j\right] \tag{1.38}$$

where $A$ is a scalar function, $B_i$ is a vector and $h_{ij}$ is a rank two tensor. According to *Helmholtz* theorem we know that any vector, $V_i$, can be decomposed as[1], $V_i \equiv \nabla_i U + U_i$, where $U$ is a *curl free* scalar and $U_i$ is a *divergence free* vector. As a result the metric perturbations $B_i$ and $h_{ij}$ can be further decomposed uniquely into three categories – *scalar, vector, tensor* as

$$B_i = \nabla_i B + \bar{B}_i, \text{ where } \nabla_i \bar{B}_i = 0 \tag{1.39}$$
$$h_{ij} = 2C\delta_{ij} + 2\nabla_i\nabla_j E + \nabla_i E_j + \nabla_j E_i + \bar{E}_{ij} \tag{1.40}$$

here $\bar{E}_{ij}$ is a *traceless transverse* tensor, $E_i$ is a *divergence free* vector, $C$, $E$ and $B$ are three scalar functions. The tensor modes represent the *gravitational waves*, vector modes decay in an expanding Universe as a result they do not play important role in the structure formation, so we shall discuss them no further here. The scalar modes are the most important, as they directly couple to the matter inhomogeneities and lead to the formation of structure in our Universe, at the same time they are the most complicated due to the *gauge* issue. Thus from Eqns.(1.38), (1.39) and (1.40) we see that there are four scalar degrees of freedom, four vector degrees of freedom and two tensor degrees of freedom corresponding to the two polarization states of gravitational waves.

---

[1] Here $\nabla$ denotes covariant derivative which reduces to the usual partial derivative for a scalar function.

The next step is to calculate the perturbations, $\delta G_{\mu\nu}$, of the *Einstein tensor* and using Eqn.(1.1) we obtain

$$\delta G_{\mu\nu} = \delta R_{\mu\nu} - \tfrac{1}{2}\delta g_{\mu\nu} R - \tfrac{1}{2} g_{\mu\nu} \delta R. \tag{1.41}$$

From Eqn.(1.1) we can now easily write down the perturbed *Einstein* equations as

$$\delta G_{\mu\nu} = 8\pi G \delta T_{\mu\nu} \tag{1.42}$$

where $\delta T_{\mu\nu}$ is the perturbation to the energy momentum tensor.

### A. Scalar Modes

It has already been mentioned that, there are two approaches to deal with the *gauge* ambiguities. One approach is to fix a *gauge*, *i.e.*, to pick conditions on the *co-ordinates* which completely eliminate the *gauge* freedom and the other one is to work with a basis of *gauge invariant* variables, which will be followed in our subsequent analyses. In the *gauge invariant* approach, we first construct two new variables – the *Bardeen potentials* [91] as follows

$$\begin{aligned} \Phi &\equiv A + \mathcal{H}(B - E') + (B - E')' \\ \Psi &\equiv -C - \mathcal{H}(B - E') \end{aligned} \tag{1.43}$$

where $\mathcal{H} \equiv \frac{a'(\eta)}{a(\eta)}$ and prime denotes derivative with respect to (w.r.t.) the conformal time. It is easy to check that those two variables are *gauge invariant* and form the basis for scalar perturbations [94, 95, 96].

From Eqn.(1.41) and Eqn.(1.43), after some straightforward but tedious algebra one can finally get the following results [1] [93]

$$\begin{aligned} a^2 \delta G^0_0 &= 2\left[\nabla^2 \Psi - 3\mathcal{H}(\mathcal{H}\Phi + \Psi') + 3\mathcal{H}\left(\mathcal{H}^2 - \mathcal{H}'\right)(B - E')\right] \\ a^2 \delta G^0_i &= 2\nabla_i \left[\Psi' + \mathcal{H}\Phi - \left(\mathcal{H}^2 - \mathcal{H}'\right)(B - E')\right] \\ a^2 \delta G^j_i &= -2\left[\left(2\mathcal{H}' + \mathcal{H}^2\right)\Phi + \mathcal{H}\Phi' + \Psi'' + 2\mathcal{H}\Psi' + \tfrac{1}{2}\nabla^2(\Phi - \Psi)\right]\delta^j_i \\ &\quad - 2\left(\mathcal{H}'' - \mathcal{H}\mathcal{H}' - \mathcal{H}^3\right)(B - E')\delta^j_i + \delta^{jk}\nabla_k \nabla_i (\Phi - \Psi). \end{aligned} \tag{1.44}$$

So, we now have the perturbed *Einstein* tensor for the scalar perturbations. In the following we shall do the same, but now for the tensor perturbations instead.

### B. Tensor Modes

The survey of tensor perturbations is comparatively easier than the scalar case, since by definition tensors are *gauge invariant*, therefore no *gauge* related confusion springs up here. Also, in the linear theory tensor modes do not couple to the scalar modes. As a result, the presence of tensor perturbations does not affect the scalars and they evolve absolutely independently. Taking into account only the tensor perturbations, the metric (1.38) can be written in the following form

$$ds^2 = a^2 \left[ -d\eta^2 + \left( \delta_{ij} + \bar{E}_{ij} \right) dx^i dx^j \right]. \tag{1.45}$$

---
[1]Here $\nabla^2$ is the 3-dimensional *Laplacian*.

The second ranked tensor $\bar{E}_{ij}$ has six *degrees of freedom*. But the tensor perturbations are *traceless*, $\bar{E}_i^i = 0$, and *transverse*, $\partial_i \bar{E}_j^i = 0$. Therefore, with these four constraints, there remain only two *degrees of freedom* or *polarizations*. Consequently, the tensor $\bar{E}_{ij}$ can be further decomposed into two linearly polarization states as [81],

$$\bar{E}_{ij} = h_\times(\eta, x) e_{ij}^\times + h_+(\eta, x) e_{ij}^+ \tag{1.46}$$

where $e_{ij}^\times$ and $e_{ij}^+$ are two fixed polarization tensors. So, under a rotation by an angle $\theta$ we have

$$(h_+ \pm i h_\times) \longrightarrow e^{\pm 2i\theta} (h_+ \pm i h_\times) \tag{1.47}$$

which shows that the tensor perturbations correspond to two *circularly* polarized states of *helicity* $\pm 2$, they are referred to as *gravitons*. Therefore, tensor perturbations are described by two functions $h_\times$ and $h_+$. Now, if we further specify the perturbations to be in the $x^1$–$x^2$ plane then $\bar{E}_{ij}$ can be rewritten in the following 3-*dimensional* matrix notation [97]

$$\bar{E}_{ij} = \begin{pmatrix} h_\times & h_+ & 0 \\ h_+ & -h_\times & 0 \\ 0 & 0 & 0 \end{pmatrix}.$$

The spatial part of the perturbed *Einstein* tensor, which provides sufficient information for the tensor modes, turns out to be

$$\delta G_j^i = \tfrac{1}{2} a^{-2} \delta^{ik} \left[ 2\mathcal{H} \bar{E}'_{kj}(\eta) + \bar{E}''_{kj}(\eta) - \nabla^2 \bar{E}_{kj}(\eta) \right]. \tag{1.48}$$

Therefore, $\delta G_2^1$ and the difference between $\delta G_1^1$ and $\delta G_2^2$ which provide details of tensor mode evolution, are given by

$$\delta G_2^1 = \tfrac{1}{2} a^{-2} \left[ 2\mathcal{H} h'_+ + h''_+ - \nabla^2 h_+ \right] \tag{1.49}$$
$$\delta G_1^1 - \delta G_2^2 = a^{-2} \left[ 2\mathcal{H} h'_\times + h''_\times - \nabla^2 h_\times \right]. \tag{1.50}$$

Thus, we now have the left hand side of the perturbed *Einstein* Eqn.(1.42), for both the scalar and tensor perturbations. But, to get their complete evolution, we also need the right hand side of Eqn.(1.42), *i.e.*, the *perturbed energy momentum tensor*.

### 1.3.2 MATTER FLUCTUATIONS

The energy momentum tensor, when matter is described by perfect fluid, can be written as

$$T^\alpha_\beta = (\rho + P) u^\alpha u_\beta - P\delta^\alpha_\beta \tag{1.51}$$

where $\rho$, $P$ and $u^\alpha$ are energy density, pressure and four-fluid velocity respectively. The first order perturbation to the energy momentum tensor is found to be

$$\delta T^0_0 = \delta\rho, \ \delta T^0_i = (\rho_0 + P_0) a^{-1}\delta u_i, \ \delta T^i_j = -\delta P \delta^i_j \tag{1.52}$$

where $\rho_0$ and $P_0$ are the background values for $\rho$ and $P$ respectively. With the perturbed *Einstein* equations in hand, we can now proceed to investigate the evolution for different fluctuation modes.

But before proceeding further, we note that the Eqn.(1.42) is not *gauge invariant*. So, we construct *gauge invariant* quantities $\delta G^{GI\ \alpha}_{\ \ \ \beta}$ and $\delta T^{GI\ \alpha}_{\ \ \ \beta}$ corresponding to perturbed *Einstein* and energy momentum tensors respectively as,

$$\begin{aligned}
\delta G^{GI\ 0}_{\ \ \ 0} &\equiv \delta G^0_0 + G^{0\prime}_0 (B - E'); & \delta T^{GI\ 0}_{\ \ \ 0} &\equiv \delta T^0_0 + T^{0\prime}_0 (B - E') \\
\delta G^{GI\ i}_{\ \ \ j} &\equiv \delta G^i_j + G^{i\prime}_j (B - E'); & \delta T^{GI\ i}_{\ \ \ j} &\equiv \delta T^i_j + T^{i\prime}_j (B - E') \\
\delta G^{GI\ 0}_{\ \ \ i} &\equiv \delta G^0_i + \{G^0_i - \tfrac{1}{3}G^k_k\}\partial_i (B - E'); & \delta T^{GI\ 0}_{\ \ \ i} &\equiv \delta T^0_i + \{T^0_i - \tfrac{1}{3}T^k_k\}\partial_i (B - E').
\end{aligned} \tag{1.53}$$

From the spatial part of the Eqns.(1.44), (1.52) and (1.53) we find that

$$2\left[(2\mathcal{H}' + \mathcal{H}^2)\Phi + \mathcal{H}\Phi' + \Psi'' + 2\mathcal{H}\Psi' + \tfrac{1}{2}\nabla^2(\Phi - \Psi)\right]\delta^j_i + \\ 2(\mathcal{H}'' - \mathcal{H}\mathcal{H}' - \mathcal{H}^3)(B - E')\delta^j_i - \delta^{jk}\nabla_k\nabla_i(\Phi - \Psi) = 8\pi G a^2 \delta P^{GI}\delta^i_j \tag{1.54}$$

where $\delta P^{GI} \equiv \delta P + P'_0 (B - E')$ is the gauge invariant pressure perturbation. The absence of *non-diagonal* space-space components in the energy momentum tensor, *i.e.*, if there is no anisotropic stress then it follows from Eqn.(1.54) that, $\Phi = \Psi$. This is precisely the case where matter is described by a scalar field with canonical form of the action. As we are interested in inflationary cosmological perturbations, we will assume $\Phi = \Psi$ from now on.

To work out the perturbed energy momentum tensor, we start with the following energy momentum tensor for a scalar field having potential $V(\phi)$

$$T_{\mu\nu} = \partial_\mu\phi\partial_\nu\phi - g_{\mu\nu}\left[\tfrac{1}{2}g^{\alpha\beta}\partial_\alpha\phi\partial_\beta\phi + V(\phi)\right]. \tag{1.55}$$

After some simple algebra and using the background *Einstein* equations we obtain the following first order components for the energy momentum tensor

$$\begin{aligned}
a^2\delta T^0_0 &= -\phi'\delta\phi' - a^2\frac{dV}{d\phi}\delta\phi + \phi'^2 A \\
a^2\delta T^0_i &= -\phi'\partial_i\delta\phi \\
a^2\delta T^j_i &= -\delta^j_i\left(a^2\frac{dV}{d\phi}\delta\phi + \phi'^2 A - \phi'\delta\phi'\right)
\end{aligned} \tag{1.56}$$

where $\delta\phi$ is the fluctuation in the inflaton. From the last of the Eqns.(1.56) we see that for a scalar field there is no anisotropic stress in the energy momentum tensor as argued previously. The variable $\delta\phi$ being *gauge dependent* has no practical usage. So we construct a new variable, *gauge-invariant* scalar field fluctuation, $\delta\phi^{\mathcal{GI}}$, as

$$\delta\phi^{\mathcal{GI}} = \delta\phi + \phi'(B - E'). \tag{1.57}$$

Now, the Eqn.(1.42) can be written as

$$\delta G^{\mathcal{GI}}{}_{\mu\nu} = 8\pi G \delta T^{\mathcal{GI}}{}_{\mu\nu}, \tag{1.58}$$

which when combined with Eqns.(1.44) and (1.56), we obtain the following perturbed *Einstein* equations where matter is described by a scalar field

$$\begin{aligned}
\nabla^2 \Phi - 3\mathcal{H}\Phi' - (\mathcal{H}' + 2\mathcal{H}^2)\Phi &= 4\pi G \left( \delta\phi^{\mathcal{GI}}\phi' + \delta\phi^{\mathcal{GI}} \frac{\partial V}{\partial \phi} a^2 \right) \\
\Phi' + \mathcal{H}\Phi &= 4\pi G \delta\phi^{\mathcal{GI}}\phi' \\
\Phi'' + 3\mathcal{H}\Phi' + (\mathcal{H}' + 2\mathcal{H}^2)\Phi &= 4\pi G \left( \delta\phi^{\mathcal{GI}}\phi' - \delta\phi^{\mathcal{GI}} \frac{\partial V}{\partial \phi} a^2 \right).
\end{aligned} \tag{1.59}$$

Now, using background *Klein-Gordon* Eqn.(1.16), from the set of Eqns.(1.59) we can derive the following equation for the *Bardeen* potential $\Phi$

$$\Phi'' + 2\left(\mathcal{H} - \frac{\phi''}{\phi'}\right)\Phi' - \nabla^2\Phi + 2\left(\mathcal{H}' - \mathcal{H}\frac{\phi''}{\phi'}\right)\Phi = 0. \tag{1.60}$$

The last term in Eqn.(1.60) describes the force due to gravity leading to gravitational instability, the second last term is the pressure term which leads to oscillation and the second term is the *Hubble* friction term. A point to be noted that here we are considering only the *adiabatic modes*, no *entropy perturbation* has been incorporated, as a result the Eqn.(1.60) does not have any source term.

As the tensor modes do not couple to matter fluctuations they obey very simple dynamical equations which can be obtained from Eqns.(1.49), (1.50) and (1.58) by setting $\delta T^{\mathcal{GI}}{}_{\mu\nu} = 0$, and in the present context they are given by

$$h''_m + 2\mathcal{H}h'_m - \nabla^2 h_m = 0; \text{ where } m = +, \times. \tag{1.61}$$

So, both the tensor modes satisfy the same equation. The solutions of the Eqn.(1.61) are known as *gravity waves*.

Thus, for both the scalar and tensor perturbations everything can be reduced to the study of a single variable.

### 1.3.3 Co-moving Curvature Perturbations

The intrinsic spatial curvature $^3R$, on the $\eta = $ constant hyper-surfaces, is given by [91]

$$^3R = -\frac{4}{a^2}\nabla^2 C. \tag{1.62}$$

The scalar $C$ in the perturbed line element (1.38) generally referred to as the *curvature perturbation*. Since, $C$ is not *gauge invariant*, it is not useful for practical purposes. So, we define a new quantity $\mathcal{R}$ by

$$\mathcal{R} \equiv C - \mathcal{H}\frac{\delta\phi}{\phi'}, \tag{1.63}$$

which is known as the *co-moving curvature perturbation*. This quantity is *gauge invariant* and related to the *gauge* dependent curvature perturbation $C$ on a generic slicing to the inflaton perturbation $\delta\phi$ in that *gauge*. The physical meaning of $\mathcal{R}$ is clear from its definition (1.63), it represents the gravitational potential on the *co-moving* hyper-surfaces where $\delta\phi = 0$.

Another important quantity is the *curvature perturbation*, $\zeta$, on the *uniform energy density hyper-surfaces*, *i.e.*, no perturbation in the energy density, which is defined as

$$\zeta \equiv C - \mathcal{H}\frac{\delta\rho}{\rho'} = \mathcal{R} - \mathcal{H}\left[\frac{\delta\rho}{\rho'} - \frac{\delta\phi}{\phi'}\right]. \tag{1.64}$$

The meaning of $\zeta$ is that, it represents gravitational potential on the slices of *uniform energy density*, *i.e.*, $\delta\rho = 0$. This quantity $\zeta$ has utmost importance in inflationary cosmological perturbations, as it remains *conserved* outside the horizon for the *adiabatic* perturbations [98]. Also, on the *super horizon scales* $\zeta \approx \mathcal{R}$ for the *adiabatic* perturbations. So, the *co-moving curvature perturbation* is also conserved outside the horizon.

In order to derive the equation of motion for $\mathcal{R}$, we first define a new quantity $v$ – the so-called *Mukhanov-Sasaki* variable [92, 93]:

$$v \equiv a\delta\phi^{GI} + z\Phi, \text{ where } z \equiv a\frac{\phi'}{\mathcal{H}}. \tag{1.65}$$

This variable $v$ is related to $\mathcal{R}$ via the relation $v = -z\mathcal{R}$. With this new variable $v$, the Eqn.(1.60) can be re-expressed as

$$v'' - \left(\nabla^2 + \frac{z''}{z}\right)v = 0. \tag{1.66}$$

The above Eqn.(1.66) is known as *Mukhanov-Sasaki equation* and carries all the information about the scalar perturbations.

The tensor modes obey a similar looking equation, given by

$$u'' - \left(\nabla^2 + \frac{a''}{a}\right)u = 0 \tag{1.67}$$

which can be obtained from Eqn.(1.61) by defining $u \equiv \frac{M_P}{\sqrt{2}} a h_m$ where $m = +, \times$.

In order to solve the Eqns.(1.66) and (1.67), we require initial conditions for scalar and tensor modes. For this, quantization of the perturbation modes (both scalar and tensor) are required. But, the main motivation behind this quantization is, it is the *vacuum quantum fluctuations* that are the source of *classical cosmological perturbations*.

### 1.3.4 Quantizing Cosmological Perturbations

To quantize the cosmological perturbation, we need the associated action. In this case, the total action of the system, *i.e.*, the *Einstein-Hilbert* action with matter represented by a single scalar field minimally coupled to gravity, is given by

$$S = -\tfrac{1}{2} \int d^4x \, M_P^2 \sqrt{-g} R - \int d^4x \sqrt{-g} \left[ \tfrac{1}{2} \partial_\mu \phi \partial^\mu \phi + V(\phi) \right]. \tag{1.68}$$

In order to derive the equations of motion for the linear perturbations, it is necessary to expand the action upto the second order in linear perturbations. Since, the first order expansion $(S^{(1)})$ vanishes when combined with background equations of motion. The term in the second order expansion $(S^{(2)})$ is the piece we are interested in, the corresponding *Euler-Lagrange* equations provide the equations of motion for the linear perturbations. After a lengthy calculation the perturbative expansion of the above action upto the second order in the metric and scalar field fluctuations turns out to be [93],

$$S^{(2)} = \tfrac{1}{2} \int d^4x \underbrace{\left[ v'^2 - \delta^{ij} \partial_i v \partial_j v + \frac{z''}{z} v^2 \right]}_{\text{Scalar Part}} + \tfrac{1}{8} \int d^4x \, M_P^2 \, \underbrace{a^2 \left[ (\bar{E}_i^j)'(\bar{E}_j^i)' - \partial_k \bar{E}_i^j \partial^k \bar{E}_j^i \right]}_{\text{Tensor Part}}. \tag{1.69}$$

The first term on the right hand side is the scalar part and the second term represents the tensorial part. The action representing tensor perturbation only, can also be written in the following form

$$S_T^{(2)} = -\tfrac{1}{2} \sum_{m=+,\times} M_P^2 \int d^4x \, \frac{a^2}{2} \, g^{\mu\nu} \partial_\mu h_m \partial_\nu h_m. \tag{1.70}$$

With the substitution $u = \frac{1}{\sqrt{2}} M_P \, a h_m$, the action (1.69) can now be re-written as,

$$S^{(2)} = \tfrac{1}{2} \int d^4x \underbrace{\left[ v'^2 - \delta^{ij} \partial_i v \partial_j v + \frac{z''}{z} v^2 \right]}_{S_v} + \tfrac{1}{2} \int d^4x \underbrace{\left[ u'^2 - \delta^{ij} \partial_i u \partial_j u + \frac{a''}{a} u^2 \right]}_{S_u} \tag{1.71}$$

where $S_v$ and $S_u$ are scalar and tensor part of the second order action respectively. So the total second order action is sum total of two second order actions corresponding to $v$ and $u$, *i.e.*, $S^{(2)} \equiv S_v + S_u$. The equations of motion for the variables $u$ and $v$ are obtained by varying the above action (1.71) w.r.t. $u$ and $v$ respectively, and are given by

$$v'' - \left(\nabla^2 + \frac{z''}{z}\right)v = 0 \tag{1.72}$$

$$u'' - \left(\nabla^2 + \frac{a''}{a}\right)u = 0, \tag{1.73}$$

which are precisely the same as Eqns.(1.66) and (1.67). Now to canonically quantize the actions $S_v$ and $S_u$, we first construct the canonical momenta $\pi_v$ and $\pi_u$, canonically conjugate to $v$ and $u$ respectively,

$$\pi_v \equiv \frac{\partial \mathcal{L}_v}{\partial v'} = v', \ \pi_u \equiv \frac{\partial \mathcal{L}_u}{\partial u'} = u'. \tag{1.74}$$

So the corresponding *Hamiltonians* turn out to be

$$H_v \equiv \int d^3x \left(v'\pi_v - \mathcal{L}_v\right) = \tfrac{1}{2}\int d^3x \left[v'^2 + \delta^{ij}\partial_i v \partial_j v - \frac{z''}{z}v^2\right] \tag{1.75}$$

$$H_u \equiv \int d^3x \left(u'\pi_u - \mathcal{L}_u\right) = \tfrac{1}{2}\int d^3x \left[u'^2 + \delta^{ij}\partial_i u \partial_j u - \frac{a''}{a}u^2\right]. \tag{1.76}$$

In quantum theory $v$, $u$ and $\pi_u$, $\pi_v$ become operators $\hat{v}$, $\hat{u}$ and $\hat{\pi}_u$, $\hat{\pi}_v$ respectively. The quantum operators satisfy the following commutation relation on $\eta = $ *constant* hyper-surface,

$$[\hat{v}(\eta,\mathbf{x}),\hat{v}(\eta,\mathbf{x}')] = [\hat{u}(\eta,\mathbf{x}),\hat{u}(\eta,\mathbf{x}')] = [\hat{\pi}_v(\eta,\mathbf{x}),\hat{\pi}_v(\eta,\mathbf{x}')] = [\hat{\pi}_u(\eta,\mathbf{x}),\hat{\pi}_u(\eta,\mathbf{x}')] = 0$$

$$[\hat{v}(\eta,\mathbf{x}),\hat{\pi}_v(\eta,\mathbf{x}')] = [\hat{u}(\eta,\mathbf{x}),\hat{\pi}_u(\eta,\mathbf{x}')] = i\delta\left(\mathbf{x}-\mathbf{x}'\right), \tag{1.77}$$

where $\delta(\mathbf{x})$ is the usual *Dirac* delta function. The operators $\hat{v}, \hat{u}$ in the *Fourier* space become

$$\hat{v}(\eta,\mathbf{x}) = \frac{1}{(2\pi)^{3/2}}\int d^3k \left[\hat{a}_k v_k(\eta)e^{i\mathbf{k}.\mathbf{x}} + \hat{a}_k^\dagger v_k^*(\eta)e^{-i\mathbf{k}.\mathbf{x}}\right]$$

$$\hat{u}(\eta,\mathbf{x}) = \frac{1}{(2\pi)^{3/2}}\int d^3k \left[\hat{b}_k u_k(\eta)e^{i\mathbf{k}.\mathbf{x}} + \hat{b}_k^\dagger u_k^*(\eta)e^{-i\mathbf{k}.\mathbf{x}}\right], \tag{1.78}$$

where the creation and annihilation operators $\hat{a}_k^\dagger, \hat{b}_k^\dagger$ and $\hat{a}_k, \hat{b}_k$ satisfy the usual commutation relations

$$[\hat{a}_{\vec{k}},\hat{a}_{\vec{k}'}] = [\hat{b}_{\vec{k}},\hat{b}_{\vec{k}'}] = [\hat{a}_{\vec{k}}^\dagger,\hat{a}_{\vec{k}'}^\dagger] = [\hat{b}_{\vec{k}}^\dagger,\hat{b}_{\vec{k}'}^\dagger] = 0 \tag{1.79}$$

$$[\hat{a}_{\vec{k}},\hat{a}_{\vec{k}'}^\dagger] = [\hat{b}_{\vec{k}},\hat{b}_{\vec{k}'}^\dagger] = i\delta(\vec{k}-\vec{k}'). \tag{1.80}$$

The equations of motion for the $k^{th}$ *Fourier* mode of the quantum fields $\hat{v}, \hat{u}$ are respectively given by

$$v_k'' + \left(k^2 - \frac{z''}{z}\right)v_k = 0 \tag{1.81}$$

$$u_k'' + \left(k^2 - \frac{a''}{a}\right)u_k = 0. \tag{1.82}$$

The above Eqns.(1.81) and (1.82) are nothing but the *Harmonic Oscillators* with time dependent frequencies. The solution of the first Eqn.(1.81) determines the *co-moving curvature perturbation* whereas the solution of Eqn.(1.82) are the *gravitational waves*. In Chapter 3 we derive the *wave-functions* for the system of cosmological perturbations, by solving the associated *Schrödinger* equation, together with a cosmological analogue of the *Berry phase* [99].

From Eqns.(1.78), (1.79) and (1.80) we obtain the following normalizations for the *Wronskian*s of Eqns.(1.81) and (1.82):

$$v_k v_k^{\star\prime} - v_k^\star v_k' = i; \quad u_k u_k^{\star\prime} - u_k^\star u_k' = i. \tag{1.83}$$

For the *slow-roll quasi-exponential* inflation, $\frac{z''}{z} \simeq \frac{a''}{a} \simeq \frac{2}{\eta^2}$, then the Eqns.(1.81) and (1.82) can be solved exactly to get

$$v_k = c_1 e^{-ik\eta}\left(1 - \frac{i}{k\eta}\right) + c_2 e^{ik\eta}\left(1 + \frac{i}{k\eta}\right) \tag{1.84}$$

$$u_k = c_3 e^{-ik\eta}\left(1 - \frac{i}{k\eta}\right) + c_4 e^{ik\eta}\left(1 + \frac{i}{k\eta}\right). \tag{1.85}$$

To completely specify the solutions, we need to supply initial conditions for $v_k$, $u_k$ and $v_k'$, $u_k'$, at some initial time $\eta = \eta_I$. Since, the *co-moving Hubble* radius shrinks during inflation, the wavelength corresponding to a given mode can be found within the *Hubble* radius if we go sufficiently back in time. Also, for wavelengths smaller than the *Hubble* radius, we can neglect the curvature effects and the modes can be treated as if they are in the *Minkowski* space-time. Therefore the natural choice for the particular solution is to take the solution that corresponds to the usual *Minkowski vacuum*, i.e.,

$$\lim_{\eta \to -\infty} v_k = \frac{e^{-ik\eta}}{\sqrt{2k}}, \quad \lim_{\eta \to -\infty} u_k = \frac{e^{-ik\eta}}{\sqrt{2k}}. \tag{1.86}$$

The above choice for the initial condition is called the *Bunch-Davies vacuum* [100, 101]. The choice of the vacuum (1.86) together with (1.83), determine the complete solution of Eqn.(1.81) and (1.82), which are given by

$$v_k = \frac{e^{-ik\eta}}{\sqrt{2k}}\left(1 - \frac{i}{k\eta}\right), \; u_k = \frac{e^{-ik\eta}}{\sqrt{2k}}\left(1 - \frac{i}{k\eta}\right). \tag{1.87}$$

Now we can compute the *correlation functions* for the operators $\hat{v}$ and $\hat{u}$ in the *Bunch-Davies vacuum*. The *Fourier transformations* of the correlation functions define the *power spectra* $P_v(k)$, $P_u(k)$ as,

$$\langle 0|\hat{v}(\vec{x_1})\hat{v}(\vec{x_2})|0\rangle = \int d^3k \, e^{i\vec{k}\cdot(\vec{x_1}-\vec{x_2})}\frac{P_v(k)}{4\pi k^3} \tag{1.88}$$

$$\langle 0|\hat{u}(\vec{x_1})\hat{u}(\vec{x_2})|0\rangle = \int d^3k \, e^{i\vec{k}\cdot(\vec{x_1}-\vec{x_2})}\frac{P_u(k)}{4\pi k^3}. \tag{1.89}$$

The assumptions of *homogeneity* and *isotropy* lead to the power spectra which are diagonal in the *Fourier space* and depend only upon the magnitude of $k$. The *dimensionless* power spectra for the *co-moving curvature perturbation* and the *gravity waves* are then defined as:

$$P_{\mathcal{R}} \equiv \frac{k^3}{2\pi^2}|\mathcal{R}|^2 = \frac{k^3}{2\pi^2}\left|\frac{v}{z}\right|^2 \qquad (1.90)$$

$$P_{\mathcal{T}} \equiv \frac{k^3}{2\pi^2}|\mathcal{T}|^2 = 2 \times \frac{k^3}{2\pi^2}\frac{2}{\mathrm{M}_P^2}\left|\frac{u}{a}\right|^2. \qquad (1.91)$$

These power spectra are most important findings of our analysis, as they can be directly predicted from the observational data. I shall now briefly discuss about the present status of different observable parameters in the context of inflation.

### 1.3.5 Inflationary Observables

So far we have theoretically demonstrated how to calculate the spectra for scalar and tensor fluctuations generated during inflation. But to compare among different classes of inflationary models, confrontations with the latest observational data are essential. The main idea is that one can predict precisely the statistics of the CMB anisotropies, once the amplitude of the primordial curvature perturbation as a function of scale, $\mathcal{R}(k)$, is supplied, provided we have some choice for other involved cosmological parameters.

The most important quantity is the dimensionless power spectrum, $P_{\mathcal{R}}$, for the *co-moving curvature perturbation*, $\mathcal{R}$, defined in Eqn.(1.90). Upto the first order in *slow-roll* parameters, the expression for $P_{\mathcal{R}}$ turns out to be

$$P_{\mathcal{R}}(k) = \frac{1}{24\pi^2 \mathrm{M}_P^4}\left(\frac{V}{\epsilon_H}\right). \qquad (1.92)$$

From *WMAP* nine years data, the present observational bound for $P_{\mathcal{R}}$ is: $10^9 P_{\mathcal{R}} = 2.464 \pm 0.072$ at $k = 0.002$ MPc$^{-1}$ [5, 47]. From *Planck* the best fit value for this parameter is: $P_{\mathcal{R}} = 2.2 \times 10^{-9}$ at $k = 0.05$ MPc$^{-1}$ [4, 102].

Another important quantity in the context of scalar perturbations is the scalar *spectral index* $n_s$, which measures the scale dependence of the power spectrum of *co-moving curvature perturbation* at the time of *horizon crossing, i.e.,* at $k = aH$, and is defined by

$$n_s(k) = 1 + \frac{d\ln P_{\mathcal{R}}}{d\ln k}\Big|_{k=aH} \simeq 1 - 4\epsilon_H + 2\eta_H. \qquad (1.93)$$

Current bound on $n_s$ at $k = 0.002$ MPc$^{-1}$ is: $n_s = 0.9608 \pm 0.008$ [5, 47] (the bound from *Planck* is $n_s = 0.9603 \pm 0.0073$ [4, 102]). The remaining one is the *running* of the scalar *spectral index* $n'_s$, which measures the scale dependence of the *spectral index* itself and has been found to be

$$n'_s(k) \equiv \frac{dn_s}{d\ln k}\Big|_{k=aH} \simeq -2\xi_V + 16\epsilon_V \eta_V - 24\epsilon_V^2, \qquad (1.94)$$

where $\xi_V \equiv M_P^4 \frac{V'(\phi)V'''(\phi)}{V^4}$. The present bound on this parameter is provided by *WMAP* [5, 47]: $n'_S = -0.023 \pm 0.011$ (from *Planck* we have $n'_S = -0.0134 \pm 0.009$ [4, 102]). The above estimates for the observational parameters have been calculated in [47] from the *WMAP–9* data setting the *pivot scale* at $k = 0.002\,\mathrm{MPc}^{-1}$ and the *slow-roll* parameters are evaluated at the time of *horizon exit*.

One can define similar quantities for the tensor perturbations also. The power spectrum for the gravity waves has been defined in Eqn.(1.91). The corresponding observational quantity is the tensor to scalar power spectrum ratio, $r$, which upto to the first order in *slow-roll* parameters is found to be

$$r = \frac{P_T|_{k=aH}}{P_\mathcal{R}|_{k=aH}} \simeq 16\epsilon_H. \tag{1.95}$$

Since, the primordial gravity waves are yet to be detected there is no precise estimate for $r$, though *WMAP* has predicted an upper limit: $r < 0.13$ (95% Confidence Level (C.L.) ) at $k = 0.002\,\mathrm{MPc}^{-1}$ [47], while recent data from *Planck* when combined with *WMAP* nine year data provides $r < 0.11$ (95% C.L.) [4, 102]. The tensor *spectral index* and its *running* are given by

$$n_T = \frac{d\ln P_T}{d\ln k}\Big|_{k=aH} \simeq -2\epsilon_H \tag{1.96}$$

$$n'_T \equiv \frac{dn_T}{d\ln k}\Big|_{k=aH} \simeq 4\epsilon_H \eta_H - 4\epsilon_H^2. \tag{1.97}$$

The predicted observational constraint on $n_T$ by *WMAP* team is: $n_T > -0.048$ (95% C.L.) at $k = 0.002\,\mathrm{MPc}^{-1}$, which has been calculated assuming $r > 0$ and $n'_T = 0$ [47].

From the Eqns.(1.95) and (1.96) we obtain a simple relation between tensor to scalar power ratio, $r$, and tensor spectral index, $n_T$, written explicitly, it looks

$$r = -8n_T. \tag{1.98}$$

Eqn.(1.98) is known as the *consistency relation* [49, 103]. This relation is obtained by considering the *slow-roll approximation* and only taking into account first order of *slow-roll* parameters. The *consistency relation* gets modified if we go beyond first order *slow-roll* approximation [104]. The relation (1.98) gets modified in the context of brane inflation [105, 106], and non-standard models of inflation [107] where generalized propagation speed (less than one) of the scalar field fluctuations relative to the homogeneous background have been considered. Generally, the observable quantities at the time of horizon crossing is evaluated by adopting perfect *de-Sitter* approximation without taking into account the effect of scalar field evolution. But, if do so the *consistency relation* takes a modified look [73, 108], which we shall discuss in Chapter 2.

## 1.4 Super Hubble Evolution of the Fluctuation Modes

The fluctuations generated during inflation left the horizon owing to the fact that, the *co-moving Hubble* radius was shrinking, whereas *co-moving* wavelengths of the perturbations remained constant. After inflation is over, the *co-moving horizon* grows, as a consequence all the modes started to *re-enter* the *horizon*, which can be understood from `Fig.1.3`.

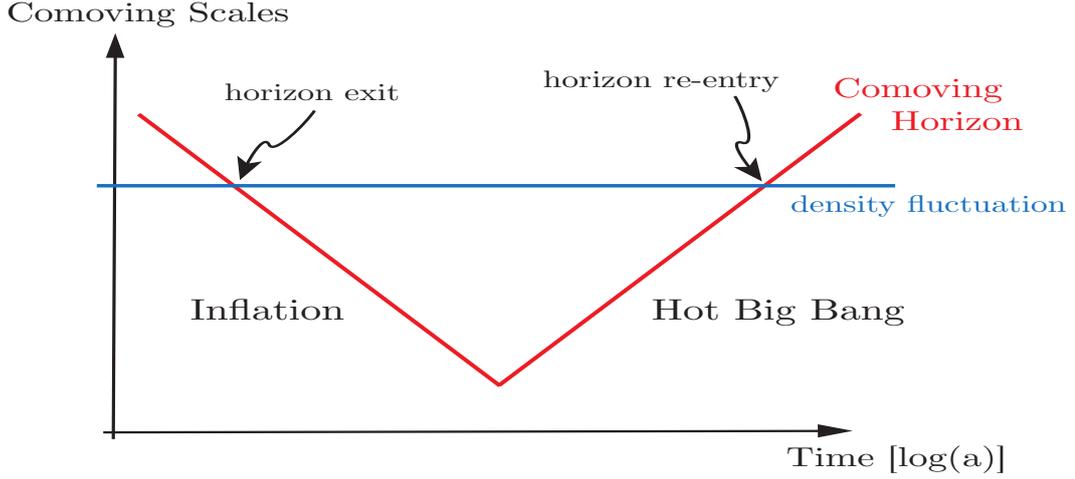

**FIG. 1.3:** *Evolution of the co-moving Hubble radius $\frac{1}{aH}$ and the co-moving wavelength (blue solid line) of the perturbations. The co-moving Hubble radius decreases first during inflation but increases after, while co-moving scales remain constant. So the perturbation modes begin to re-enter the horizon as soon as inflation is over. Figure adopted from [62].*

The constancy of $\mathcal{R}$ outside the horizon for *adiabatic* perturbations [98] allows us predict cosmological observables. After horizon re-entry, $\mathcal{R}$ determines the perturbations to the cosmic fluid observed in CMB and in the large scale structure of the matter. Here we shall only deal with *super-Hubble* evolution of the *adiabatic* perturbations during matter domination. To proceed further, we now consider the energy-momentum tensor in the following form,

$$T^{\mu\nu} = (\rho + P)u^\mu u^\nu + Pg^{\mu\nu} \qquad (1.99)$$

where $u^\mu \equiv (1, v^i)$ is the *four velocity* of the fluid and $v^i$ is the associated velocity perturbation. Therefore, for $\Phi = \Psi$, the perturbed *Einstein* Eqns.(1.59), with the difference that we have now cosmological fluid instead of scalar field, are given by

$$\begin{aligned}
\nabla^2 \Phi - 3\mathcal{H}\Phi' - 3\mathcal{H}^2\Phi &= 4\pi G a^2 \delta\rho^{\mathcal{GI}} \\
\partial_i (\Phi' + \mathcal{H}\Phi) &= 4\pi G a^2 (\rho + P) v_i^{\mathcal{GI}} \\
\Phi'' + 3\mathcal{H}\Phi' + (2\mathcal{H}' + \mathcal{H}^2)\Phi &= 4\pi G a^2 \delta P^{\mathcal{GI}}.
\end{aligned} \qquad (1.100)$$

For *adiabatic* perturbations we can always write $\delta P = c_s^2 \delta \rho$, where $c_s$ is the speed of sound. Therefore combining Eqns.(1.100) we can obtain a simple equation

$$\Phi'' + 3\mathcal{H}\left(1 + c_s^2\right)\Phi' - c_s^2 \nabla^2 \Phi + \left[2\mathcal{H}' + \left(1 + c_s^2\right)\mathcal{H}^2\right]\Phi = 0. \qquad (1.101)$$

The right hand side vanishes as we did not take into account *entropy perturbation*. Now by defining the following *gauge invariant* variable [93, 109, 110]

$$\zeta \equiv -\left(\frac{2}{3}\frac{\mathcal{H}^{-1}\Phi' + \Phi}{1 + \omega} + \Phi\right), \tag{1.102}$$

in *super-Hubble* regime, which is equivalent to neglect the term $c_S^2 \nabla^2 \Phi$ in Eqn.(1.101), we can write

$$\zeta' = 0. \tag{1.103}$$

On the *co-moving gauge*, *i.e.*, where $v = B = 0$, $\zeta$ is the curvature perturbation. In the *Fourier* space, using the definition (1.102) and *super-Hubble* equivalence of $\zeta$ and $\mathcal{R}$, we have

$$\Phi_k = -\frac{3 + 3\omega}{5 + 3\omega}\zeta_k \simeq -\frac{3 + 3\omega}{5 + 3\omega}\mathcal{R}_k, \tag{1.104}$$

where $\omega$ has been treated as a constant. Therefore, the fluctuation in the gravitational potential, $\Phi$, is also conserved on the *super-Hubble* regime. Further, if initially the equation of state is given by the constant $\omega_I$ and later changes to another constant $\omega_F$, then [111]

$$\Phi_{k,F} \approx \frac{1 + \omega_F}{1 + \omega_I}\frac{5 + 3\omega_I}{5 + 3\omega_F}\Phi_{k,I}. \tag{1.105}$$

Therefore, during the transition from radiation ($\omega = 1/3$) to matter ($\omega = 0$) domination, the potential $\Phi_{k,I}$ changes to $\Phi_{k,F} = \frac{9}{10}\Phi_{k,I}$, on the *super-Hubble* scales.

The matter density fluctuation can be deduced from Eqns.(1.100) and using the background *Friedmann* Eqn.(1.4) for $\kappa = 0$ we obtain

$$3\mathcal{H}\Phi'_k + 3\mathcal{H}^2\Phi_k + k^2\Phi_k = -\tfrac{3}{2}\mathcal{H}^2\delta_m(k). \tag{1.106}$$

where $\delta_m \equiv \delta\rho^{GI}/\rho_m$ is the matter density contrast. So given a specific form for the primordial curvature perturbation, $\mathcal{R}_k$, we can determine the fluctuations in matter. Now, during matter domination ($\delta P^{GI} = 0$) $\mathcal{H} = 2/\eta$. Consequently, from the last equation of (1.100), in the *super-Hubble* regime, we get

$$\Phi_k = C_1 + C_2\eta^{-5}, \tag{1.107}$$

where $C_1$, $C_2$ are independent of time but may have spatial dependences. Ignoring the decaying part we can now conclude that $\Phi_k$ does not evolve with time on the *super-horizon* scales. As a consequence from Eqn.(1.106), for $k \ll \mathcal{H}$, we have

$$\delta_m(k) = -2\Phi_k = \tfrac{6}{5}\mathcal{R}_k. \tag{1.108}$$

This relation shows that perturbation with a given wavelength generated during inflation evolves with a slightly rescaled amplitude on the *super-horizon* scales. Not only that, when a given wavelength re-enters the horizon, the amplitude of the gravitational potential depends upon the time of re-entry. Therefore, given an early *universe* inflationary model, we can calculate the *co-moving curvature* perturbation and its impact on matter density contrast therefrom. In other words, measurement of the matter density fluctuations helps us predict correct amplitude of the perturbation generated during inflation. I shall come back to this in a later chapter. Let's now move on to the most powerful and engrossing source that caries the best information about the early Universe – the *cosmic microwave background radiation*.

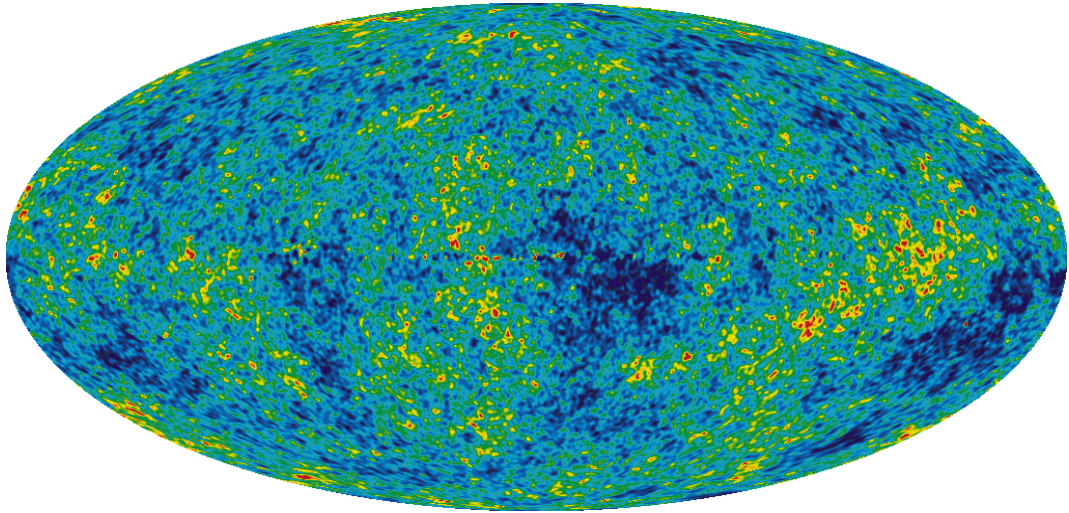

**Fig. 1.4:** *Fluctuations in the CMB temperature, as measured by WMAP, about the average temperature of 2.725 K. Figure taken from [62].*

## 1.5 The Cosmic Microwave Background

The *Cosmic Microwave Background* [112, 113], which is an *electromagnetic-radiation* filling our Universe, provides a *snapshot* of the Universe when it was about 400,000 years old. At that time, baryons and photons were tightly coupled and all the electrons were free. Due to the scattering from these electrons the mean free path of photons was very short compared to the cosmological distance scales, as a result the Universe was opaque. Later, when the plasma became cool enough to produce helium and hydrogen atoms the photons were released and Universe finally turned transparent. This phenomenon is known as *recombination* and the surface corresponding to that redshift is called the *last scattering surface* (LSS). After recombination, the primordial radiation does not interact with matter and photons come to us without being further scattered. The CMB is very close to isotropic with anisotropies being of the order of $10^{-5}$, so we conclude that at the time of last scattering our Universe was almost homogeneous and isotropic.

The small anisotropies in CMB are observed as the angular fluctuations of its temperature [2, 3, 4, 5, 114, 115]. Since, outside the horizon the primordial curvature perturbations remain frozen [98], the measurement of CMB temperature fluctuations at angular scales $> 1°$, which correspond to the *super-Hubble* size at *recombination*, directly probes the primordial spectrum. On the other hand, temperature fluctuations on the scales $< 1°$ are very sensitive to various cosmological parameters. So, precise measurements of small scale temperature fluctuations help us determine those parameters.

Apart from temperature anisotropies, CMB is also linearly polarized which was first detected by *Degree Angular Scale Interferometer* (DASI) [116]. So, CMB provides two additional observable parameters beside temperature anisotropy in the form of linear polarizations. The CMB temperature fluctuations and polarizations are directly related to various cosmological parameters and also to the spectrum of primordial perturbations generated during inflation.

In this section, I shall mostly discuss CMB temperature fluctuations broadly following the trail as given in [73, 117].

### 1.5.1 Temperature Fluctuations

The *last scattering surface* is spherical around us, so it is convenient to expand the temperature field into *spherical harmonics*, $Y_{lm}$, as

$$\frac{\delta T(\hat{\boldsymbol{n}})}{\bar{T}} = \sum_{\ell m} T_{\ell m} Y_{\ell m}(\hat{\boldsymbol{n}}) \tag{1.109}$$

where $\bar{T}$ is the background temperature of CMB, $\hat{\boldsymbol{n}}$ is the unit vector on the surface of the last scattering sphere and $\delta T(\hat{\boldsymbol{n}}) \equiv T(\hat{\boldsymbol{n}}) - \bar{T}$, is the temperature fluctuation in the direction of $\hat{\boldsymbol{n}}$. Using the orthonormal property of the *spherical harmonics* we get

$$T_{\ell m} = \int d\hat{n} \frac{\delta T(\hat{n})}{\bar{T}} Y_{\ell m}^*(\hat{\boldsymbol{n}}). \tag{1.110}$$

The angular power spectrum for the temperature fluctuations in CMB is then defined as

$$\langle T_{\ell m} T_{\ell' m'}^* \rangle = \delta_{mm'} \delta_{\ell\ell'} C_\ell^{TT}. \tag{1.111}$$

These $C_\ell^{TT}$ contain all the informations about CMB temperature anisotropies, provided the fluctuations are *Gaussian*.

There are mainly two reasons for temperature fluctuations in CMB – inhomogeneities present at LSS and redshift variations of the photons during their journey from LSS to the observer, which has been roughly shown in `Fig.1.5`.

Before decoupling, frequent photon collisions kept the photon distribution isotropic in its rest frame. As a result, an observer just after recombination sees no anisotropy in CMB except for monopole due to fluctuations in photon energy density $\delta_\gamma \equiv \frac{\rho_\gamma - \bar{\rho}}{\bar{\rho}_\gamma}$ at LSS and a dipole which corresponds to relative velocity $\mathbf{v}_\gamma$ of the CMB with respect to the observer. So, in the *sudden*

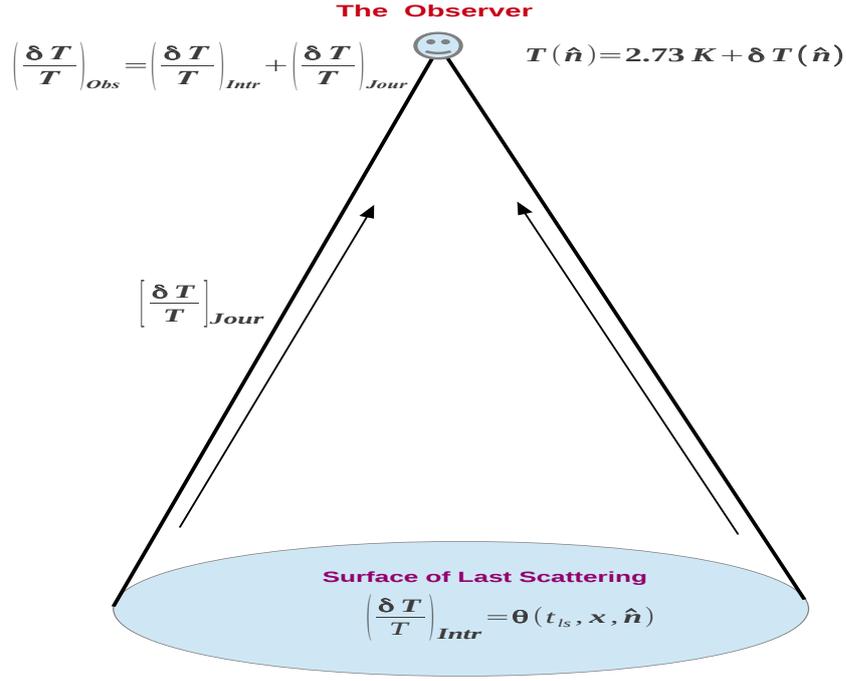

**FIG. 1.5:** *The CMB anisotropies:– intrinsic, i.e., anisotropy present at LSS and anisotropy that arises as photons travel from the LSS to us.*

*decoupling approximation, i.e.*, considering decoupling to be an instantaneous process which is equivalent to snub the effect of finite thickness of LSS, the temperature fluctuation just after decoupling turns out to be [81]

$$\Theta(\eta, \mathbf{x}, \hat{\boldsymbol{n}})|_{\text{ls}} \equiv \frac{\delta T(\eta, \mathbf{x}, \hat{\boldsymbol{n}})}{\bar{T}(\eta)}|_{\text{ls}} = \left(\tfrac{1}{4}\delta_\gamma + \hat{\boldsymbol{n}}.\mathbf{v}_\gamma\right)_{\text{ls}}, \qquad (1.112)$$

where $\hat{\boldsymbol{n}}$ is the unit vector along the direction of photon propagation and the subscript "ls" stands for that the quantity is evaluated at the time of *last scattering*. During their journey from LSS to observer, the fractional temperature fluctuation in CMB, $\Theta(\eta, \mathbf{x}, \hat{\boldsymbol{n}})$, gets another contribution from the *Sachs-Wolfe* effect [118], $\Theta(\eta, \mathbf{x}, \hat{\boldsymbol{n}})_{\text{SW}}$. So, the total CMB anisotropy in the sudden decoupling approximation is given by

$$\Theta(\eta, \mathbf{x}, \hat{\boldsymbol{n}}) = \left(\tfrac{1}{4}\delta_\gamma + \hat{\boldsymbol{n}}.\mathbf{v}_\gamma\right)_{\text{ls}} + \Theta(\eta, \mathbf{x}, \hat{\boldsymbol{n}})_{\text{SW}}. \qquad (1.113)$$

### A. Large Scale Anisotropy: Sachs-Wolfe Effect

The large scale ($\theta \gg 1°$) CMB anisotropies, *i.e.*, where $\ell \ll 180$, are induced by the perturbations having wavelengths exceeding the *Hubble radius* at LSS and they did not get enough time to evolve from their *primordial state*. Consequently, measurement of the large scale temperature fluctuations provides pristine information about the primordial perturbations.

The large scale CMB temperature anisotropies are dominated by the *Sachs-Wolfe* effect [118]. By defining $q(\mathbf{x}, \eta) \equiv a(\eta)\hat{p}(\mathbf{x}, \eta)$, $\hat{p}$ being photon 4-momentum, the *Sachs-Wolfe* effect in the sudden decoupling approximation turns out to be

$$\Theta_{\text{SW}}(\hat{\bm{n}}) \equiv \int_{\eta_{\text{ls}}}^{\eta_0} \frac{dq}{q} \;=\; \Psi_{\text{ls}} - \Psi_0 + \int_{\eta_{\text{ls}}}^{\eta_0} \frac{\partial}{\partial \eta}(\Psi + \Phi) d\eta = \Psi_{\text{ls}} + \int_{\eta_{\text{ls}}}^{\eta_0} \frac{\partial}{\partial \eta}(\Psi + \Phi) d\eta \quad (1.114)$$

where $\hat{\bm{n}}$ is the unit vector along the line of sight and $\eta_0$ is the present value of the *conformal time*. The term $\Psi_0$ has been dropped as it only affects the unobservable monopole. The contribution from the integral is known as *integrated Sachs-Wolfe effect*. Now the integral in Eqn.(1.114) is non-zero only if the gravitational potentials are time dependent. We assume that Universe is completely matter dominated from LSS to present, so that $\frac{\partial \Psi}{\partial \eta} = \frac{\partial \Phi}{\partial \eta} = 0$, and there will not be any *integrated Sachs-Wolfe effect*. Therefore, the total *Sachs-Wolfe* contribution, considering negligible *integrated Sachs-Wolfe* effect and $\Phi = \Psi$, is

$$\Theta_{\text{SW}}(\hat{\bm{n}}) = \Phi_{\text{ls}} \quad (1.115)$$

and the total CMB temperature anisotropy assuming complete matter domination at last scattering in the sudden decoupling approximation turns out to be

$$\Theta(\hat{\bm{n}}) = \left[ \left( \tfrac{1}{4}\delta_\gamma + \Phi \right) + \hat{\bm{n}}.\mathbf{v}_\gamma \right]_{\text{ls}}. \quad (1.116)$$

Now, for *adiabatic* perturbations $\frac{1}{4}\delta_\gamma = \frac{1}{3}\delta_m = -\frac{2}{3}\Phi$, and in this case the velocity perturbation, $v_\gamma$, at *super-horizon* scales is negligible. So, using Eqn.(1.108) we finally get total CMB anisotropy at large angular scales as,

$$\Theta(\hat{\bm{n}}) \;=\; \left( \tfrac{1}{4}\delta_\gamma + \Phi \right)_{\text{ls}} = \tfrac{1}{3}\Phi_{\text{ls}} = -\tfrac{1}{5}\mathcal{R}_{\text{ls}}. \quad (1.117)$$

This result is known as the *Sachs-Wolfe effect*, which is basically independent of cosmological parameters and only depends on the *primordial curvature perturbation*. The above result is indeed a very good approximation for the CMB anisotropies upto $\ell \sim 30$ or so. The corresponding spectrum is given by

$$\begin{aligned} C_\ell^{\textbf{SW}} &= \frac{4\pi}{25} \int_0^\infty \frac{dk}{k} j_\ell^2(k\eta_0) \underbrace{P_\mathcal{R}(k)}_{\text{Inflation}} \\ &\simeq \frac{4\pi}{25} \frac{1}{\ell(\ell+1)} P_\mathcal{R}\left( \ell/\eta_0 \right), \end{aligned} \quad (1.118)$$

where $j_\ell$ is *spherical Bessel* function of oder $\ell$ and in the second line we have neglected the scale dependence of $P_\mathcal{R}$. This is the most dominant contribution to the CMB angular power spectrum for angular separations $\theta \gg 1°$ or $\ell \ll 180$, which correspond to *super-horizon* scales at *last scattering surface*.

### B. Small Scale Anisotropy: Baryon Acoustic Oscillation

In the smaller scales, *i.e.*, for $\ell > 180$, the main contribution to $C_\ell$ comes from the perturbations which have entered the *horizon* before *recombination*. As a result, they have evolved significantly from their *primordial state* before we observe them.

The *Fourier* transform of Eqn.(1.116) gives

$$\Theta(\hat{\bm{n}}) = \frac{1}{(2\pi)^{3/2}} \int \left[\left\{\tfrac{1}{4}\delta_\gamma(\mathbf{k}) + \Phi(\mathbf{k})\right\} + \hat{\bm{n}}.\mathbf{v}_\gamma(\mathbf{k})\right]_{\text{ls}} e^{\mathbf{k}.\hat{\bm{n}}\eta_0} d^3k \qquad (1.119)$$

where we have used $\mathbf{x}_{\text{ls}} \equiv (\eta_0 - \eta_{\text{ls}})\hat{\bm{n}} \approx \eta_0\hat{\bm{n}}$. Taking into account the following relation, between *3-dimensional* plane waves and spherical harmonics,

$$e^{i\mathbf{k}.\mathbf{x}} = 4\pi \sum_{\ell' m'} i^{\ell'} j_\ell(kx) Y_{\ell'm'}(\mathbf{x}) Y^*_{\ell'm'}(\mathbf{k}), \qquad (1.120)$$

we get the following expression for the CMB multipoles [81]:

$$\Theta_\ell(\mathbf{k}) = \left[\tfrac{1}{4}\delta_\gamma(\eta_{\text{ls}}, \mathbf{k}) + \Phi(\eta_{\text{ls}}, \mathbf{k})\right] j_\ell(k\eta_0) + \mathbf{v}_\gamma(\eta_{\text{ls}}, \mathbf{k}) \frac{dj_\ell(k\eta_0)}{d(k\eta_0)}. \qquad (1.121)$$

In the following we shall derive an approximate analytical expression for $\Theta_\ell$.

In the tight coupling approximation, the photon-baryon system can be characterized by a single fluid velocity $\mathbf{v}_b = \mathbf{v}_\gamma \equiv \mathbf{v}$. The gravitational attraction of the over-dense regions causes the photon-baryon fluid to fall into their gravitational wells. However, this increases the radiation pressure, as a result fluid starts to oscillate. This phenomena dominates CMB anisotropies on the smaller scales, $\theta < 1°$. The acoustic oscillation of the baryon-photon fluid is governed by the following equation [81, 97]

$$\tfrac{1}{4}\delta''_{\gamma,k} + \tfrac{1}{4}\frac{\mathsf{R}'}{1+\mathsf{R}}\delta'_{\gamma,k} + \tfrac{1}{4}k^2 c_s^2 \delta_{\gamma,k} = -\tfrac{1}{3}k^2 \Phi_k(\eta) + \frac{\mathsf{R}'}{1+\mathsf{R}}\Phi'_k(\eta) + \Phi''_k(\eta) \qquad (1.122)$$

where $\mathsf{R} \equiv \frac{3\rho_b}{4\rho_\gamma}$, is the ratio of baryon to photon energy density and $c_s^2 = \frac{1}{3(1+\mathsf{R})}$, is the square of the sound speed in the baryon-photon fluid. In tight coupling approximation $\mathsf{R}$ changes very slowly so that $c_s$ is almost constant and we may ignore the small damping of the oscillation in (1.122). Assuming complete matter domination at LSS, so that $\Phi_k$ is constant, the general solution of Eqn.(1.122) is given by

$$\tfrac{1}{4}\delta_{\gamma,k} = -(1+\mathsf{R})\Phi_k + A_k \cos(kr_s) + B_k \sin(kr_s) \qquad (1.123)$$

where $r_s \equiv \int_0^\eta c_s(\eta) d\eta$, is the *co-moving* size of the sound horizon and $A_k$, $B_k$ are constants. It is easy to check that $A_k = \tfrac{1}{2}\Phi_k \equiv \tfrac{1}{2}T_k^0 \Phi_k^0$, where $\Phi_k^0 = -\tfrac{2}{3}\mathcal{R}|_{k=1/\eta}$ are the modes which have reentered the horizon during radiation domination and $T_k^0$ is the *transfer function* associated with them, and $B_k = 0$. So, the particular solution is now given by

$$\tfrac{1}{4}\delta_{\gamma,k} = -(1+\mathsf{R})\Phi_k + \tfrac{1}{2}T_k^0\Phi_k^0\cos(kr_s). \tag{1.124}$$

In order to obtain photon velocity perturbation, $v_{\gamma,k}$, we differentiate (1.124) w.r.t. conformal time $\eta$ and use the relation $\tfrac{1}{4}\delta'_{\gamma,k} = -\tfrac{1}{3}\,kv_{\gamma,k} + \Phi'_k$ to get

$$v_{\gamma,k} = \tfrac{3}{2}c_S T_k^0 \Phi_k^0 \sin(kr_s). \tag{1.125}$$

In the smaller scales, photon diffusion lowers the inhomogeneities, which is known as *Silk damping* [119]. Taking this into account, the acoustic oscillation at LSS turns out to be

$$\begin{aligned}
\tfrac{1}{4}\delta_\gamma(\eta_{\mathrm{ls}},k) &= -(1+\mathsf{R})\Phi(\eta_{\mathrm{ls}},k) + \tfrac{1}{2}T_k^0\Phi_k^0 \left[e^{-k^2/k_D^2}\cos(kr_S)\right]_{\eta=\eta_{\mathrm{ls}}} \\
v_\gamma(\eta_{\mathrm{ls}},k) &= \tfrac{3}{2}c_S T_k^0 \Phi_k^0 \left[e^{-k^2/k_D^2}\sin(kr_S)\right]_{\eta=\eta_{\mathrm{ls}}}
\end{aligned} \tag{1.126}$$

where $k_D^{-2} \equiv \left[\tfrac{2}{5}\int_0^\eta c_S^2 \tfrac{\tau_\gamma}{a} d\eta\right]$, defines the *Silk damping* scale and $\tau_\gamma$ is the mean free time of photon due to *Thompson* scattering.

So far, we have worked in the tight coupling approximation by neglecting the thickness of LSS which is an idealistic assumption. Now, if we include finite thickness of LSS, then Eqn.(1.121) can be cast as [117],

$$\Theta_\ell = \left(\left[\tfrac{1}{4}\delta_\gamma(\eta_{\mathrm{ls}},\mathbf{k}) + \Phi(\eta_{\mathrm{ls}},\mathbf{k})\right] j_\ell(k\eta_0) + v_\gamma(\eta_{\mathrm{ls}},\mathbf{k})\frac{dj_\ell(k\eta_0)}{d(k\eta_0)}\right)e^{-(\sigma k\eta_{\mathrm{ls}})^2} \tag{1.127}$$

where $\sigma \equiv 1.49 \times 10^{-2}\left[1+(1+z_{eq}/z_{\mathrm{ls}})^{-1/2}\right]$, $z_{eq}$ and $z_{\mathrm{ls}}$ are redshifts corresponding to matter-radiation equality and LSS respectively. So, the CMB spectrum for small angular scales now turns out to be

$$\begin{aligned}
C_\ell &= \frac{2}{\pi}\int_0^\infty k^2 dk |\Theta_\ell|^2 \\
&\simeq \frac{2}{\pi}\int_0^\infty k^2 dk \left[\left|\tfrac{1}{4}\delta_\gamma + \Phi\right|^2_{\eta_{\mathrm{ls}}} + |v_\gamma|^2_{\eta_{\mathrm{ls}}}\left(1 - \frac{\ell[\ell+1]}{k^2\eta_0^2}\right)\right] e^{-2(\sigma k\eta_{\mathrm{ls}})^2} j_\ell^2(k\eta_0).
\end{aligned} \tag{1.128}$$

The *super-Hubble* evolution of the gravitational potential $\Phi$ during matter dominated era can be calculated from Eqn.(1.104) by setting $\omega = 0$. But, to get its small scale behavior we need to evolve $\Phi$ to the smaller scales, which can be incorporated by suitably defining the associated *transfer function*. So, we define

$$\Phi_k \equiv -\tfrac{3}{5}T_k \mathcal{R}_k, \tag{1.129}$$

where $T_k$ is the transfer function which approaches to 1 for the modes having wavelengths comparable to the contemporary size of the *horizon*. Then with further simplifications, the above Eqn.(1.128) boils down to [73, 117]

$$\begin{aligned}C_\ell &\approx 4\pi P_{\Phi^0}\int_0^\infty \frac{dx}{x}j_\ell^2(x\ell)\left[\tfrac{81}{100}\mathsf{R}^2 T_k^2(x)e^{-\frac{\ell^2 x^2}{\ell_f^2}} - \tfrac{9}{10}\mathsf{R}T_k(x)T_k^0(x)e^{-\tfrac{1}{2}\frac{\ell^2 x^2}{\ell_f^2+l_s^2}}\cos(\rho\ell x)\right.\\ &\quad + \left.\tfrac{1}{4}T_k^{0^2}(x)e^{-\frac{\ell^2 x^2}{\ell_s^2}}\cos^2(\rho\ell x) + \tfrac{9}{4}c_S^2 T_k^{0^2}(x)e^{-\frac{\ell^2 x^2}{\ell_s^2}}\left(1-\frac{\ell(\ell+1)}{\ell^2 x^2}\right)\sin^2(\rho\ell x)\right]\end{aligned}\quad(1.130)$$

where $x = \frac{k\eta_0}{\ell}$, $\ell_f^{-2} \equiv 2\sigma^2\left(\frac{\eta_{\text{ls}}}{\eta_0}\right)^2$, $\ell_s^{-2} \equiv 2\left[\sigma^2 + (k_D\eta_{\text{ls}})^{-2}\right]\left(\frac{\eta_{\text{ls}}}{\eta_0}\right)^2$, $\rho \equiv \frac{1}{\eta_0}\int_0^{\eta_{\text{ls}}} c_S d\eta$ and $P_{\Phi^0} \equiv \frac{k^3}{2\pi^2}|\Phi_{k=aH=1/\eta}^0|^2$. All the statistical informations for the *small scale* CMB temperature anisotropy are encoded in the Eqn.(1.130). The dependency on various cosmological parameters is implicit here, but are encoded in the *transfer functions*.

In the present context of precision cosmology the study of *CMB polarization, Gravitational Lensing of CMB, dark energy* etc. are very important, which are also very much related to the current discussion. But I don't want to prolong this introductory review, the discussions on *CMB polarization, Gravitational Lensing of CMB* are upheld for the Chapter 5, though I shall not enter into *dark energy* related aspects. The main idea is that, the quantum fluctuations of the inflaton field generate cosmological perturbations of all wavelengths and inflation then stretched those perturbation modes outside the *horizon*. Long after inflation they re-enter the *horizon* and create temperature anisotropies in CMB. The radiation from LSS has to travel a long distance before we can observe them. In between that, the paths of the CMB photons get distorted by the fluctuations in the gravitational potential, and the power in CMB spectra are changed, which is known as *Gravitational Lensing of CMB*. Now let's stop here, and turn our attention to the organization of this thesis.

## 1.6 Plan of the Thesis

The thesis can broadly be divided into two sections – (i) *theoretical aspects of cosmological inflation* and (ii) *observational aspects of cosmological inflation*. Chapters 2 and 3 mostly deal with (i) whereas Chapters 4 and 5 deal with (ii), though there are some overlap of these two aspects as they are not entirely distinct from each other.

In the next Chapter 2 we have discussed a new model of inflation – *mutated hilltop inflation* [72, 73]. Here, the analytical expressions for most of the quantities involved have been derived. The observable parameters are evaluated at the time of *horizon crossing* by taking into account the direct effect of scalar field evolution, which yields a modified *consistency relation*. The inflationary observable parameters as estimated from the model are then confronted with observational data from *WMAP*. We then continue with our analytical framework to analyze post-inflationary evolution of the perturbations and its impact on CMB temperature anisotropies. Precisely, we have derived expressions for matter power spectrum, *Sachs-Wolfe* effect and baryon acoustic oscillations based on our model and confronted the results with observation following a semi-analytical approach.

Then in Chapter 3 we have derived a cosmological analogue of the *Berry phase* [99] in the context of inflationary cosmological perturbations [120]. We first develop the framework and then derive the *wave-function* for the inflationary cosmological perturbations by solving the *Schrödinger* equation using *dynamical invariant operator* technique [121, 122, 123]. Then we calculate the *Berry phase* for both the scalar and tensor modes. We have also shown that the cosmological analogue of the *Berry phase* is a new parameter made of corresponding spectral indices. Finally, we establish a naive link between the *Berry phase* thus obtained and inflationary observable parameters together with its physical significance.

Next, in Chapter 4 we have discussed *quasi-exponential models of inflation* [108] using *Hamilton-Jacobi* technique. Starting with a phenomenological *Hubble parameter*, we derive the condition under which the model yields quasi-exponential solution. Then we confront our analysis with the *WMAP* seven years data using the numerical code CAMB [124].

In Chapter 5 we start with a brief review on *weak gravitational lensing of CMB* and *CMB polarization*. We then provide a new technique for subtracting out lensing effect from CMB spectra to get *delensed CMB spectra*, using a simple *matrix inversion technique* [125]. To compare our results with the theory, we have developed a FORTRAN–90 code that takes the lensed CMB power spectra and the CMB lensing potential as inputs and provides us delensed CMB power spectra. For the calibration of our code, we have used *WMAP* seven years data and compare our results with CAMB outputs.

Finally, I finish with outcomes of this thesis together with future prospects in Chapter 6.

# CHAPTER 2

## Mutated Hilltop Inflation

This Chapter is based on our following works:

- Barun Kumar Pal, Supratik Pal and B. Basu, `Mutated hilltop inflation: a natural choice for early universe`, **JCAP 1001**, 029 (2010).
- Barun Kumar Pal, Supratik Pal and B. Basu, `A semi-analytical approach to perturbations in mutated hilltop inflation`, **IJMPD 21**, 1250017 (2012).

## 2.1 INTRODUCTION

Inflationary scenario first came into existence through a pioneering work by *Guth* in 1980 [1]. It is the ability to bring forth quantum seeds for the observed cosmological fluctuations leading to a nearly *scale-free* spectrum, that has turned inflationary scenario into an integral part of modern cosmology. But, the large observational window is still allowing a number of inflationary models. So, inflationary model building is still alive with added complicacy arising from the current precision level in measuring various cosmological parameters [4, 5].

In the most simple case inflation is driven by a single scalar field, *inflaton*, which evolves slowly along a nearly flat potential generally very difficult to achieve in the context of particle physics motivated models [8, 9]. But this flatness condition can be accomplished without much effort by postulating inflation occurring near the local maximum of the potential which was first shown in [71, 126]. These types of models are dubbed as *hilltop inflation* and has drawn a lot of attention of late, not only for its ability to supply flat potentials but many existing inflationary models can be converted to hilltop [71, 126] by suitable tuning of parameters. Here we would like to investigate a variant of hilltop inflation model, "*mutated hilltop inflation*" [72, 73], the crucial characteristic feature of which is that, here the modification to the flat potential is not mere addition of one or two terms from a power series, rather a hyperbolic function which contains infinite number of terms in the power series expansion, thereby making the theory more concrete and accurate at the same time. The associated potential vanishes at its *absolute minimum* which makes inflation to end so that it is not *eternal*. Also, during inflation, the second derivative of the potential is *negative*, which is recently preferred by *Planck* [4, 102]. The name "*mutated hilltop inflation*" has been inspired from "*mutated hybrid inflation*" [127] by noticing the similarity of the inflaton potential shape, which is *inverted*, in both the scenarios. Also, the energy scale for mutated hilltop inflation is well below the *Planck scale* which is also in accord with the mutated hybrid inflation. The mutated hybrid inflation involve two real scalar fields. In contrast with ordinary hybrid inflation, the waterfall field (field other than inflaton) for the involved potential has no mass term during inflation and is not fixed at the origin, instead it adjusts to minimize the potential at each value of the inflaton.

In this chapter we would like to see how far one can proceed analytically based on a particular model of inflation, adopting least number of approximations. This will not only allow us confront inflationary predictions directly with the observational data, but may also provide few interesting results. In the process we shall provide analytical results for most of the quantities involved. In estimating inflationary observables at the time of *horizon crossing*, we have adopted a new route by taking into account the effect of scalar field evolution which leads to a modified *consistency relation*. The analytical framework also allows us compute post inflationary perturbations.

## 2.2 The Model

Let us propose a form of the potential

$$V(\phi) = V_0 \left[1 - \text{sech}(\alpha\phi)\right], \tag{2.1}$$

where $V_0$ represents the typical energy scale for hilltop inflation and $\alpha$ is a parameter having dimension of inverse *Planck* mass. `Fig.2.1` shows explicit behavior of the potential with the inflaton for different values of $\alpha$. It will be revealed later on that $\alpha = (2.9 - 3.1) \, \text{M}_P^{-1}$ gives the best fit model from the observational ground. So, here, and throughout the rest of the chapter, we adhere to this range for $\alpha$. The nature of the potential in `Fig.2.1` is a characteristic feature of mutated hilltop inflation models, which resembles *mutated hybrid inflation models* [127].

The characteristic feature of the earlier models of hilltop inflation is that the inflation occurs near the maximum of the potential. The presence of an infinite number of terms in the power series expansion of the potential (2.1) makes *mutated hilltop inflation* different from other hilltop inflation where mostly two-term approximations have been incorporated [71, 126]. Not only that, mutated hilltop inflaton potential satisfies $V(\phi_{min}) = V'(\phi_{min}) = 0$, at its absolute mini-mum $\phi_{min}$, which characterizes a significant difference from the usual hilltop potential [128]. The vanishing of the potential at its *absolute minimum* guarantees that inflation is not *eternal*.

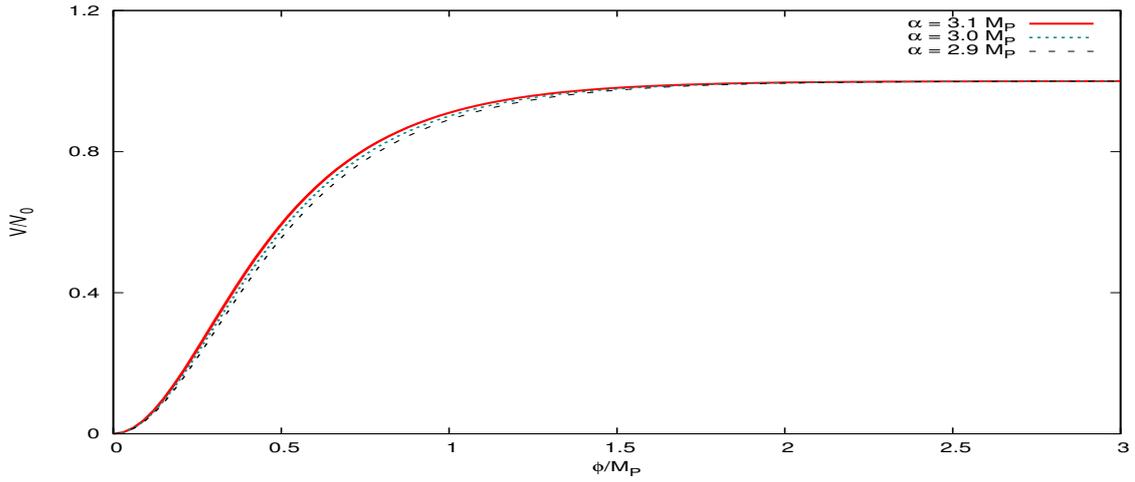

**FIG. 2.1:** *Variation of the mutated hilltop inflation potential with the scalar field for three sets of values for $\alpha$. For super-Planckian inflaton value the potential has a sufficiently flat region.*

## 2.3 SOLUTION TO DYNAMICAL EQUATIONS

The *Friedmann* Eqn.(1.17) for the potential (2.1), turns out to be

$$H^2 = \tfrac{1}{3M_P^2} V_0[1 - \text{sech}(\alpha\phi)]. \qquad (2.2)$$

In the present context the *Hubble slow-roll parameters*, defined in Eqn.(1.21), are given

$$\begin{aligned}
\epsilon_H &= \tfrac{M_P^2}{2} \alpha^2 \frac{\text{sech}^2(\alpha\phi)\tanh^2(\alpha\phi)}{[1 - \text{sech}(\alpha\phi)]^2} \\
\eta_H &= \tfrac{M_P^2}{2} \alpha^2 \left\{ \frac{2\,\text{sech}(\alpha\phi)[\text{sech}^2(\alpha\phi) - \tanh^2(\alpha\phi)]}{[1 - \text{sech}(\alpha\phi)]} - \frac{\text{sech}^2(\alpha\phi)\tanh^2(\alpha\phi)}{[1 - \text{sech}(\alpha\phi)]^2} \right\}.
\end{aligned} \qquad (2.3)$$

The basic intention of using *Hubble slow-roll parameters* is not to restrict ourselves in the usual *slow-roll* approximation which is somewhat limited in a generic *super-gravity* theory [129, 130] but to use a more accurate version of the same.

The equation supplementary to (2.2), *i.e.*, the *Klein-Gordon* Eqn.(1.16), which in the *slow-roll* approximation is given by Eqn.(1.18) and in the present context turns out to be

$$\frac{\sqrt{1 - \text{sech}(\alpha\phi)}}{\text{sech}(\alpha\phi)\tanh(\alpha\phi)} d\phi + \alpha\sqrt{\frac{V_0}{3}} M_P\, dt = 0. \qquad (2.4)$$

An exact solution for the above equation can indeed be obtained by direct integration. Written explicitly, the solution looks

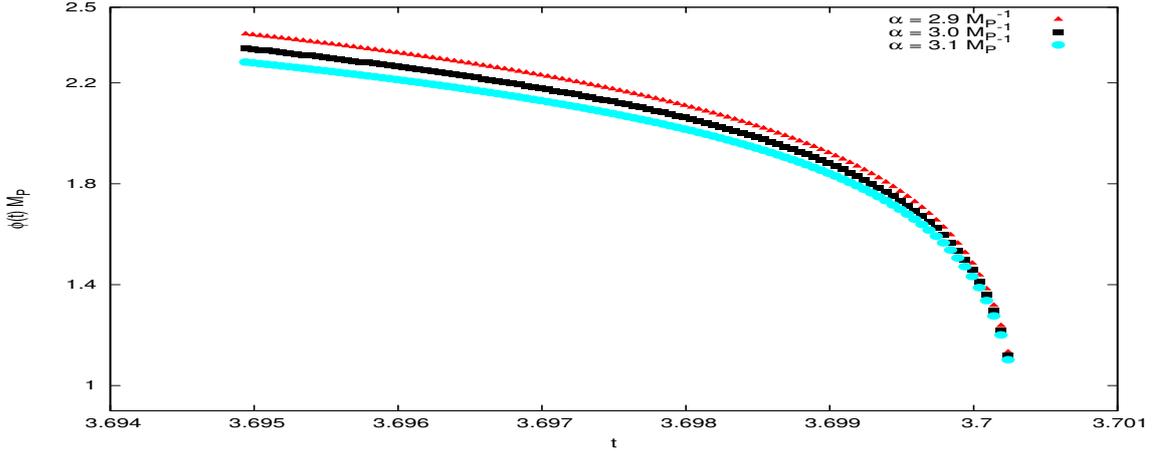

**FIG. 2.2:** *Variation of the inflaton field with time (in units of $10^{10}\ M_P^{-1}$) for three set of values for $\alpha$.*

$$\frac{1}{\sqrt{2}\alpha}\left[-\sqrt{2}\sinh^{-1}\left(\sqrt{2}\sinh\frac{\alpha\phi}{2}\right)+2\left(\tanh^{-1}\left[\frac{\sinh\frac{\alpha\phi}{2}}{\sqrt{\cosh\alpha\phi}}\right]+\sqrt{\cosh\alpha\phi}\sinh\frac{\alpha\phi}{2}\right)\right]$$
$$=-\alpha\sqrt{\frac{V_0}{3}}M_P t + \textit{constant}. \quad (2.5)$$

However, though exact, this solution is not very useful for practical purpose because of its complexity. Precisely, in order to deal with estimates of the observable quantities, one has to use this expression for further analytical and numerical calculations, which is not easy to handle. Instead, one can search for an approximate analytic solution of Eqn.(2.4) which will help us derive analytical expressions for most of the parameters involved with the theory of inflation and perturbations therefrom. This will, in turn, help us visualize the pros and cons of the scenario both analytically and numerically for quantitative estimation at a later stage. We foresee more merit in this route and will follow it subsequently.

We first note that Eqn.(2.4) can be rewritten as

$$\left[\text{sech}^2(\alpha\phi)+\text{sech}^3(\alpha\phi)\right]^{-\frac{1}{2}}d\phi + \alpha\sqrt{\frac{V_0}{3}}M_P\ dt = 0. \quad (2.6)$$

Now neglecting the term $\text{sech}^3(\alpha\phi)$ in Eqn.(2.6) as $\text{sech}(\alpha\phi)$ is small in the present case, the analytic solution of the Eqn.(2.6) is found to be

$$\sinh(\alpha\phi) = -\alpha^2\sqrt{\frac{V_0}{3}}M_P\ t + \textit{constant}, \quad (2.7)$$

where the constant can be calculated from the condition that at the end of inflation, $t = t_E$, the scalar field has the value $\phi = \phi_E$. This readily gives

$$\sinh(\alpha\phi) = \alpha^2\sqrt{\frac{V_0}{3}}M_P(d-t),\ \text{with}\ d = t_E + \sinh(\alpha\phi_E)\left[\alpha^2\sqrt{\frac{V_0}{3}}M_P\right]^{-1} \quad (2.8)$$

where $\phi_E$ is estimated from the relation: $\epsilon_H(\phi_E) = 1$. In `Fig.2.2` we have shown the variation of the scalar field with the cosmic time for $\alpha = (2.9 - 3.1)\, M_P^{-1}$. The plot clearly shows that the scalar field gradually decays as it approaches towards the end of inflation, finally reaching a value $\phi_E$.

Using the expression (2.8) for the scalar field as a solution of the *Klein-Gordon* equation in the *slow-roll* regime, we arrive at the following equation

$$\frac{da}{a} = \frac{\alpha^2 \frac{V_0}{3}(d-t)}{\sqrt{1 + \alpha^4 \frac{V_0}{3} M_P^2 (d-t)^2}} dt. \tag{2.9}$$

Consequently, the solution for the scale factor turns out to be

$$a(t) = a_I\, \exp\left[-(\alpha M_P)^{-2} \sqrt{1 + \alpha^4 \frac{V_0}{3} M_P^2 (d-t)^2}\right] \tag{2.10}$$

where $a_I \equiv a(t_E)\, \exp\left[(\alpha M_P)^{-2} \cosh(\alpha \phi_E)\right]$ is the scale factor at the end of inflation, scaled by the exponential term.

The expression for the number of *e-foldings* can also be found analytically using Eqn.(1.23), which turns out to be

$$N \sim (\alpha^2 M_p^2)^{-1} \left[\cosh(\alpha\phi) - \ln \cosh^2\left(\tfrac{1}{2}\alpha\phi\right)\right]_{\phi_E}^{\phi_I}. \tag{2.11}$$

The `Table 2.1` summarizes estimates of the involved parameters corresponding to three different values of $\alpha$.

| $\alpha$ $M_P^{-1}$ | $\epsilon_H < 1$ $\phi \geq M_P$ | $|\eta_H| < 1$ $\phi \geq M_P$ | $\phi_E$ $M_P$ | $\phi_I$ $M_P$ | $N$ | $\phi_{hc}$ $M_P$ |
|---|---|---|---|---|---|---|
| 2.9 | 0.5989 | 1.0219 | 1.0219 | 2.4463 | 70 | 2.3943 |
|     |        |        |        | 2.3943 | 60 | 2.3330 |
|     |        |        |        | 2.3711 | 56 | 2.3052 |
| 3.0 | 0.5876 | 1.0080 | 1.0080 | 2.3871 | 70 | 2.3369 |
|     |        |        |        | 2.3369 | 60 | 2.2777 |
|     |        |        |        | 2.3145 | 56 | 2.2507 |
| 3.1 | 0.5768 | 0.9944 | 0.9944 | 2.3311 | 70 | 2.2825 |
|     |        |        |        | 2.2825 | 60 | 2.2252 |
|     |        |        |        | 2.2608 | 56 | 2.1990 |

**TAB. 2.1:** *Table for different parameters related to the mutated hilltop inflation.*

Here $\phi_{hc}$ corresponds to the value of the inflaton at *horizon crossing*. From the above table it is clear that *mutated hilltop inflation* occurs while the inflaton remains above the *Planck* scale. The range for the model parameter $\alpha$, $2.9\, M_P^{-1} \leq \alpha < 3.1\, M_P^{-1}$, has been constrained in such a way that the inflaton scale remains *super-Planckian* during inflation. It can also be seen that $\eta_H > 1$ happens earlier than $\epsilon_H > 1$, so the *slow-roll approximation* may become inaccurate towards the end of inflation. So, we have considered the end of inflation to be at $\eta_{_H} = 1$. But, as it occurs during negligible number of *e-foldings*, the observable predictions for mutated hilltop inflation remain unaffected.

## 2.4 IMPLICATIONS TO COSMOLOGICAL PERTURBATIONS

Before going into the details of perturbations we note that in the expression for the scale factor (2.10), the term $\alpha^4 \frac{V_0}{3} M_P^2 (d-t)^2$ is much larger than unity during inflation, so the following approximation holds good for mutated hilltop model of inflation

$$a(t) \sim a_{_I} \exp\left[M_P^{-1} \sqrt{\tfrac{V_0}{3}}(t-d)\right]. \tag{2.12}$$

Hence the scale factor behaves pretty close to *de-Sitter* so that we can use near *de-Sitter* approximation wherever necessary. We shall employ this argument in the following analysis.

The *conformal time*, $\eta$, which is more useful than the *cosmic time*, $t$, to study cosmological perturbation, in the present context turns out to be,

$$\eta = -M_P \sqrt{\tfrac{3}{V_0}} a(\eta)^{-1}. \tag{2.13}$$

### 2.4.1 CURVATURE PERTURBATION

The *Mukhanov-Sasaki* equation is given by (1.81) and for the present case we have

$$z \approx \alpha^{-1} M_P \sqrt{\tfrac{3}{V_0}} |\eta|^{-1} \left[\ln f(\eta)\right]^{-1}, \tag{2.14}$$

where we have defined $f(\eta) \equiv a_{_I} M_P^{-1} \sqrt{\tfrac{V_0}{3}} |\eta|$, also during *slow-roll* inflation we know that $\frac{z''}{z} \approx \frac{a''}{a}$. Therefore *Mukhanov-Sasaki* Eqn.(1.81) turns out to be

$$v_k'' + \left(k^2 - \tfrac{2}{\eta^2}\right) v_k = 0, \tag{2.15}$$

which has the solution as given by Eqn.(1.87). Thus the *co-moving curvature perturbation*, which is defined by $\mathcal{R}_k = -\frac{v_k}{z}$, in this context has the following form

$$\mathcal{R}_k = -\alpha M_P^{-1} \sqrt{\tfrac{V_0}{3}} \frac{e^{-ik\eta}}{\sqrt{2k}} |\eta| \left(1 - \tfrac{i}{k\eta}\right) [\ln f(\eta)]. \tag{2.16}$$

Consequently, the spectrum for the *co-moving curvature perturbation* turns out to be

$$P_\mathcal{R}(k) = \frac{\alpha^2 V_0}{12\pi^2 M_P^2} \left(1 + k^2 \eta^2\right) [\ln f(\eta)]^2. \tag{2.17}$$

Now, in order to evaluate the spectrum at the time of *horizon exit* during inflation, *i.e.*, at $k = aH$, we proceed along two different paths. First we shall discard the effect of scalar field evolution and adopt perfect *de-Sitter* approximation so that at *horizon crossing* $k = aH = \eta^{-1}$. Then we take into account the effect of scalar field evolution in our estimations, *i.e.*, we do not put $k = aH = \eta^{-1}$ a priori. In due course it will be shown that the results thus obtained differ by a small amount and lead to a modified *consistency relation*.

To proceed through the second trail, we first note that at *horizon crossing*

$$k = aH = a_I M_P^{-1} \sqrt{\tfrac{V_0}{3}} \, e^{M_P^{-1} \sqrt{\tfrac{V_0}{3}}(t-d)}, \quad \text{and} \quad d\ln k = M_P^{-1} \sqrt{\tfrac{V_0}{3}} dt \tag{2.18}$$

which can be derived from Eqn.(2.12). Therefore, at *horizon crossing*, for the inflationary model under consideration we get

$$1 + k^2 \eta^2 \approx 2 - \frac{1}{\alpha^2 \sqrt{\tfrac{V_0}{3}} M_P (d-t)}. \tag{2.19}$$

The result is slightly different from the usual relation $1 + k^2\eta^2 = 2$. Our approach is a bit similar to [105], where variable speed of the scalar field fluctuation has been taken into account for evaluating observable quantities at *horizon exit*. In the following we shall use *superscript* "1" to denote observable parameters estimated following the *usual trail* and *superscript* "2" to refer observables estimated using our *new trail*.

Now the spectrum for $\mathcal{R}_k$ at the time of *horizon crossing* by discarding the effect of scalar field evolution turns out to be

$$P_\mathcal{R}^{(1)}|_{k=aH} = \frac{\alpha^2 V_0}{6\pi^2 M_P^2} \left[\ln f(\eta_k)\right]^2, \tag{2.20}$$

where $\eta_k$ represents the *conformal time* when a particular mode with *wavenumber* $k$ leaves the *horizon*. In the new trail, using Eqn.(2.19), at *horizon exit* we obtain

$$P_\mathcal{R}^{(2)}|_{k=aH} = \frac{\alpha^2 V_0}{6\pi^2 M_P^2} \left[1 - \frac{\alpha^{-2} M_P^{-2}}{2\ln f(\eta_k)}\right] \left[\ln f(\eta_k)\right]^2. \tag{2.21}$$

Therefore from Eqns.(2.20) and (2.21), we see that if one considers the direct effect of scalar field evolution in estimating the power spectrum at the time of *horizon exit*, the result will differ slightly from the usual one.

Once the spectrum is known, we can immediately calculate corresponding *spectral index*, which in the standard path yields,

$$n_S^{(1)} = 1 + \frac{d\ln P_\mathcal{R}(k)}{d\ln k}\bigg|_{k=aH} = 1 - 2\left[\ln f(\eta_k)\right]^{-1}, \tag{2.22}$$

and in our non-standard approach ends up with

$$n_S^{(2)} = 1 - \left[\frac{4\ln f(\eta_k) - \alpha^{-2} M_P^{-2}}{2\{\ln f(\eta_k)\}^2 - \alpha^{-2} M_P^{-2} \ln f(\eta_k)}\right]. \tag{2.23}$$

Again we see that, there is slight difference between two results. We have also succeeded in deriving an expression for the *running of the spectral index* from our model. The corresponding expression in the usual approach turns out to be

$$n'_S{}^{(1)} \equiv \frac{dn_S}{d\ln k}|_{k=aH} = -2\left[\ln f(\eta_k)\right]^{-2} \tag{2.24}$$

and our new trail yields

$$n'_S{}^{(2)} = -\left[\frac{8\{\ln f(\eta_k)\}^2 - 4\alpha^{-2}M_P^{-2}\ln f(\eta_k) + \alpha^{-4}M_P^{-4}}{\left\{2\left[\ln f(\eta_k)\right]^2 - \alpha^{-2}M_P^{-2}\ln f(\eta_k)\right\}^2}\right]. \tag{2.25}$$

Clearly, though the quantities within the parentheses in Eqns.(2.24) and (2.25) are quite small, but still yield nonzero values resulting in a negative running of the spectral index consistent with *WMAP* three years data set and the following data coming in subsequent years (like *WMAP–5, WMAP–9* and *Planck*).

Comparing Eqns.(2.20, 2.21), Eqns.(2.22, 2.23) and Eqns.(2.24, 2.25), we see that the correction terms that arise due to the effect of scalar field evolution though very small, but may have some significant role from the theoretical point of view.

### 2.4.2 Tensor Perturbations

The tensor modes are given by Eqn.(1.87) and the corresponding spectrum in the present context is given by

$$P_\mathcal{T} = \frac{V_0}{3\pi^2 M_P^4}(1 + k^2\eta^2). \tag{2.26}$$

Now proceeding similar to scalar perturbations, we first evaluate the spectrum (2.26) at *horizon exit* which in the standard approach yields

$$P_\mathcal{T}^{(1)}|_{k=aH} = \frac{2V_0}{3\pi^2 M_P^4}. \tag{2.27}$$

Thus we see that tensor power spectrum is a constant, *i.e.*, scale independent. This was expected as we are using the scale factor (2.12) which is *de-Sitter*. But following our new trail, *i.e*, by taking into account scalar field evolution effect at the time of *horizon crossing*, we obtain

$$P_\mathcal{T}^{(2)}|_{k=aH} = \frac{V_0}{3\pi^2 M_P^4}\left[2 - \frac{\alpha^{-2}M_P^{-2}}{\ln f(\eta_k)}\right] \tag{2.28}$$

which is certainly not scale invariant. In the standard lore, from our model, the *tensor spectral index* $(n_\mathcal{T})$ and its *running* $(n'_\mathcal{T})$ turn out to be identically zero. Following our approach we have succeeded in obtaining non-trivial expressions for them,

$$n_\mathcal{T}^{(2)} = -\frac{1}{\alpha^2 M_P^2}\frac{1}{\left\{2\left[\ln f(\eta_k)\right]^2 - \alpha^{-2}M_P^{-2}\ln f(\eta_k)\right\}} \tag{2.29}$$

$$n'^{(2)}_{\mathcal{T}} = -\frac{1}{\alpha^2 \mathrm{M}_P^2} \frac{4 \ln f(\eta_k) - \alpha^{-2} \mathrm{M}_P^{-2}}{\left\{ 2 \left[ \ln f(\eta_k) \right]^2 - \alpha^{-2} \mathrm{M}_P^{-2} \ln f(\eta_k) \right\}^2}. \quad (2.30)$$

Now using Eqns.(2.21, 2.28), for mutated hilltop inflation, the tensor to scalar power spectra ratio, $r$, turns out to be

$$r = 4 \left[ \alpha^2 \mathrm{M}_P^2 f(\eta_k)^2 \right]^{-1}. \quad (2.31)$$

This result is same no matter whether or not we take into account the effect of scalar field evolution in estimating the spectra at *horizon crossing*. Using Eqns.(2.29) and (2.31) we derive the following *consistency relation* for mutated hilltop model of inflation

$$r = -8 n_{\mathcal{T}} \left( 1 - \frac{\sqrt{r}}{4\alpha \mathrm{M}_P} \right) \quad (2.32)$$

The spectrum $P_{\mathcal{T}|_{k=aH}}$ of tensor perturbation conveniently specified by the tensor fraction $r = \frac{P_{\mathcal{T}|_{k=aH}}}{P_{\mathcal{R}|_{k=aH}}}$ yields the relation $r = -8 n_{\mathcal{T}}$ in the *slow-roll* approximation [49, 103]. But, Eqn.(2.32) shows that when the explicit effect of the scalar field evolution is taken into account in evaluating the observable parameters at horizon exit we obtain a consistency relation which is slightly modified. Of course, there exist in the literature other ways of obtaining a modified consistency relation. Such a modified consistency relation can be found in any analysis where higher order terms in the expansion of *slow-roll* parameters are taken into account [104]. The consistency relation is also modified in the context of brane inflation [105, 106] and non-standard models of inflation [107] where generalized propagation speed (less than one) of the scalar field fluctuations relative to the homogeneous background have been considered. Further, deviation from the usual consistency relation can be found in [131] where *tensor to scalar ratio* has been shown to be a function of *tensor spectral index, scalar spectral index* and *running of the tensor spectral index*. Our intention here is to show that these results reflect in our analysis by taking into account the evolution of the scalar field directly in the analytical evaluations.

### 2.4.3 Quantitative Estimation

In Table 2.2 we have estimated the observable parameters directly from the theory of fluctuations for three different values of $\alpha$.

| $\alpha$ $\mathrm{M}_P^{-1}$ | $a_I$ $10^{25}\mathrm{M}_P^{-1}$ | $V_0^{1/4}$ $10^{15}\mathrm{GeV}$ | $\eta_k$ | $P_{\mathcal{R}}^{1/2}$ $10^{-5}$ | $n_s$ | $r$ $10^{-4}$ |
|---|---|---|---|---|---|---|
| 2.9 | 6.7091 | 3.4104 | 216.649 | 3.7762 | 0.9609 | 1.8176 |
| 3.0 | 6.6492 | 3.4104 | 216.649 | 3.9059 | 0.9609 | 1.6990 |
| 3.1 | 6.5945 | 3.4104 | 216.649 | 4.0377 | 0.9609 | 1.5917 |

**Tab. 2.2:** *Table for the observable quantities as obtained from the theory of fluctuations. To evaluate the parameters at the time of horizon crossing we have applied our newly developed technique.*

For the above estimation, we have taken the following representative values for the quantities involved[1] (for $N = 60$): $V_0^{1/4} = 1.4 \times 10^{-3}$ M$_P$, $t_E \simeq 3.70028 \times 10^{10}$ M$_P^{-1}$, $t_I \simeq 3.69493 \times 10^{10}$ M$_P^{-1}$ and $a_I \simeq 1.85184 \times 10^{-1}$ M$_P^{-1}$. Also, we have assumed that the cosmological scales leave the horizon during first 10 *e-foldings* to get, $d \simeq 3.70038 \times 10^{10}$ M$_P^{-1}$, $\eta_I \simeq -4.77201 \times 10^6$ and $\eta_E \simeq -4.17816 \times 10^{-20}$. From the `Table 2.2`, it is quite clear that the observable parameters related to perturbations are in tune with the observations.

## 2.5 Matter Power Spectrum

So far, we have surveyed fluctuations generated during inflation and evaluated inflationary observable parameters at the time of horizon crossing from a different perspective. The results show small corrections compared to that obtained in a standard way. To relate these initial perturbation seeds with the cosmological observables we now follow their evolutions during matter domination.

In Chapter 1 we have seen that for adiabatic perturbation the *co-moving curvature perturbation*, $\mathcal{R}(k)$, remains constant in the *supper-horizon scales*. Therefore taking the limit $k|\eta| \to 0$ of the Eqn.(2.16) we get *supper-Hubble* behavior of the *co-moving curvature perturbation* in mutated hilltop inflation, and for the modes re-entering the *horizon* during matter dominated era we have

$$\mathcal{R}|_{k=aH=\frac{2}{\eta}} \simeq \tfrac{1}{\mathrm{M}_P} \alpha \sqrt{\tfrac{V_0}{3}} \tfrac{i}{\sqrt{2k^3}} \ln g(k) \tag{2.33}$$

where $g(k) \equiv 2a_I \mathrm{M}_P^{-1} \sqrt{\tfrac{V_0}{3}} k^{-1}$. The fluctuation of the gravitational potential $\Phi$ governs the matter density fluctuation, $\delta_m$, through the Eqn.(1.106). The relation between $\Phi_k$ and $\mathcal{R}_k$ in the *super-Hubble* regime during matter domination is given by

$$\Phi_k \approx -\tfrac{3}{5} \mathcal{R}_k. \tag{2.34}$$

Consequently, we now have at the time of *horizon* re-entry during matter domination

$$\Phi|_{k=aH} \approx -\tfrac{3}{5} \mathcal{R}|_{k=aH} \approx -\tfrac{3}{5\mathrm{M}_P} \alpha \sqrt{\tfrac{V_0}{3}} \tfrac{i}{\sqrt{2k^3}} \ln g(k). \tag{2.35}$$

Therefore, substituting $k = aH = \tfrac{2}{\eta}$ in Eqn.(1.106) we can determine matter density contrast at the time of horizon re-entry during matter dominated era and using Eqn.(2.35) corresponding matter power spectrum turns out to be

$$P_{\delta_m}|_{k=aH} \equiv \tfrac{k^3}{2\pi^2} |\delta_m|^2 \approx \tfrac{16\alpha^2 V_0}{75\pi^2 \mathrm{M}_P^2} g(k)^2. \tag{2.36}$$

In `Fig.2.3` we have shown the typical behavior of the *dimensionless matter power spectrum* as a function of scale.

---
[1] Here we have assumed that the scale factor $a$ has dimension of length so that $k$ and $\eta$ are dimensionless.

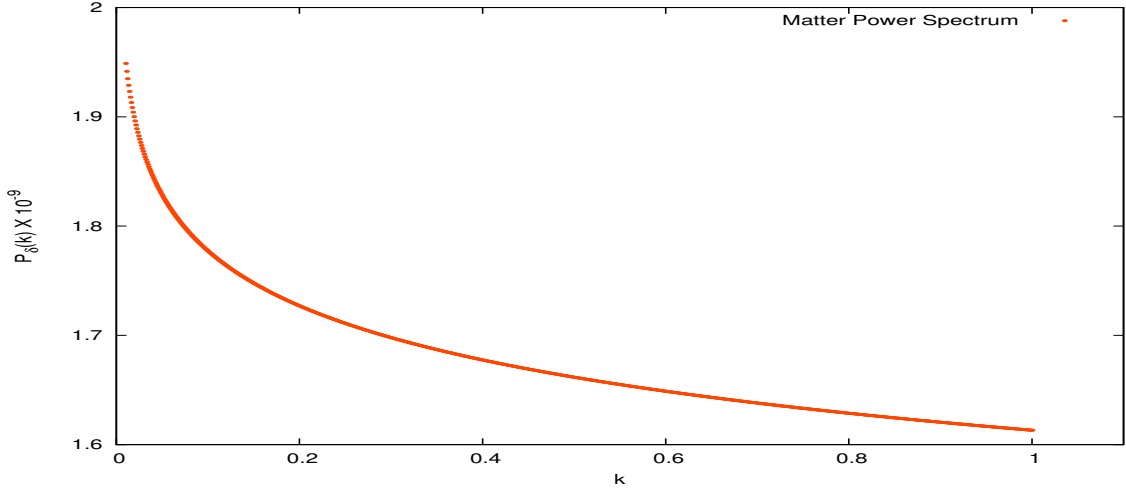

**FIG. 2.3:** *Variation of the matter power spectrum with the scale.*

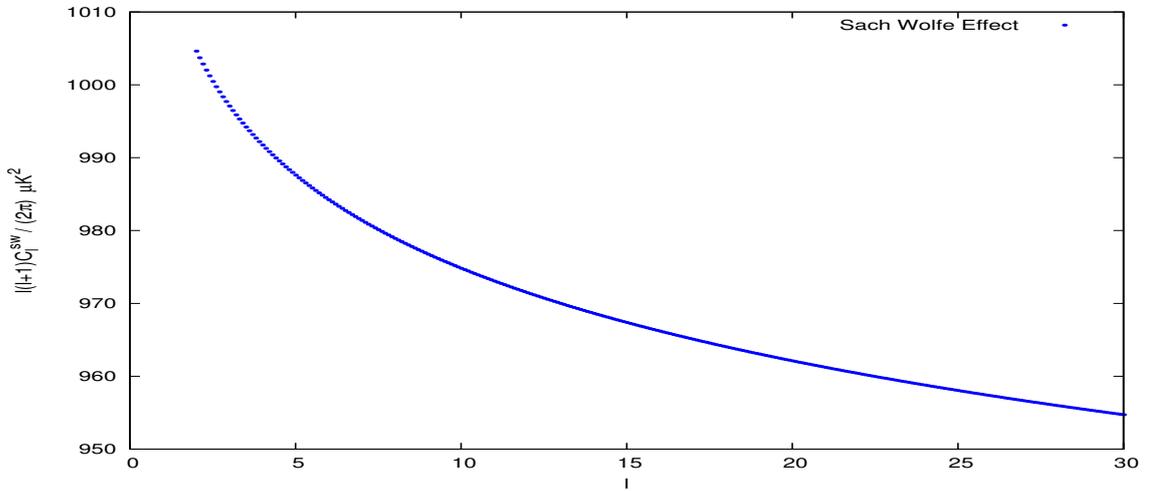

**FIG. 2.4:** *Variation of the Sachs-Wolfe spectrum with the multipoles $\ell$, for the mutated hilltop inflation for $\alpha = 3 \, M_P^{-1}$.*

## 2.6 TEMPERATURE ANISOTROPIES IN CMB

Let us now proceed further with this analytical framework and derive CMB angular power spectrum resulting from temperature anisotropies due to scalar curvature perturbations for mutated hilltop inflation. The dominant contribution to the large scale anisotropy comes from the *Sachs-Wolfe* effect [118] and the corresponding spectrum in the *sudden decoupling* approximation is roughly given by

$$C_\ell^{\text{SW}} \approx \frac{\alpha^2 V_0}{75\pi M_P^2} \left( \ln \left[ a_I M_P^{-1} \sqrt{\frac{V_0}{3}} \frac{2\eta_0}{\ell} \right] \right)^2 \frac{1}{2\ell(\ell+1)}, \tag{2.37}$$

where complete matter domination at LSS together with adiabatic nature of the perturbations have been assumed. Fig.2.4 shows variation of the *Sachs-Wolfe* effect with the multipoles $\ell$ in the units of $\mu K^2$, where we have taken the following representative values for the quantities involved, $V_0^{1/4} = 1.4 \times 10^{-3}\,M_P$, $\alpha = 3.0\,M_P^{-1}$, $a_I = 6.6492 \times 10^{25}\,M_P^{-1}$ and $\eta_0 = 7.42438 \times 10^{26}$. `Fig.2.4` reveals that the *Sachs-Wolfe* plateau is not exactly flat but slightly tilted towards larger $\ell$. This is not surprising, since the primordial curvature perturbation in this case is not strictly scale-invariant. However, as of now this is an interesting result which reveals the credentials of analytical calculations of post-inflationary perturbations from a typical model of inflation from the very first principle.

In the smaller scales CMB spectrum is dominated by acoustic oscillation of the baryon-photon fluid. The approximate solution to the acoustic oscillation Eqn.(1.122) is given by Eqn.(1.124). Now, here we have

$$\Phi_k = -\frac{3\alpha}{5M_P}T_k\sqrt{\frac{V_0}{3}}\frac{i}{\sqrt{2k^3}}\ln g(k) \tag{2.38}$$

as the small scale solution of gravitational potential for the modes re-entering the *horizon* during matter dominated era and $T_k$ is the *transfer function* associated with them. We also have

$$\Phi_k^0 \equiv -\frac{2\alpha}{3M_P}\sqrt{\frac{V_0}{3}}\frac{1}{\sqrt{2k^3}}\left[\ln\left(a_I M_P^{-1}\sqrt{\frac{V_0}{3}}k^{-1}\right)\right] \tag{2.39}$$

as the initial fluctuation in the gravitational potential, *i.e.*, the gravitational potential for the modes that re-entered the *horizon* during radiation dominated era and $T_k^0$ is the corresponding *transfer function*. The photon velocity perturbation, $v_{\gamma,k}$, in this case turns out to be

$$v_{\gamma,k} = \tfrac{3}{2}c_s T_k^0 \Phi_k^0 \sin(kr_s). \tag{2.40}$$

In the small scale regime, CMB angular power spectrum is given by Eqn.(1.130), which can be further simplified by taking large $\ell$ limit of the *spherical Bessel* functions that yields

$$\begin{aligned}C_\ell &\approx \frac{2\pi P_{\Phi^0}}{l^2}\int_1^\infty \frac{dx}{x^2\sqrt{x^2-1}}\left\{\tfrac{81}{100}R^2 T_k^2(x)e^{-\frac{\ell^2 x^2}{\ell_f^2}} - \tfrac{9}{10}R T_k(x)T_k^0(x)e^{-1/2\frac{\ell^2 x^2}{\ell_f^2+\ell_s^2}}\cos(\rho\ell x)\right.\\ &\quad \left. + \tfrac{1}{4}T_k^{0\,2}(x)e^{-\frac{\ell^2 x^2}{\ell_s^2}}\cos^2(\rho\ell x) + \tfrac{9}{4}c_s^2 T_k^{0\,2}(x)e^{-\frac{\ell^2 x^2}{\ell_s^2}}\left[1-\frac{\ell(\ell+1)}{\ell^2 x^2}\right]\sin^2(\rho\ell x)\right\}.\end{aligned} \tag{2.41}$$

The power spectrum, $P_{\Phi^0}$, for $\Phi_k^0$ in the context of inflationary model under consideration has the following from

$$P_{\Phi^0} \equiv \tfrac{1}{27\pi^2 M_P^2}\alpha^2 V_0 \left[\ln\left(a_I M_P^{-1}\sqrt{\frac{V_0}{3}}\frac{\eta_0}{\ell}\right)\right]^2. \tag{2.42}$$

The above results may not be exact, but still qualitative behavior of the associated physical quantities can be extracted from them very easily. The *transfer functions* used for the calculations have been taken from [111, 117] with appropriate modifications, written explicitly, they look

$$T^0(x) = 1.20 + 0.09\ln\left(\frac{I_\Lambda \ell\, x}{250\sqrt{\Omega_M}}\right)$$

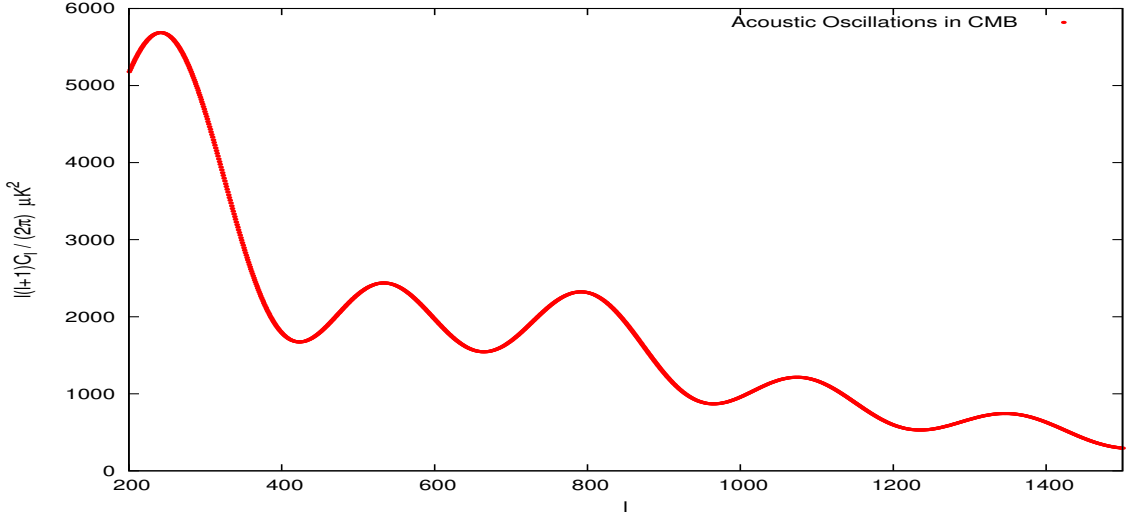

**FIG. 2.5:** *Variation of the CMB angular power spectrum with the multipoles $\ell$ in the large $\ell$ limit.*

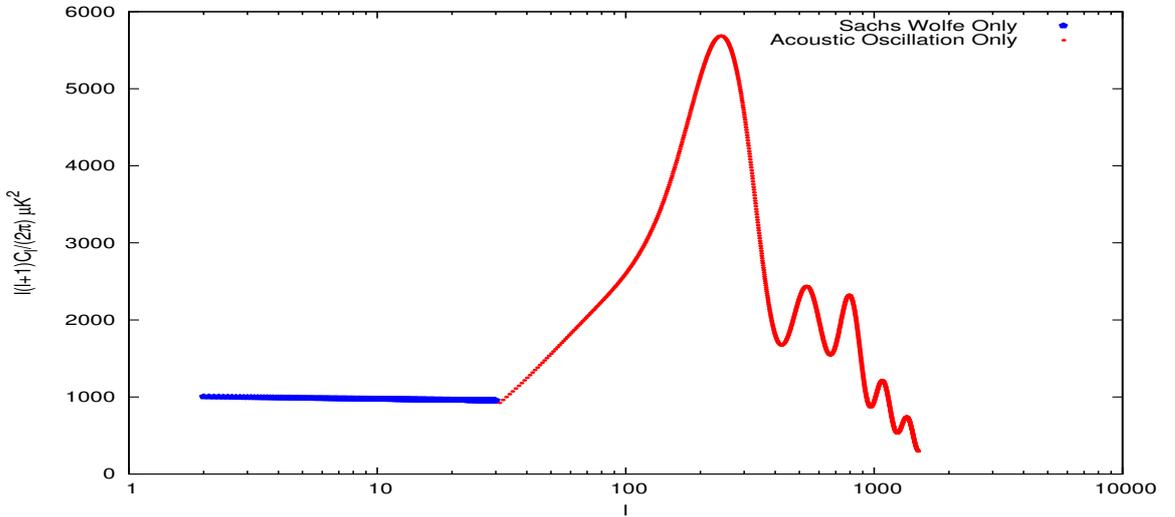

**FIG. 2.6:** *Variation of the CMB angular power spectrum with the multipoles $\ell$, considering Sachs-Wolfe effect up to $\ell = 30$ and only baryon acoustic oscillations for the rest.*

$$T(x) = 0.52 - 0.21 \ln\left(\frac{I_\Lambda \ell \, x}{250\sqrt{\Omega_M}}\right), \text{ with } I_\Lambda \equiv 3\left(\frac{\Omega_{DE}}{\Omega_M}\right)^{\frac{1}{6}} \left[\int_0^{\sinh^{-1}\sqrt{\Omega_{DE}/\Omega_M}} \frac{dx}{(\sinh x)^{2/3}}\right]^{-1},$$

where the late time effect due to dark energy has been incorporated via $I_\Lambda$. The integral is evaluated with the following set of numerical values for the involved parameters: $h = 0.71, \Omega_B = 0.0449, \Omega_M = 0.2669, \Omega_{DE} = 0.734, \Omega_\gamma h^2 = 2.48 \times 10^{-5}, z_{ls} = 1090, z_{eq} = 3300, R_{ls} = 0.627405, \sigma = 2.23245 \times 10^{-2}, \ell_f = 1617, \ell_s = 1254, \rho = 0.0112647, \eta_0 = 7.42438 \times 10^{26}$.

In Fig.2.5 we have plotted the variation of temperature anisotropy spectrum with the multipoles for $\ell \geq 200$ which depicts true characteristics of baryon acoustic oscillation in CMB. In Fig.2.6, resulting from our analysis, we see the first and the most prominent peak of acoustic oscillation arises at $\ell \sim 241$ and at a height of $\sim 5900\mu K^2$. The first peak is followed by two nearly equal height peaks at $\ell \sim 533$ and $\sim 791$. After the third peak there is a damping tail in the oscillation with successive peaks with lower heights. This happens due to taking into account *Silk damping* effect in our analysis, which is also consistent with CMB observations. So mutated hilltop inflation conforms well with the recent observations.

## 2.7 CHAPTER SUMMARY

In this Chapter we have discussed a variant of hilltop inflation models – the *mutated hilltop inflation*, driven by a hyperbolic potential for the scalar field. To this end, we have derived the expressions for the scalar field, the scale factor, the number of e-foldings from our model. The *inflationary observables* have been estimated at the time of *horizon crossing* during inflation following a non-standard trail by taking into account the effect of scalar field evolution, which results in a modified *consistency relation*. The results have then been subjected to observational tests by finding out the values of observable quantities. We have studied quantum fluctuations and corresponding classical perturbations as well as allied observational aspects by a semi-analytical treatment, based on mutated hilltop inflation. The analytical expressions for most of the observable parameters have also been found. An approximated expression for the fluctuations in the gravitational fluctuations has been successfully obtained using relativistic perturbations. We have also obtained an expression for the scale dependent matter power spectrum. We have explicitly shown that the *Sachs-Wolfe plateau* is not strictly flat but slightly tilted towards the larger multipoles for mutated hilltop inflation. Moreover, we have also studied the baryon acoustic oscillation based on our model analytically, which reduces numerical complications to a great extent and at the same time provides physical insight of the scenario. Finally, we have employed certain simplistic numerical techniques and found that the positions of the acoustic peaks conform well within the estimated values of different cosmological parameters.

# CHAPTER 3

# Quantum States of Inflationary Perturbations & The Berry Phase

This Chapter is based on our following work:

1. Barun Kumar Pal, Supratik Pal and B. Basu, `The Berry phase in inflationary cosmology`, **CQG 30**, 125002 (2013).

## 3.1 INTRODUCTION

The *Berry phase,* is a phase acquired over the course of a cycle when the system is subjected to cyclic *adiabatic* process, which results from the geometrical properties of the parameter space of the *Hamiltonian*. The phenomenon was first discovered in 1956 which is known as *Pancharatnam phase* [132], and rediscovered in 1984 by *Berry* [99]. Although, the *geometric phase* [133] was known long ago *a la Aharonov-Bohm effect* [134], the general context of a quantum-mechanical state developing *adiabatically* in time under a slowly varying parameter dependent *Hamiltonian* has been analyzed by *Berry* [99], who argued that when the parameters return *adiabatic*ally to their initial values after traversing a closed path, the wave-function acquires a *geometric phase factor* depending on the path, in addition to the well-known *dynamical phase factor*. Since its inception the *Berry* phase has been the subject of a variety of theoretical and experimental investigations [133]. Some analyses have been made to study this phase in the area of cosmology and gravitation also [135, 136]. Investigations were also made to study the behavior of a scalar particle in a class of stationary space-time backgrounds and the emergence of the *Berry* phase in dynamics of a particle in the presence of a rotating cosmic string [137]. The gravitational analogue of the *Aharonov-Bohm* effect in the spinning cosmic string space-time background was also obtained [138]. Within a typical framework of cosmological model

*Theoretical and Observational Aspects of Cosmological Inflation*     **48**     ©*Barun Kumar Pal*

the *Berry* phase has been shown to be associated with the decay width of the state in case of some well known examples of vacuum instability [139].

Among other motivations, the inflationary scenario [1] is successful to a great extent in explaining the origin of cosmological perturbation seeds [93]. The accelerated expansion converts the initial vacuum quantum fluctuations into macroscopic cosmological perturbations. So, measurement of any quantum property which reflects on classical observables serve as a supplementary probe of inflationary cosmology, complementing the well-known CMB polarization measurements [4, 5]. Though this is an important issue, we notice that there has been very little study in the literature which deals with proposing measurable quantities which may measure the genuine quantum property of the seeds of classical cosmological perturbations. The only proposition that has drawn our attention is via violation of the *Bell*'s inequality [140]. This has led us to investigate for the potentiality of the *Berry phase* in providing a measurable quantum property which is inherent in the macroscopic character of classical cosmological perturbations.

In this chapter we shall see that, for *slow-roll* inflation, the total accumulated phase gained by each mode during *sub-Hubble* oscillations (in the *adiabatic* limit) is a new parameter made of corresponding (scalar and tensor) spectral indices. So in principle, measurement of the *Berry phase* of the quantum cosmological perturbations provides us an indirect route for estimating spectral indices and other observable parameters therefrom. Further, since tensor spectral index is related to the tensor to scalar amplitude ratio through the consistency relation, the *Berry phase* may be utilized to act as a supplementary probe of inflationary cosmology. This is precisely where our goal lies in this chapter.

## 3.2 PRELIMINARIES

Before going into the details, we shall first derive mathematical expression for the *Berry phase* and discuss briefly about the *Lewis-Riesenfeld invariant formulation* [121] which will be employed in our subsequent analyses. Then we shall deduce the *Berry phase* associated with a *generalized time dependent harmonic oscillator*. From now on we shall use the terms '*geometric phase*' and the '*Berry phase*' interchangeably.

### 3.2.1 BERRY PHASE: DEFINITION

Let's consider a *Hamiltonian* $\mathbf{H}(\mathbf{R})$ that depends on $R_1,\ R_2,\ ...\ R_n$ – components of a vector $\mathbf{R}$. We assume that $\mathbf{H}(\mathbf{R})$ has at least one discrete non-degenerate *eigenvalue* $E_i(\mathbf{R})$ associated with the *eigenstate* $|\Psi_i(\mathbf{R})\rangle$. Now if $\mathbf{R}$ varies with time, then $|\Psi_i(\mathbf{R})\rangle$ is not an exact solution of the following time dependent *Schrödinger* equation

$$\hat{\mathbf{H}}|\Psi\rangle = i\frac{\partial}{\partial \eta}|\Psi\rangle. \tag{3.1}$$

But if **R** changes *slowly*, then the system adjusts itself with the changing *Hamiltonian*, instead of jumping to another *eigenstate*. Let $\mathbf{R}(\eta/\Gamma)$ evolves over a time interval $0 \leq \eta \leq \Gamma$, larger the value of $\Gamma$ slower is the evolution. If at a time $\eta = 0$ the system is in the state $|\Psi_i[\mathbf{R}(0)]\rangle$, then at time $\eta = \Gamma$ the state will be $e^{i\phi_i(\Gamma)}|\Psi_i[\mathbf{R}(1)]\rangle$ with probability approaching 1 as $\Gamma \to \infty$ according to *quantum adiabatic theorem* [141]. The phase $\phi_i(\eta)$ is obtained from the *Schrödinger* Eqn.(3.1),

$$\hat{\mathbf{H}}(\mathbf{R})e^{i\phi_i(\eta)}|\Psi(\mathbf{R})\rangle = i\frac{\partial}{\partial \eta}e^{i\phi_i(\eta)}|\Psi(\mathbf{R})\rangle. \tag{3.2}$$

Now multiplying both sides of Eqn.(3.2) by $e^{-i\phi_i(\eta)}\langle\Psi(\mathbf{R})|$ and using $\hat{\mathbf{H}}|\Psi_i\rangle = E_i|\Psi_i\rangle$ we obtain

$$\frac{\partial \phi_i}{\partial \eta} = -E_i(\mathbf{R}) + i\langle\Psi_i(\mathbf{R})|\tfrac{\partial}{\partial \eta}|\Psi_i(\mathbf{R})\rangle. \tag{3.3}$$

So the integration between $\eta = 0$ to $\eta = \Gamma$ yields

$$\phi_i(\Gamma) - \phi_i(0) = -\int_0^\Gamma E_i(\mathbf{R})d\eta + i\int_0^\Gamma \langle\Psi_i(\mathbf{R})|\tfrac{\partial}{\partial \eta}|\Psi_i(\mathbf{R})\rangle d\eta. \tag{3.4}$$

The first term on the *r.h.s.* of Eqn.(3.4) is the usual *dynamical phase* and the second term is known as the *geometric phase* or most commonly the *Berry phase*.

### 3.2.2 LEWIS-RIESENFELD INVARIANT OPERATOR TECHNIQUE

Over the past few decades the *Lewis-Riesenfeld invariant* formulation [121] has become an effective tool in the treatment of time dependent quantum systems. In this formulation, we first look for a nontrivial *Hermitian* operator $I(\eta)$ satisfying the *Liouville von-Neumann* equation

$$\frac{dI}{d\eta} = -i[I, \mathbf{H}] + \frac{\partial I}{\partial \eta} = 0, \tag{3.5}$$

where *H* is the *Hamiltonian* of the system. Let's assume that there exists an operator *I* satisfying Eqn.(3.5) which does not contain time differentiation. Then the solutions of the *Schrödinger* Eqn.(3.1) can be written in the following form [142]

$$\Psi_n = e^{i\alpha_n(\eta)}\Theta_n, \quad n = 0, 1, 2... \tag{3.6}$$

where $\Theta_n$ are the eigenfunctions of *I* and $\alpha_n$ are called *Lewis phases*. The *Lewis phase* is calculated from its definition [121, 123, 142] which is

$$\frac{d\alpha_n}{d\eta} = \left\langle \Theta_n \left| i\frac{\partial}{\partial \eta} - \mathbf{H} \right| \Theta_n \right\rangle. \tag{3.7}$$

Therefore, given any system we can find the associated *Schrödinger* wave-function using the *Lewis-Riesenfeld* invariant operator formulation.

### 3.2.3 TIME DEPENDENT HARMONIC OSCILLATOR AND THE BERRY PHASE

We now apply *dynamical invariant operator technique* to a time dependent *harmonic oscillator* in order to calculate the corresponding *Berry phase*. The *Hamiltonian* describing a *generalized time dependent harmonic oscillator* has the following form

$$\boldsymbol{H} = \tfrac{1}{2}\left[X(\eta)q^2 + Y(\eta)(pq + qp) + Z(\eta)p^2\right] \tag{3.8}$$

where $q$ and $p$ are position and momentum operators satisfying usual commutation relation $[p, q] = i$ and $X$, $Y$, $Z$ are time dependent parameters. The above *Hamiltonian* describes a *harmonic oscillator* with frequency $\omega \equiv (XZ - Y^2)^{\frac{1}{2}}$, provided $XZ - Y^2 > 0$. The above system (3.8) possesses a well known *dynamical invariant* [143] given by

$$I = \tfrac{1}{2}\left\{\frac{q^2}{r^2} + \left[r\left(p + \frac{Y}{Z}q\right) - \frac{r'}{Z}q\right]^2\right\} \tag{3.9}$$

where $r \equiv Z^{1/2}\rho$ and $\rho$ satisfies the following *Milne-Pinney* equation,

$$\rho'' + \Omega^2\rho = \rho^{-3}, \text{ with } \Omega^2 \equiv \omega^2 + \tfrac{1}{2}\frac{Z''}{Z} - \tfrac{3}{4}\left(\frac{Z'}{Z}\right)^2 - Z\left(\frac{Y}{Z}\right)'. \tag{3.10}$$

The matrix elements necessary to calculate the *Lewis* phase are given by [142]

$$\langle\Theta_n|\partial/\partial\eta|\Theta_n\rangle = \tfrac{i}{2}\left(n + \tfrac{1}{2}\right)\left(\rho\rho'' - \rho'^2\right) \tag{3.11}$$

$$\langle\Theta_n|\boldsymbol{H}|\Theta_n\rangle = \tfrac{1}{2}\left(n + \tfrac{1}{2}\right)\left(\rho'^2 + \Omega^2\rho^2 - \rho^{-2}\right). \tag{3.12}$$

So, from Eqn.(3.7) we can immediately calculate the *Lewis phases*, written explicitly, they look

$$\alpha_n = -\left(n + \tfrac{1}{2}\right)\int\frac{d\eta}{\rho(\eta)^2}. \tag{3.13}$$

Once *Lewis phase* is known, we can straightway write down solutions of the *Schrödinger* equation. Therefore, *Lewis-Riesenfeld* technique provides an elegant way to solve time dependent systems in a very simple manner.

Now if we assume the *invariant operator I* is $\Gamma$ periodic and its *eigenvalues* are non-degenerate, then the *Berry phase* associated with the system (3.8) turns out to be

$$\gamma_n \equiv i\int_0^\Gamma\left\langle\Theta_n\left|\frac{\partial}{\partial\eta}\right|\Theta_n\right\rangle d\eta = -\tfrac{1}{2}\left(n + \tfrac{1}{2}\right)\int_0^\Gamma\left(\rho_k^{-2} - \rho_k^2\Omega^2 - \rho_k'^2\right)d\eta. \tag{3.14}$$

In the following we shall use the result (3.14) to derive a cosmological analogue of the *Berry phase* in the context of inflationary cosmological perturbations.

## 3.3 LINEAR COSMOLOGICAL PERTURBATIONS

The quantum fluctuations in the inflaton field are realized by *Mukhanov-Sasaki* Eqn.(1.66) which is analogous to a *time dependent harmonic oscillator*. The associated physical mechanism for cosmological perturbations can be reduced to the quantization of a parametric oscillator leading to particle creation due to interaction with the gravitational field and may be termed as cosmological *Schwinger effect* [144]. The relation between the *Berry phase* and its classical analogue, the *Hannay angle* [145], has been studied for the generalized time dependent harmonic oscillator [146, 147]. Naturally, one may expect to derive the cosmological analogue of the *Berry phase* in the context of inflationary perturbations and search for possible consequences via observable parameters. To this end, we first derive an exact wave-function for the system of inflationary cosmological perturbations by solving the associated *Schrödinger* equation. The relation [143, 148] between the dynamical invariant [121, 122, 123, 142] and the *geometric phase* will then be utilized to derive the corresponding *Berry phase*.

The *Hamiltonian* corresponding to the scalar part of the action (1.69) is given by Eqn.(1.75). Now promoting the fields to the operator and taking *Fourier* transformation we obtain

$$\begin{aligned} \hat{\mathbf{H}}_{v_{\mathbf{k}}} &= \tfrac{1}{2}\left[\hat{\pi}_{v_{\mathbf{k}},1}^2 + \left(k^2 - \frac{z''}{z}\right)\hat{v}_{\mathbf{k},1}^2\right] + \tfrac{1}{2}\left[\hat{\pi}_{v_{\mathbf{k}},2}^2 + \left(k^2 - \frac{z''}{z}\right)\hat{v}_{\mathbf{k},2}^2\right] \\ &\equiv \hat{\mathbf{H}}_{v_{\mathbf{k}},1} + \hat{\mathbf{H}}_{v_{\mathbf{k}},2}, \end{aligned} \quad (3.15)$$

where we have decomposed $\hat{v}_{\mathbf{k}} \equiv \hat{v}_{\mathbf{k},1} + i\,\hat{v}_{\mathbf{k},2}$ and $\hat{\pi}_{v_{\mathbf{k}}} \equiv \hat{\pi}_{v_{\mathbf{k}},1} + i\,\hat{\pi}_{v_{\mathbf{k}},2}$ into their real and imaginary parts. Similarly, the *Hamiltonian* corresponding to the tensorial part of the action (1.69) is

$$\begin{aligned} \hat{\mathbf{H}}_{u_{\mathbf{k}}} &= \tfrac{1}{2}\left[\hat{\pi}_{u_{\mathbf{k}},1}^2 + \left(k^2 - \frac{a''}{a}\right)\hat{u}_{\mathbf{k},1}^2\right] + \tfrac{1}{2}\left[\hat{\pi}_{u_{\mathbf{k}},2}^2 + \left(k^2 - \frac{a''}{a}\right)\hat{u}_{\mathbf{k},2}^2\right] \\ &\equiv \hat{\mathbf{H}}_{u_{\mathbf{k}},1} + \hat{\mathbf{H}}_{u_{\mathbf{k}},2}, \end{aligned} \quad (3.16)$$

where we have decomposed $\hat{v}_{\mathbf{k}} \equiv \hat{u}_{\mathbf{k},1} + i\,\hat{u}_{\mathbf{k},2}$ and $\hat{\pi}_{u_{\mathbf{k}}} \equiv \hat{\pi}_{u_{\mathbf{k}},1} + i\,\hat{\pi}_{u_{\mathbf{k}},2}$ into their real and imaginary parts.

Thus, for both the scalar and tensor perturbations, *Hamiltonians* are sum of two time dependent harmonic oscillators, each of them having the following form

$$\hat{\mathbf{H}}_{j\mathbf{k}} = \tfrac{1}{2}\left[\hat{p}_{j\mathbf{k}}^2 + \omega^2 \hat{q}_{j\mathbf{k}}^2\right], \; j = 1, 2 \quad (3.17)$$

where $\hat{q}_{j\mathbf{k}} \equiv \hat{v}_{\mathbf{k},j},\, \hat{u}_{\mathbf{k},j}$; $\hat{p}_{j\mathbf{k}} \equiv \hat{\pi}_{v_{\mathbf{k}},j},\, \hat{\pi}_{u_{\mathbf{k}},j}$; $j = 1, 2$ and $\omega = \sqrt{k^2 - \frac{z''}{z}},\, \sqrt{k^2 - \frac{a''}{a}}$ for scalar and tensor respectively. It should be noted that for complete solution of the *Schrödinger* Eqn.(3.1) for the *Hamiltonian* (3.17) we have to deal with two situations:

1. $k^2 > \frac{z''}{z}, \frac{a''}{a}$ where $\omega$ is real, which corresponds to the sub-Hubble modes and
2. $k^2 < \frac{z''}{z}, \frac{a''}{a}$ which corresponds to super-Hubble modes having imaginary frequency.

For the later case, the *Hamiltonian* can be re-written as

$$\hat{\mathbf{H}}_{j\mathbf{k}} = \tfrac{1}{2}\left[\hat{p}_{j\mathbf{k}}^2 - \omega_I^2 \hat{q}_{j\mathbf{k}}^2\right] \quad (3.18)$$

which represents an *inverted harmonic oscillator* with time dependent frequency given by $i\omega_I = i\sqrt{\frac{z''}{z} - k^2}, i\sqrt{\frac{a''}{a} - k^2}$ for scalar and tensor respectively. Though in further derivations we will not be concerned with the *super-Hubble* modes, but at this stage we are ready to provide the solutions for the whole spectrum.

We find the solution to the *Schrödinger* equation for the *Hamiltonians* (3.17) and (3.18) using the *Lewis-Riesenfeld* invariant formulation [142]. To be precise, we want to analyze the situation by solving the associated *Schrödinger* equation

$$\hat{\mathbf{H}}_{\mathbf{k}}\Psi \equiv (\hat{\mathbf{H}}_{1\mathbf{k}} + \hat{\mathbf{H}}_{2\mathbf{k}})\Psi = i\frac{\partial}{\partial\eta}\Psi, \tag{3.19}$$

were $\mathbf{H}_k \equiv \mathbf{H}_{1\mathbf{k}} + \mathbf{H}_{2\mathbf{k}}$. We shall solve the *Schrödinger* equation in two different regimes – *sub-Hubble* and *super-Hubble*.

### 3.3.1 SUB-HUBBLE MODES

The associated *Hamiltonian* for the *sub-Hubble* modes is given by (3.17). Following the usual path [121, 123, 142, 146] we obtain

$$I_k = \tfrac{1}{2}\left[\frac{q_{1\mathbf{k}}^2}{\rho_k^2} + (\rho_k p_{1\mathbf{k}} - \rho_k' q_{1\mathbf{k}})^2\right] + \tfrac{1}{2}\left[\frac{q_{2\mathbf{k}}^2}{\rho_k^2} + (\rho_k p_{2\mathbf{k}} - \rho_k' q_{2\mathbf{k}})^2\right] \equiv I_1 + I_2 \tag{3.20}$$

where $\rho_k$ is a time dependent real function satisfying the following *Milne-Pinney* equation

$$\rho_k'' + \omega^2(\eta, k)\rho_k = \rho_k^{-3}. \tag{3.21}$$

In order to solve the *Schrödinger* Eqn.(3.19), we also need the eigenstates of the operator $I_k$ governed by the eigenvalue equation

$$I_{\mathbf{k}}\Theta_{n_1,n_2}(q_{1\mathbf{k}}, q_{2\mathbf{k}}, \eta) = \lambda_{n_1,n_2}\Theta_{n_1,n_2}(q_{1\mathbf{k}}, q_{2\mathbf{k}}, \eta), \tag{3.22}$$

which turns out to be [136]

$$\Theta_{n_1,n_2} = \frac{\bar{H}_{n_1}\left[\frac{q_{1\mathbf{k}}}{\rho_k}\right]\bar{H}_{n_2}\left[\frac{q_{2\mathbf{k}}}{\rho_k}\right]}{\sqrt[4]{\pi^2 2^{2(n_1+n_2)}(n_1!n_2!)^2\rho_k^4}} \times \exp\left[\frac{i}{2}\left(\frac{\rho_k'}{\rho_k} + \frac{i}{\rho_k^2}\right)(q_{1\mathbf{k}}^2 + q_{2\mathbf{k}}^2)\right] \tag{3.23}$$

where $\bar{H}_n$ are the *Hermite polynomials* of order $n$ and the associated eigenvalues are

$$\lambda_{n_1,n_2} = \left(n_1 + \tfrac{1}{2}\right) + \left(n_2 + \tfrac{1}{2}\right). \tag{3.24}$$

The *Lewis phase* can be found from its definition (3.7), which in this case turns out to be

$$\alpha_{n_1,n_2} = -(n_1 + n_2 + 1)\int\frac{d\eta}{\rho_k^2} \tag{3.25}$$

So, eigenstates of the *Hamiltonian* (3.17) are now completely known and are given by

$$\Psi_{n_1,n_2} = \frac{e^{i\alpha_{n_1,n_2}(\eta)}\bar{H}_{n_1}\left[\frac{q_{1\mathbf{k}}}{\rho_k}\right]\bar{H}_{n_2}\left[\frac{q_{2\mathbf{k}}}{\rho_k}\right]}{\sqrt[4]{\pi^2 2^{2(n_1+n_2)}(n_1!n_2!)^2\rho_k^4}} \times \exp\left[\frac{i}{2}\left(\frac{\rho_k'}{\rho_k}+\frac{i}{\rho_k^2}\right)(q_{1\mathbf{k}}^2+q_{2\mathbf{k}}^2)\right]. \quad (3.26)$$

Eqn.(3.26) represents the wave-function for the *sub-Hubble* modes of the inflationary cosmological perturbations.

### 3.3.2 SUPER-HUBBLE MODES

The *super-Hubble* evolution of the perturbation modes are governed by the *Hamiltonian* (3.18), which is nothing but time dependent *inverted harmonic oscillator*. The time dependent *inverted harmonic oscillator* is exactly solvable just like the standard time dependent harmonic oscillator. However, the physics of the time dependent inverted oscillator is very different [149, 150, 151], it has a wholly continuous energy spectrum varying from minus to plus infinity; its energy eigenstates are no longer square-integrable and they are doubly degenerate with respect to either the incident direction or, alternatively, the parity.

In this case the invariant operator can be worked out in a similar way as before which for the system under consideration turns out to be [150],

$$I_k = \frac{1}{2}\left[-\frac{q_{1\mathbf{k}}^2}{\rho_k^2}+(\rho_k p_{1\mathbf{k}}-\rho_k' q_{1\mathbf{k}})^2\right]+\frac{1}{2}\left[-\frac{q_{2\mathbf{k}}^2}{\rho_k^2}+(\rho_k p_{2\mathbf{k}}-\rho_k' q_{2\mathbf{k}})^2\right] \equiv I_1+I_2 \quad (3.27)$$

where $\rho_k$ now satisfies following auxiliary equation

$$\rho_k'' - \omega_I^2(\eta,k)\rho_k = -\rho_k^{-3}. \quad (3.28)$$

The eigenstates of the operator $I_k$ is governed by the eigenvalue equation

$$I_{\mathbf{k}}\Theta_{\lambda_1,\lambda_2}(q_{1\mathbf{k}},q_{2\mathbf{k}},\eta) = \lambda_{\lambda_1,\lambda_2}\Theta_{\lambda_1,\lambda_2}(q_{1\mathbf{k}},q_{2\mathbf{k}},\eta). \quad (3.29)$$

Following the steps as in [150, 151], the eigenstates of the operator $I_k$ turns out to be

$$\Theta_{\lambda_1,\lambda_2} = \frac{1}{\rho_k}\exp\left[\frac{i}{2}\frac{\rho_k'}{\rho_k}(q_{1\mathbf{k}}^2+q_{2\mathbf{k}}^2)\right] \times W_{\lambda_1}\left(\frac{\sqrt{2}q_{1\mathbf{k}}}{\rho_k},\lambda_1\right)W_{\lambda_2}\left(\frac{\sqrt{2}q_{2\mathbf{k}}}{\rho_k},\lambda_2\right) \quad (3.30)$$

where $W_{\lambda_1}$ and $W_{\lambda_2}$ are parabolic cylinder or *Weber functions*. Consequently, the solution to the *Schrödinger* equation (3.19) for the *Hamiltonian* (3.18) is now completely known –

$$\Psi_{\lambda_1,\lambda_2} = e^{i\alpha_{\lambda_1,\lambda_2}}\Theta_{\lambda_1,\lambda_2}, \text{ with } \alpha_{\lambda_1,\lambda_2} = -(\lambda_1+\lambda_2)\int\frac{d\eta}{\rho_k^2}. \quad (3.31)$$

Thus, we have now calculated the wave-functions for both the *sub-Hubble* and *super-Hubble* evolutions of the cosmological perturbation modes generated during inflation.

In order to quantify the fluctuations produced during inflation, the theory of general relativistic perturbations has to be employed. However, in practice the field equations cannot be solved with their full generalities, some approximations has to be made. The linear order approximation to the cosmological perturbation has been developed to a high degree of sophistication during the last few decades [91, 93]. Although, there are discussions on the deviation from the first order approximation from the observation [152] and theoretical approaches [87, 88, 98, 153, 154] through the *non-Gaussianity*, *non-adiabaticity* and so on. But, at a first go linear order perturbation theory has remained as the primary tool for investigating the behavior of the fluctuations produced during inflation. The developments in the observations were also supported by the theoretical sophistication of the linear order cosmological perturbation theory. At this juncture we also remind that our objective here is to see whether a link can be established between the quantum property of the seeds of classical cosmological perturbations and inflationary observables through the derivation of the associated *Berry phase*, for which consideration of the *sub-Hubble* modes are appropriate. Keeping all these points under consideration, we concentrate on the evolution of the *sub-Hubble* perturbation modes within the framework of linearized perturbation theory.

## 3.4 BERRY PHASE FOR THE SUB-HUBBLE MODES

In order to calculate the *Berry phase* for the *sub-Hubble* modes we shall first make use of the following identities

$$\frac{z''}{z}v^2 = \left(\frac{z'}{z}\right)^2 v^2 - 2\frac{z'}{z}vv' + \left[\frac{z'}{z}v^2\right]', \quad \frac{a''}{a}u^2 = \left(\frac{a'}{a}\right)^2 u^2 - 2\frac{a'}{a}uu' + \left[\frac{a'}{a}u^2\right]'. \quad (3.32)$$

Then the action (1.69), for the scalar perturbation only, can be expressed as [155]

$$\mathbf{S}^S = \tfrac{1}{2} \int d\eta d\mathbf{x} \left[ v'^2 - \delta^{ij}\partial_i v \partial_j v - 2\left(\frac{z'}{z}\right)^2 vv' + \frac{z'}{z}v^2 \right] \quad (3.33)$$

which we find more convenient to work with. The *Hamiltonian* associated with the above action turns out to be

$$\mathbf{H}^S = \tfrac{1}{2} \int d^3x \left[ \pi_v^2 + \delta^{ij}\partial_i v \partial_j v + 2\frac{z'}{z} v \pi_v \right] \quad (3.34)$$

where now $\pi_v = v' - \frac{z'}{z}v$. Promoting the fields to operators and taking the *Fourier* decomposition we find the *Hamiltonian* density operator corresponding to the action (3.33):

$$\begin{aligned}
\hat{\mathbf{H}}^S_{\mathbf{k}} &= \tfrac{1}{2} \left[ \hat{\pi}^2_{v_{\mathbf{k}},1} + \frac{z'}{z}\left(\hat{\pi}_{v_{\mathbf{k}},1}\hat{v}_{\mathbf{k},1} + \hat{v}_{\mathbf{k},1}\hat{\pi}_{\mathbf{k},1}\right) + k^2\hat{v}^2_{\mathbf{k},1} \right] \\
&+ \tfrac{1}{2} \left[ \hat{\pi}^2_{v_{\mathbf{k}},2} + \frac{z'}{z}\left(\hat{\pi}_{v_{\mathbf{k}},2}\hat{v}_{\mathbf{k},2} + \hat{v}_{\mathbf{k},2}\hat{\pi}_{\mathbf{k},2}\right) + k^2\hat{v}^2_{\mathbf{k},2} \right] \equiv \hat{\mathbf{H}}^S_{\mathbf{k},1} + \hat{\mathbf{H}}^S_{\mathbf{k},1}. \quad (3.35)
\end{aligned}$$

Similarly, the *Hamiltonian* operator corresponding to tensor perturbations is found to be

$$\hat{\mathbf{H}}_{\mathbf{k}}^{T} = \tfrac{1}{2}\left[\hat{\pi}_{u_{\mathbf{k},1}}^{2} + \frac{a'}{a}\left(\hat{\pi}_{u_{\mathbf{k},1}}\hat{u}_{\mathbf{k},1} + \hat{u}_{\mathbf{k},1}\hat{\pi}_{\mathbf{k},1}\right) + k^{2}\hat{u}_{\mathbf{k},1}^{2}\right]$$
$$+ \tfrac{1}{2}\left[\hat{\pi}_{u_{\mathbf{k},2}}^{2} + \frac{a'}{a}\left(\hat{\pi}_{u_{\mathbf{k},2}}\hat{u}_{\mathbf{k},2} + \hat{u}_{\mathbf{k},2}\hat{\pi}_{\mathbf{k},2}\right) + k^{2}\hat{u}_{\mathbf{k},2}^{2}\right] \equiv \hat{\mathbf{H}}_{\mathbf{k},1}^{T} + \hat{\mathbf{H}}_{\mathbf{k},1}^{T}. \quad (3.36)$$

In compact notation, general form of the *Hamiltonians* can be written as a sum of two generalized time dependent harmonic oscillators as

$$\hat{\mathbf{H}}_{\mathbf{k},j} = \tfrac{1}{2}\left[k^{2}\hat{q}_{j\mathbf{k}}^{2} + Y(\eta)\left(\hat{p}_{j\mathbf{k}}\hat{q}_{j\mathbf{k}} + \hat{q}_{j\mathbf{k}}\hat{p}_{j\mathbf{k}}\right) + \hat{p}_{j\mathbf{k}}^{2}\right] \quad (3.37)$$

where $\hat{q}_{j\mathbf{k}} = \hat{v}_{\mathbf{k},j},\ \hat{u}_{\mathbf{k},j};\ \hat{p}_{j\mathbf{k}} = \hat{\pi}_{v_{\mathbf{k}},j},\ \hat{\pi}_{u_{\mathbf{k}},j}$ and $Y = \frac{z'}{z},\ \frac{a'}{a}$ for the scalar and tensor modes respectively with the frequency given by $\omega = \sqrt{k^{2} - Y^{2}}$ and $j = 1, 2$.

Following the usual trail as given in [121, 123, 142] we find

$$I_{k} = \tfrac{1}{2}\left[\frac{q_{1\mathbf{k}}^{2}}{\rho_{k}^{2}} + \{\rho_{k}\left[p_{1\mathbf{k}} + Y q_{1\mathbf{k}}\right] - \rho_{k}'q_{1\mathbf{k}}\}^{2}\right] + \tfrac{1}{2}\left[\frac{q_{2\mathbf{k}}^{2}}{\rho_{k}^{2}} + \{\rho_{k}\left[p_{2\mathbf{k}} + Y q_{2\mathbf{k}}\right] - \rho_{k}'q_{2\mathbf{k}}\}^{2}\right]$$
$$= I_{1} + I_{2} \quad (3.38)$$

where $\rho_{k}$ now satisfies the following equation

$$\rho_{k}'' + \Omega^{2}(\eta, k)\rho_{k} = \rho_{k}^{-3}, \text{ with } \Omega^{2} = \omega^{2} - \tfrac{dY}{d\eta}. \quad (3.39)$$

The eigenstates of the operator $I_{k}$ turn out to be

$$\Theta_{n_{1},n_{2}} = \frac{\bar{H}_{n_{1}}\left[\frac{q_{1\mathbf{k}}}{\rho_{k}}\right]\bar{H}_{n_{2}}\left[\frac{q_{2\mathbf{k}}}{\rho_{k}}\right]}{\sqrt[4]{\pi^{2}2^{2(n_{1}+n_{2})}(n_{1}!n_{2}!)^{2}\rho_{k}^{4}}} \times \exp\left[\frac{i}{2}\left(\frac{\rho_{k}'}{\rho_{k}} - Y(\eta) + \frac{i}{\rho_{k}^{2}}\right)\left(q_{1\mathbf{k}}^{2} + q_{2\mathbf{k}}^{2}\right)\right] \quad (3.40)$$

So the eigenstates of the *Hamiltonian* are now given by

$$\Psi_{n_{1},n_{2}} = e^{i\alpha_{n_{1},n_{2}}(\eta)}\Theta_{n_{1},n_{2}} \quad (3.41)$$

where the *Lewis* phase is given by: $\alpha_{n_{1},n_{2}} = -(n_{1} + n_{2} + 1)\int d\eta/\rho_{k}^{2}$. The phase $\alpha_{n_{1},n_{2}}(\eta)$ is the combination of the *dynamical phase* and the *geometric phase* which is very clear from the Eqn.(3.7). Once the *Lewis* phase is known, we can utilize it to derive the geometric phase associated with the system corresponding to the particle creation through the vacuum quantum fluctuations during inflation.

But before proceeding in this direction, we would like to present the general wave-function for the vacuum state of the inflationary cosmological perturbations. To this end, let us consider the parametric harmonic oscillator's *Hamiltonian*

$$\hat{\mathbf{H}}_{\mathbf{k}}^{S} = \tfrac{1}{2}\left[\hat{\pi}_{v_{\mathbf{k}}}^{2} + \frac{z'}{z}\left(\hat{\pi}_{\mathbf{k}}\hat{v}_{\mathbf{k}} + \hat{v}_{\mathbf{k}}\hat{\pi}_{v_{\mathbf{k}}}\right) + k^{2}\hat{v}_{\mathbf{k}}^{2}\right] \quad (3.42)$$

which refers to (3.35) and can be solved analytically for the vacuum. By the following similarity transformation $\hat{\tilde{H}}_{\mathbf{k}} \equiv A\tilde{H}_{\mathbf{k}}A^{-1}$ with $A = e^{-i\frac{z'}{2z}v_{\mathbf{k}}^2}$, the *Hamiltonian* (3.42) can be reduced to the following form

$$\tilde{\mathbf{H}}_{\mathbf{k}}^S = \tfrac{1}{2}\left[\hat{\pi}_{v_{\mathbf{k}}}^2 + \left(k^2 - \frac{z'^2}{z^2}\right)\hat{v}_{\mathbf{k}}^2\right]. \tag{3.43}$$

The wave-function for the *Hamiltonian* (3.43) is well known and has the form [144, 156]

$$\tilde{\Psi}_k = N_k e^{-\Omega_k v_k^2}, \text{ where } |N_k| = \left(\frac{2Re\Omega_k}{\pi}\right)^{1/4}, \quad \Omega_k = -\frac{i}{2}\frac{f_k'}{f_k} \text{ with } f_k'' + \left(k^2 - \frac{z'^2}{z^2}\right)f_k = 0. \tag{3.44}$$

For the vacuum state we know

$$f_k = \frac{1}{\sqrt{2k}}e^{ik\eta} \tag{3.45}$$

which provides us

$$\tilde{\Psi}_k = \left(\frac{k}{\pi}\right)^{1/4} e^{-\frac{k}{2}v_k^2}. \tag{3.46}$$

Hence the vacuum state wave-function for the inflationary cosmological scalar perturbations turns out to be

$$\Psi_k^S = \left(\frac{k}{\pi}\right)^{1/2} e^{-\left(i\frac{z'}{z}+k\right)\left(v_{k,1}^2+v_{k,2}^2\right)}. \tag{3.47}$$

Similarly, we can write the vacuum state wave-function for the tensor perturbation as

$$\Psi_k^T = \left(\frac{k}{\pi}\right)^{1/2} e^{-\left(i\frac{z'}{z}+k\right)\left(u_{k,1}^2+u_{k,2}^2\right)}. \tag{3.48}$$

The phase part, which is the combination of the *dynamical phase* and the *geometric phase* of the wave-function is now explicit but it is not easy to separate out the *geometric phase* from those expressions. But, in our present framework this can be done using the results already derived, which is what we do in the following.

Now, assuming the invariant $I_k(\eta)$ is $\Gamma$ periodic and its eigenvalues are non-degenerate, from Eqns.(3.40) and (3.25) we obtain the following relation for the *Berry phase*

$$\begin{aligned}\gamma_{n_1,n_2,k} &\equiv i\int_0^\Gamma \left\langle \Theta_{n_1,n_2}\left|\frac{\partial}{\partial \eta}\right|\Theta_{n_1,n_2}\right\rangle d\eta \\ &= -\tfrac{1}{2}(n_1+n_2+1)\int_0^\Gamma \left(\frac{1}{\rho_k^2} - \rho_k^2\omega^2 - (\rho_k')^2\right)d\eta.\end{aligned} \tag{3.49}$$

To get a deeper physical insight the quantitative estimation of the *Berry* phase is very important. Eqn.(3.49) tells us that for this estimation, the knowledge of $\rho_k$ is essential but the solution of Eqn.(3.21) is very difficult to obtain. Another point to be carefully handled is to set the value of the parameter $\Gamma$. Keeping all these in mind and considering compatible physical conditions we proceed as follows.

First we note that in the *adiabatic limit* (which is quite justified for *sub-Hubble* modes) Eqn.(3.39) can be solved [121] by a series of powers in *adiabatic* parameter, $\delta (\ll 1)$. To this end, we define a slowly varying time variable as $\tau = \delta\eta$ and write the solution to *Milne-Pinney*

Eqn.(3.39) as
$$\rho_k = \rho_0 + \delta\rho_1 + \delta^2\rho_2 + ... \tag{3.50}$$

Inserting this expansion into the *Milne-Pinney* equation we get

$$\delta^2 \rho_0 \ddot{\rho}_0 + \rho_0^2 \left[1 + 2\delta\rho_0\rho_1 + \delta^2\rho_1^2 + 2\delta^2\rho_0\rho_2\right]\left(\omega^2 - \delta\dot{Y}\right)$$
$$= \frac{1}{\rho_0^2 + 2\delta\rho_0\rho_1 + \delta^2\rho_1^2 + 2\delta^2\rho_0\rho_2} + O(\delta^3). \tag{3.51}$$

Here dot represents derivative w.r.t. new time variable $\tau$. Collecting the zeroth order terms from both sides we obtain $\rho_0^2 = \omega^{-\frac{1}{2}}$. Now the integrand of Eqn.(3.49) can be rewritten as

$$\frac{1}{\rho_k^2} - \rho_k^2\omega^2 - (\rho_k')^2 = \rho_k\rho_k'' - (\rho_k')^2 - \rho_k^2 Y' \simeq \delta\rho_0^2\dot{Y} + \delta^2\left[\rho_0\ddot{\rho}_0 - \dot{\rho}^2 - 2\rho_0\rho_1\dot{Y}\right] + O(\delta^3)$$
$$= \delta\omega^{-\frac{1}{2}}\dot{Y} + O(\delta^2) \tag{3.52}$$

Thus for the ground state of the system, in the *adiabatic limit*, the *Berry phase* for a particular perturbation mode can be evaluated up to the first order in $\delta$, which is given by

$$\gamma_k^{(S,\,T)} = -\frac{1}{2}\int_0^\Gamma \frac{\delta\dot{Y}}{\sqrt{k^2 - Y^2}}d\eta = -\frac{1}{2}\int_0^\Gamma \frac{Y'}{\sqrt{k^2 - Y^2}}d\eta \tag{3.53}$$

where the superscripts $S$ and $T$ stand for scalar and tensor modes respectively. One may note that our result (3.53) coincides with that of *Berry* [157].

Our next task is to fix the value of the parameter $\Gamma$. To this end we shall calculate the *total Berry phase* accumulated by each mode during *sub-Hubble* evolution in the inflationary era. For the ground state of the system this turns out to be

$$\gamma_{k\ sub}^{S,T} = -\frac{1}{2}\lim_{\eta'\to-\infty}\int_{\eta'}^{\eta_0^{S,T}} \frac{Y'}{\sqrt{k^2 - Y^2}}d\eta \tag{3.54}$$

where $\eta_0^{S,T}$ is the conformal time which satisfies the relation $k^2 = \left[Y(\eta_0^{S,T})\right]^2$ so that the modes are within the horizon and oscillating with real frequencies. A non-zero value of the parameter $\gamma_{k\ sub}^{S,T}$ will ensure that there are some nontrivial effects of the curved space-time background on the evolution of the quantum fluctuations and may play an important role in the growth of inflationary cosmological perturbations.

## 3.5 BERRY PHASE AND THE COSMOLOGICAL PARAMETERS

The most fascinating part of our analysis is that, we can set up a direct link between this *cosmological analogue of the Berry phase* and the cosmological observables. From now on we shall drop the subscript '*sub*' keeping in mind that the calculations are for *sub-horizon* modes only.

Now, If we neglect the time variation in *slow-roll* parameters then Eqn.(3.54) can be inte-grated analytically. Then the *accumulated Berry phase* during *sub-Hubble* evolution of the scalar modes, in terms of the *slow-roll* parameters, turns out to be

$$\begin{aligned}
\gamma_k^S &= \tfrac{1}{2} \lim_{\eta' \to -\infty} \int_{\eta'}^{-\frac{\sqrt{1+6\epsilon_V - 2\eta_V}}{k}} \frac{\frac{z''}{z} - \left(\frac{z'}{z}\right)^2}{\sqrt{k^2 - \left(\frac{z'}{z}\right)^2}} d\eta \\
&= \tfrac{1}{2} \lim_{\eta' \to -\infty} \int_{\eta'}^{-\frac{\sqrt{1+6\epsilon_V - 2\eta_V}}{k}} \frac{\eta^{-2}(1+3\epsilon_V - \eta_V)}{\sqrt{k^2 - \eta^{-2}(1+6\epsilon_V - 2\eta_V)}} d\eta + O(\epsilon_V^2, \eta_V^2, \epsilon_V \eta_V) \\
&\approx -\frac{\pi}{4} \frac{1 + 3\epsilon_V - \eta_V}{\sqrt{1 + 6\epsilon_V - 2\eta_V}}.
\end{aligned} \tag{3.55}$$

For brevity, we have restricted our analysis up to the first order in *slow-roll* parameters and neglected time variations of $\epsilon_V$, $\eta_V$ defined in Eqn.(1.19). Similarly, for the tensor modes we have

$$\begin{aligned}
\gamma_k^T &= \tfrac{1}{2} \lim_{\eta' \to -\infty} \int_{\eta'}^{-\frac{\sqrt{1+2\epsilon_V}}{k}} \frac{\frac{a''}{a} - \left(\frac{a'}{a}\right)^2}{\sqrt{k^2 - \left(\frac{a'}{a}\right)^2}} d\eta \\
&= \tfrac{1}{2} \lim_{\eta' \to -\infty} \int_{\eta'}^{-\frac{\sqrt{1+2\epsilon_V}}{k}} \frac{\eta^{-2}(1+\epsilon_V)}{\sqrt{k^2 - \eta^{-2}(1+2\epsilon_V)}} d\eta + O(\epsilon_V^2, \eta_V^2, \epsilon_V \eta_V) \\
&\approx -\frac{\pi}{4} \frac{1 + \epsilon_V}{\sqrt{1 + 2\epsilon_V}}.
\end{aligned} \tag{3.56}$$

Here also we have neglected any time variation in the *slow-roll* parameters, $\epsilon_V$, $\eta_V$, and restricted our analysis up to the first order in them. For the estimation of $\gamma_k^{S,T}$, the *slow-roll* parameters are to be evaluated at the start of inflation. But during inflation the *slow-roll* parameters do not evolve significantly from their initial values for first few *e*-folds, which is relevant for the present day observable modes as they are supposed to leave the horizon during first 10 *e-folds*. So, in the above estimates for $\gamma_k^{S,T}$ we can consider $\epsilon_V$ and $\eta_V$ as their values at horizon crossing without committing any substantial error.

We are now in a position to relate this phase with the observable parameters. At the horizon exit the fundamental observables can be expressed in terms of *slow-roll* parameters (up to the first order in $\epsilon_V$, $\eta_V$) as in Eqns.(1.92, 1.93, 1.95, 1.96) [9, 49, 104]. Consequently, the *accumulated Berry phase* associated with the *sub-Hubble* oscillations of the scalar fluctuations during inflation can be expressed in terms of the observable parameters (and vice versa) as follows

$$\gamma_k^S \approx -\frac{\pi}{8} \frac{3 - n_s(k)}{\sqrt{2 - n_s(k)}} \iff n_s(k) \approx 3 - \frac{8\gamma_k^S}{\pi} \left( \frac{4\gamma_k^S}{\pi} - \sqrt{\frac{16[\gamma_k^S]^2}{\pi^2} - 1} \right). \tag{3.57}$$

Therefore *accumulated Berry phase* for the scalar modes is related to the scalar spectral index.

From the above relation it is also very clear how the *Berry phase* is related to the cosmological curvature perturbations. For the tensor modes using (3.56) the corresponding expressions turn out to be

$$\gamma_k^T \approx -\frac{\pi}{8}\frac{2-n_T(k)}{\sqrt{1-n_T(k)}} \iff n_T(k) \approx 2 - \frac{8\gamma_k^T}{\pi}\left(\frac{4\gamma_k^T}{\pi} - \sqrt{\frac{16[\gamma_k^T]^2}{\pi^2}-1}\right). \qquad (3.58)$$

Eqns.(3.57) and (3.58) reveal that the *Berry phases* due to scalar and tensor modes basically correspond to new parameters made of corresponding spectral indices. Here we note that the relations (3.57), (3.58) are not exact in general but they are in the linearized theory of cosmological perturbation. Had we taken into account the second and higher order contributions of the cosmological fluctuations the relations (3.57), (3.58) would have been different.

Further, the accumulated *Berry phase* associated with the total gravitational fluctuations (a sum-total of $\gamma_k^S$ and $\gamma_k^T$) can be expressed in terms of the other observable parameter as well, giving

$$\gamma_k \equiv \gamma_k^S + \gamma_k^T \approx -\frac{\pi}{8}\left[\frac{3-n_S(k)}{\sqrt{2-n_S(k)}} + \frac{2+\frac{r}{8}}{\sqrt{1+\frac{r}{8}}}\right] \approx -\frac{\pi}{8}\left[\frac{3-n_S(k)}{\sqrt{2-n_S(k)}} + \frac{2+\frac{V}{12\pi^2 M_P^4 P_\mathcal{R}}}{\sqrt{1+\frac{V}{12\pi^2 M_P^4 P_\mathcal{R}}}}\right].$$

Therefore the *accumulated Berry phase* for *sub-Hubble* oscillations of the perturbation modes during inflation can be completely envisioned through the observable parameters. We also see that the total *Berry phase* of a single mode can be characterized by the curvature perturbation. The estimation of the *Berry phase* gives a deeper physical insight of the quantum property of the inflationary perturbation modes. As a result, at least in principle, we can claim that measurement of the *Berry phase* can serve as a probe of quantum properties reflected on classical observables.

## 3.6 PHYSICAL SIGNIFICANCE

The physical implication of the *Berry phase* in cosmology is already transparent from our above analyses. In a nutshell, the classical cosmological perturbation modes (both scalar and tensor) having quantum origin picks up a phase during their advancement through the curved space-time background that depends entirely on the background geometry and may be, at least in principle, estimated quantitatively by measuring the corresponding spectral indices. So the *Berry phase* for the quantum counterpart of the classical cosmological perturbations endows us the measure of spectral index. Also, the existing literature suggests that there may be an intriguing direct link of the cosmological *Berry phase* with the CMB. The interpretation of the *Wigner* rotation matrix as the *Berry phase* [158] has already been elaborated by the proposal of an optical demonstration [159]. On the other-hand, *Wigner* rotation matrix can be represented as a measure of statistical isotropy violation of the temperature fluctuations in CMB [160].

Though we did not proceed further in this direction.

The current observation from *WMAP* [5, 47] has put stringent constraint on $n_s$ ($n_s = 0.9608 \pm 0.008$) but only an upper bound for $r$ has been reported so far ($r < 0.13$ at $95\%$ C.L.). Given this status, any attempt towards the measurement of cosmological *Berry phase* may thus reflect observational credentials of this parameter in inflationary cosmology. For example, it is now well-known that any conclusive comment on the energy scale of inflation provides crucial information about fundamental physics. However, in CMB polarization experiments, the energy scale cannot be conclusively determined because there is a confusion between $E$ and $B$ modes in presence of *lensing*, which can only be sorted out once $r$ is measured conclusively. But, $B$ mode polarized state can be contaminated with *cosmic strings*, *primordial magnetic field* etc, thereby making it difficult to measure $r$ conclusively (for a lucid discussion see [161]). So, cosmological *Berry phase* may have the potentiality to play some important role in inflationary cosmology, since it is related to both $r$ and $V$.

## 3.7 CHAPTER SUMMARY

In this chapter we have demonstrated how the exact wave function for the quantum cosmological perturbations can be analytically obtained by solving the associated *Schrödinger* equation employing the *dynamical invariant technique*. This helps us derive an expression for cosmological analogue of *Berry phase*. Finally, we demonstrate how this quantity is related to cosmological parameters and show the physical significance of the cosmological *Berry phase*.

So far as the detection of cosmological *Berry phase* is concerned, we are far away from quantitative measurements. A possible theoretical aspect of detection [162] of the analogue of cosmological *Berry phase* may be developed in squeezed state formalism [155]. In principle, the *Berry phase* can be measured from an experiment dealing with phase difference (e.g. interference). Recently, an analogy between phonons in an axially time-dependent ion trap and quantum fields in an expanding / contracting universe has been derived and corresponding detection scheme for the analogue of cosmological particle creation has been proposed which is feasible with present-day technology [163]. Besides, there exists [155] a scheme for measuring the *Berry phase* in the vibrational degree of freedom of a trapped ion. We hope that these types of detection schemes may be helpful for the observation of the cosmological analogue of the *Berry phase* in laboratory in future.

# Hamilton-Jacobi Formulation: Application to Quasi-Exponential Inflation

This Chapter is based on our following work:

1. Barun Kumar Pal, Supratik Pal and B. Basu, `Confronting quasi-exponential inflation with WMAP seven`, **JCAP 1204**, 009 (2012).

## 4.1 INTRODUCTION

Present-day cosmology is inflowing into an era where it is becoming more and more possible to constrain the models of the early universe by precise data coming from highly sophisticated observational probes like *WMAP* [5], *Planck* [4], QUaD [114], ACBAR [115], *SPT* [164], *ACT* [165]. Such observations are gradually leading theoretical cosmology towards the details of physics at very high energies, and the possibility of testing some of the speculative ideas of recent years. Inflation – the most fascinating among them – was first proposed back in 1981 by Guth [1]. So far we were mostly employing analytical techniques to explore the pros and cons of inflationary cosmology. In the present chapter and in the next one, we shall explore the credentials of numerical cosmology in the context of inflation and CMB.

With the level of precision set by recent observations in measuring various cosmological parameters, theoretical predictions are no more sufficient to discriminate among different classes of cosmological models. Of course, one has to rely on analytical framework upto some level but at the end, employment of highly sophisticated numerical codes, in the context of CMB the likes of CAMB [124], COSMOMC [166], are indispensable. Before the detection of CMB anisotropies by COBE [2], cosmological observations had limited range of scales to access, and it was suffi-cient to predict from an inflationary scenario a scale-invariant spectrum of density perturbations and a negligible amplitude of gravitational waves. However, since COBE [3] and, of late, *WMAP* [5] and *Planck* [4] came forward, the spectra are now well-constrained over a wide range of scales ranging from 1 Mpc upto 10,000 Mpc. So the inflationary predictions should now be very precise to incorporate latest observations, say, from *WMAP* seven year run. The recent data from *Planck* [4] has put further constraint on the theoretical predictions. The current bound on the ratio of the tensor to scalar amplitudes, $r < 0.11$ (95% C.L. ) at $k = 0.002 \, \text{Mpc}^{-1}$, has been set by *Planck* [102]. Recently *SPT* has detected CMB B-mode polarizations produced by gravitational lensing [167] which may be used as a probe of both structure formation and the inflationary epoch. So this is high time to confront different class of inflationary models with latest data and forthcoming predictions.

Following the *Hamilton-Jacobi* formalism [50, 51] and using a phenomenological *Hubble* parameter, here we intend to confront *quasi-exponential* inflationary models with *WMAP*–7 data [5]. The absence of time dependence in the *Hubble* parameter resulting in *de-Sitter* inflation was very appealing from theoretical point of view but its acceptability is more or less limited considering present day observations. So, certain deviation from an exact exponential inflation turns out to be a good move so as to go along with latest as well as forthcoming data. In general these models are called *quasi-exponential* inflation. Our primary intention here is to confront this *quasi-exponential* inflation with *WMAP* seven using the publicly available code CAMB [124]. In the following, we shall first model the *quasi-exponential* inflation with a phenomenological *Hubble* parameter using *Hamilton-Jacobi* formalism. The model parameter will then be constrained by demanding successful inflation and from the observational ground. Then we shall use CAMB to get matter power spectrum and CMB angular power spectra for *quasi-exponential* inflation. Nevertheless, as it will turn out, the analysis also predicts tensor to scalar ratio of the order of $10^{-2}$ which is in tune with the recent data from *Planck*.

## 4.2 HAMILTON-JACOBI FORMULATION

The usual technique used to solve the inflationary dynamical equations is the *slow-roll approximations* [42]. But it is not the only possibility for successfully implementing models of inflation and solutions outside the *slow-roll approximations* have been found [168]. Rather, considering the precision level today's telescopes are probing / will probe in near future, the need to go beyond *slow-roll* approximation is slowly but steadily licking in. To incorporate all the models irrespective of *slow-roll* approximations, *Hamilton-Jacobi* formalism [50, 51] has turned out to be a very useful tool. The formulation is imitative by considering the *inflaton* field itself to be the evolution parameter. The key advantage of this formalism is that here we only need the *Hubble parameter* $H(\phi)$, to be specified rather than the inflaton potential $V(\phi)$. Since $H$ is a geometric quantity, unlike $V$, inflation is more naturally described in this language [50, 51]. Further, be-ing first order in nature, these equations are easily tractable to explore the underlying physics. Once $H(\phi)$ has been specified, one can, in principle, derive a relation between $\phi$ and $t$ which will enable him to get hold of $H(t)$ and the scale factor, $a(t)$, therefrom. So, *Hamilton-Jacobi* formalism provides a straightforward way of exploring inflationary scenario and related observational aspects.

In order to derive the *Hamilton-Jacobi* equations in the context of inflation, we first take derivative of Eqn.(1.14) with respect to $t$, and substitute it into Eqn.(1.16) to get

$$2\dot{H} = - M_P^{-2} \dot{\phi}^2 \tag{4.1}$$

where an *over-dot* represents a time derivative. Now dividing both sides of the above equation by $\dot{\phi}$, we obtain

$$\dot{\phi} = -2 M_P^2 H'(\phi) \tag{4.2}$$

where a *prime* denotes derivative with respect to the scalar field $\phi$. Finally substituting back Eqn.(4.2) into Eqn.(1.14) we get the following first order differential equations [50, 51]

$$[H'(\phi)]^2 - \tfrac{3}{2M_P^2} H(\phi)^2 = -\tfrac{1}{2M_P^4} V(\phi) \tag{4.3}$$

$$\dot{\phi} = -2M_P^2 H'(\phi). \tag{4.4}$$

The above two equations govern the inflationary dynamics in *Hamilton-Jacobi* formalism. The shape of the associated potential can be obtained by rearranging the terms of Eqn.(4.3) to give

$$V(\phi) = 3M_P^2 H^2(\phi) \left[1 - \tfrac{1}{3}\left(2 M_P^2 \frac{H'(\phi)^2}{H(\phi)^2}\right)\right] = 3M_P^2 H^2(\phi)\left[1 - \tfrac{1}{3}\epsilon_H\right] \tag{4.5}$$

where $\epsilon_H$ is defined in Eqns.(1.21). Combining Eqns.(4.4), (4.5) and (1.15) we further have

$$\frac{\ddot{a}}{a} = H(\phi)^2 \left[1 - \epsilon_H\right] \tag{4.6}$$

So, in *Hamilton-Jacobi* formulation the accelerated expansion takes place when $\epsilon_H < 1$ is satisfied and the precise end of inflation is given by the condition $\epsilon_H = 1$.

The above formulation forms the backbone of *Hamilton-Jacobi* algorithm which is indispensable in *Monte-Carlo* simulation in COSMOMC. Though we will not directly employ the algorithm in this thesis, let us briefly discuss it for completion. We shall follow the trail as given in [169].

The *Hamilton Monte Carlo* (HMC) [170, 171] is a *Markov chain Monte Carlo* (MCMC) technique built upon the basic principle of *Hamiltonian* mechanics. HMC works by sampling from a larger parameter space than we want to explore, by introducing $M$ auxiliary variables, one for each parameter in the model. Here we consider each of the parameters in the problem as coordinates. HMC regards the target distribution which we seek as an effective potential in this coordinate system, and for each coordinate it generates a generalized momentum, i.e., it expands the parameter space from its original *coordinate space* to a phase space, in which there is a so-called *extended* target distribution. If the target density in $M$ dimensions is $p(\boldsymbol{\theta})$, then we can define a potential as

$$U(\boldsymbol{\theta}) \equiv -\ln p(\boldsymbol{\theta}). \tag{4.7}$$

For each coordinate, $\theta_\alpha$, we generate a momentum, $u_\alpha$, conveniently from a *normal distribution* with zero mean and unit variance, so the $M$–dimensional momentum distribution is a simple *multivariate Gaussian* which we denote by $\mathcal{N}(\boldsymbol{u})$. Now, we define the kinetic energy as

$$K(\boldsymbol{u}) = \tfrac{1}{2}\boldsymbol{u}^T\boldsymbol{u}, \tag{4.8}$$

so that the *Hamiltonian* is given by

$$\mathcal{H}(\boldsymbol{\theta},\ \boldsymbol{u}) \equiv U(\boldsymbol{\theta}) + K(\boldsymbol{u}). \tag{4.9}$$

The trick is that we generate chains to sample the extended target density

$$p(\boldsymbol{\theta},\ \boldsymbol{u}) = \exp\left[-\mathcal{H}(\boldsymbol{\theta},\ \boldsymbol{u})\right] = \exp\left[-U(\boldsymbol{\theta})\right]\exp\left[-K(\boldsymbol{u})\right] \propto p(\boldsymbol{\theta})\mathcal{N}(\boldsymbol{u}) \tag{4.10}$$

and if we then marginalize over $\boldsymbol{u}$ by simply ignoring the $\boldsymbol{u}$ coordinates attached to each point in the chain, the resulting *marginal distribution* samples the desired target distribution $p(\boldsymbol{\theta})$. The key is that if we can solve exactly the *Hamiltonian* equations

$$\dot{\theta}_\alpha = u_\alpha,\ \ \dot{u}_\alpha = \frac{\partial \mathcal{H}}{\partial u_\alpha} \tag{4.11}$$

then $\mathcal{H}$ remains invariant, so the extended target density is always the same, and the acceptance is unity. Furthermore, we can integrate the equations for a long time if we wish, de-correlating the points in the chain. The last point is that if we change the momentum only with *Hamilton*'s equations of motion, we will restrict ourselves to a locus in phase space, and the target distribution will not be properly explored. To avoid this, a new momentum is generated randomly when each point in the chain is generated. The art is to choose a good step in the integration, and the number of steps to take before generating a new point. Perhaps unsurprisingly, choosing these such that the new point differs from the previous one by about the size of the target peak works well. In a nutshell, *Hamiltonian* dynamics allows the chain to move along trajectories of constant energy, taking large jumps in the parameter space with relatively inexpensive computations [171].

## 4.3 THE SETUP: QUASI-EXPONENTIAL INFLATION

Let us consider the following phenomenological *Hubble* parameter involving a single scalar field

$$H(\phi) = H_{inf}\ \exp\left[\frac{\mathtt{M}_P^{-1}\phi}{p\left(1 + \mathtt{M}_P^{-1}\phi\right)}\right] \tag{4.12}$$

where $p$ is a dimensionless parameter and $H_{inf}$ is a constant having dimension of *Planck* mass. The value of the constants can be fixed from the conditions for successful inflation and observational bounds. As it will turn out in subsequent analyses this *Hubble* parameter can indeed be cast into a form of *quasi-exponential* inflation for some choice in the parameter space.

Two parameters $\epsilon_{H}$ and $\eta_{H}$ in the present context have the following form

$$\epsilon_{H} = 2\left[p^2\left(1+\mathrm{M}_P^{-1}\phi\right)^4\right]^{-1}, \quad \eta_{H} = -2\left(-1+2p+2p\,\mathrm{M}_P^{-1}\phi\right)\left[p^2\left(1+\mathrm{M}_P^{-1}\phi\right)^4\right]^{-1}. \tag{4.13}$$

Now the condition for inflation may also be put forward using equation of state parameter which in the present context turns out to be

$$\omega(\phi) = 4\left[3p^2\left(1+\mathrm{M}_P^{-1}\phi\right)^4\right]^{-1} - 1. \tag{4.14}$$

From the constraint $\omega < -\frac{1}{3}$ during inflation, together with Eqn.(4.14) we obtain a lower bound for the otherwise free parameter $p$, which is given by

$$|p| > \sqrt{2}\left(1+\mathrm{M}_P^{-1}\phi\right)^{-2}. \tag{4.15}$$

Thus, for a wide range of values of $p$, the above form (4.12) of the *Hubble* parameter renders inflationary solution. However, at the end of inflation $\omega \geq -\frac{1}{3}$ allows us to estimate an upper bound for the parameter $p$, which turns out to be

$$|p| \leq \sqrt{2}\left(1+\mathrm{M}_P^{-1}\phi_{E}\right)^{-2} \tag{4.16}$$

where $\phi_{E}$ is the value of the inflaton at the end of inflation. Now we note that, if $p$ is negative then from Eqn.(4.4) and Eqn.(4.12) we have $\dot\phi > 0$, i.e., $\phi(t)$ increases with time, which incorporates the so called *graceful exit problem* [56, 57] as then $\epsilon_{H}$ remains always less than one. We discard the values of $p$ greater or equal to $\sqrt{2}$ as well, since $p \geq \sqrt{2}$ would imply $\epsilon_{H} < 1$ for any value of $\phi$ giving rise to the same problem. As a result, the feasible range for $p$ would be: $0 < p < \sqrt{2}$. Further, for sufficient inflation we also need $|\eta_{H}| < 1$ during inflation and violation of that condition after inflation drags the inflaton towards its potential minimum quickly. But if $p < 0.586$ then $|\eta_{H}|$ will always be less than one. Imposing this condition further restricts the range of the otherwise free parameter $p$ within $0.586 < p < \sqrt{2}$. We shall use a representative value for the parameter $p$ within the above range later on while confronting CAMB [124] outputs with *WMAP* seven [5].

The expression for the scale factor is obtained by making use of the following relation

$$\dot\phi\frac{a'(\phi)}{a(\phi)} = H(\phi), \tag{4.17}$$

which when combined with the Eqns.(4.12) and (4.4) yield

$$a(\phi) = a_{E}\exp\left[-\frac{p}{6}\left(1+\mathrm{M}_P^{-1}\phi\right)^3\right] \tag{4.18}$$

where $a_E \equiv a(\phi_E) \exp\left[-\frac{p}{6}\left(1 + M_P^{-1}\phi_E\right)^3\right]$, $a(\phi_E)$ is the scale factor at the end of inflation.

| $p$ | $\epsilon_H < 1$ $\phi \geq M_P$ | $|\eta_H| < 1$ $\phi \geq M_P$ | $\phi_E$ $M_P$ | $\phi_{in}$ $M_P$ | N | $V(\phi_{in})^{1/4}$ $10^{16}$ GeV |
|---|---|---|---|---|---|---|
| 0.60 | 0.535 | 0.385 | 0.535 | 7.260 | 56 | 1.005 |
|  |  |  |  | 7.894 | 70 | 1.012 |
| 0.70 | 0.421 | 0.414 | 0.421 | 6.845 | 56 | 0.901 |
|  |  |  |  | 7.448 | 70 | 0.907 |

**TAB. 4.1:** *Different parameters for two different values of $p$ within its allowed range.*

The primary quantities related to inflation have been summarized in Table 4.1. From the Table 4.1, we see that $|\eta_H| \approx 1$ after the end of inflation. So, *slow-roll* would be a very good approximation throughout the inflationary period, though we do not utterly require them in this formalism [172].

We shall now see that the *Hubble* parameter given by Eqn.(4.12), indeed corresponds to *quasi-exponential* inflation. For this demonstration, we first expand the *Hubble* parameter, $H(\phi)$, into a power series of $\phi$

$$H(\phi) = H_I\left[1 - \frac{1}{p}\left(1 + M_P^{-1}\phi\right)^{-1} + \frac{1}{2!}\frac{1}{p^2}\left(1 + M_P^{-1}\phi\right)^{-2} - ...\right], \qquad (4.19)$$

where we have defined $H_I \equiv H_{inf}\, e^{\frac{1}{p}}$. The above expansion, at next to leading order, when combined with Eqn.(4.4) gives rise to the following expression for the scalar field

$$\phi \approx M_P\left[\left(6/p H_I [t - t_E]\right)^{1/3} - 1\right]. \qquad (4.20)$$

Here $t_E \equiv \frac{6H_I}{p}t_{end} + (1 + M_P^{-1}\phi)^3$ and $t_{end}$ corresponds to the time at the end of inflation.

The time evolution of the scale factor is obtained by combining Eqns.(4.18) and (4.20), written explicitly, it looks

$$a(t) \approx a_E \exp\left[H_I(t - t_E)\right]. \qquad (4.21)$$

So the expression for the conformal time turns out to be

$$\eta \approx -H_I^{-1} a(t)^{-1}. \qquad (4.22)$$

Thus, our analysis indeed deals with *quasi-exponential* inflation. The higher order terms in the expansion (4.19) will, in principle, measure further corrections to the scale factor but the analytical solutions are neither always obtainable nor utterly required. Rather, one can directly confront the observable parameters with *WMAP* seven in order to constrain *quasi-exponential* inflation, as done in the rest of the chapter.

## 4.4　PERTURBATIONS AND OBSERVABLE PARAMETERS

In this section we shall survey inflationary cosmological perturbations based on the inflationary model under consideration. The *co-moving curvature perturbation*, $\mathcal{R}_k$, is related to the *Mukhanov* variable by $v_k = -z\mathcal{R}_k$, where in the present context $z$ has the following form

$$z = -\frac{2\,M_P\,a(\phi)}{p\left(1+M_P^{-1}\phi\right)^2} = 2M_P\left\{pH_I\eta\left[6/p\ln\left(H_I a_E|\eta|\right)\right]^{2/3}\right\}^{-1}. \tag{4.23}$$

So, the $k^{th}$ *Fourier* mode of the *co-moving curvature perturbation* is approximately given by

$$\mathcal{R}_k \approx -\frac{pH_I\eta\left[6/p\ln\left(H_I a_E|\eta|\right)\right]^{2/3}}{2\,M_P}\frac{e^{-ik\eta}}{\sqrt{2k}}\left(1-\frac{i}{k\eta}\right). \tag{4.24}$$

Consequently, the dimensionless power spectrum for $\mathcal{R}_k$ turns out be,

$$P_{\mathcal{R}}(k) = \frac{p^2 H_I^2}{16 M_P^2 \pi^2}\left(1+k^2\eta^2\right)\left[6/p\ln\left(H_I a_E|\eta|\right)\right]^{4/3}. \tag{4.25}$$

For the evaluation of $P_{\mathcal{R}}(k)$ at horizon exit, we proceed similar to that in Chapter 2 to get

$$1+k^2\eta^2 = 1+a^2H^2\eta^2 \approx 2\left(1-\left[6H_I p^2(t_e-t)\right]^{-\frac{1}{3}}\right). \tag{4.26}$$

So, the power spectrum for $\mathcal{R}_k$ when evaluated at horizon crossing turns out to be

$$P_{\mathcal{R}}(k)|_{k=aH} = \frac{p^2 H_I^2}{8\pi^2 M_P^2}\left(\left[\frac{6A_k}{p}\right]^{4/3} - \frac{6A_k}{p^2}\right) \tag{4.27}$$

where we have defined $A_k \equiv \ln\left(H_I a_E k^{-1}\right)$. Also at horizon crossing $d\ln k = H_I dt$, consequently, the expressions for the scalar spectral index and its running are given by

$$n_S(k) = 1 - \left(\tfrac{4}{3}\left[\tfrac{6}{p}\right]^{4/3} A_k^{1/3} - \tfrac{6}{p^2}\right)\left(\left[\tfrac{6A_k}{p}\right]^{4/3} - \tfrac{6A_k}{p^2}\right)^{-1} \tag{4.28}$$

$$n'_S(k) = -\left\{\tfrac{4}{3}\left(\tfrac{6}{p}\right)^{8/3} A_k^{2/3} - \tfrac{20}{9}\tfrac{6^{7/3}A_k^{1/3}}{p^{10/3}} + \tfrac{36}{p^4}\right\}\left(\left[\tfrac{6A_k}{p}\right]^{4/3} - \tfrac{6A_k}{p^2}\right)^{-2}. \tag{4.29}$$

Here note that, we would have got $n_S = 1 - \tfrac{4}{3A_k}$ and $n'_S = -\tfrac{4}{3A_k^2}$ as the expressions for the scalar spectral index and its running respectively, had we put $k = -\eta^{-1}$ directly into Eqn.(4.25).

The power spectrum for the tensor modes representing primordial gravitational waves in the present context turns out to be

$$P_{\mathcal{T}}(k) = \frac{H_I^2}{\pi^2 M_P^2}\left(1+k^2\eta^2\right). \tag{4.30}$$

Therefore, using Eqn.(4.26) we can immediately write

$$P_{\mathcal{T}}(k)|_{k=aH} = \frac{2H_I^2}{\pi^2\,M_P^2}\left[1-\left[6p^2 A_k\right]^{-1/3}\right]. \tag{4.31}$$

Corresponding spectral index and running are derived from Eqn.(4.31), and are given by

$$n_{_{\mathrm{T}}}(k) = -\tfrac{1}{3}\left\{(6p^2)^{1/3}A_k^{4/3} - A_k\right\}^{-1} \quad (4.32)$$

$$n'_{_{\mathrm{T}}}(k) = -\tfrac{1}{9}\left\{4[6p^2 A_k]^{1/3} - 3\right\}\left\{(6p^2)^{1/3}A_k^{4/3} - A_k\right\}^{-2}. \quad (4.33)$$

The tensor spectral index, $n_{_{\mathrm{T}}}$, and its running, $n'_{_{\mathrm{T}}}$, would have been zero, had we substituted $k = -\eta^{-1}$ into Eqn.(4.30). The results (4.32) and (4.33) brought out the effect of scalar field evolution in estimating observable quantities at *horizon-crossing*.

In the present context, the tensor to scalar ratio is found to be

$$r = 16\,[6A_k]^{-4/3}p^{-2/3}. \quad (4.34)$$

The *consistency* relation is obtained from Eqn.(4.32) and (4.34), which is given by

$$r = -8n_{_{\mathrm{T}}}\left[1 - \frac{r^{1/4}}{2p^{1/2}}\right]. \quad (4.35)$$

The deviation from the usual relation, $r = -8n_{_{\mathrm{T}}}$, was anticipated by considering the fact that the effect of scalar field evolution has been directly taken into account in estimating the observable parameters at *horizon* crossing.

## 4.5 Confrontation with WMAP Seven

Recently, SPT has detected [167] CMB $B$-mode polarization produced by gravitational lensing. But to comment on primordial gravity waves or on the energy scale of inflation we need large scale $B$-mode signal. The current bound on the tensor to scalar ratio, $r < 0.11$, is set by *Planck* [4, 102]. We have found $r$ to be of the order of $10^{-2}$, which is consistent with the *Planck*, making the present discussion fascinating from the observational point of view.

### 4.5.1 Direct numerical estimation

In Table 4.2 we have estimated the observable parameters from the first principle of the theory of fluctuation as derived in the previous section, for two different values of $p$ within its allowed range, $0.586 < p < \sqrt{2}$. For the estimation we have taken $H_{inf} = 2.27 \times 10^{-6}$ M$_P$, $a_E = 7.5 \times 10^{-31}$ and set the pivot scale at $k_0 = 0.002$ Mpc$^{-1}$. Table 4.2 reveals that the observable parameters as derived from our analysis are in excellent agreement with the current observations as given by *WMAP* seven years data for $\Lambda$CDM background [5].

### 4.5.2 CAMB Output and Comparison

In what follows, we shall make use of the publicly available code CAMB [124] in order to confront our results directly with observational data. For CAMB, the Eqn.(4.27) has been set as initial power spectrum and the values of the initial parameters associated with inflation are taken from the Table 4.2 for $p = 0.60$. Also, *WMAP* seven years dataset for $\Lambda$CDM background has been used in CAMB to generate matter power spectrum and CMB angular power spectra. We have set the pivot scale at $k_0 = 0.002$ Mpc$^{-1}$ in CAMB.

| $p$ | $P_{\mathcal{R}}^{1/2}$ $10^{-5}$ | $n_S$ | $n_\mathcal{T}$ $10^{-3}$ | $n'_S$ $10^{-4}$ | $n'_\mathcal{T}$ $10^{-5}$ | $r$ $10^{-2}$ |
|---|---|---|---|---|---|---|
| 0.60 | 4.912 | 0.9746 | -1.516 | -4.253 | -3.853 | 0.967 |
| 0.70 | 4.114 | 0.9747 | -1.343 | -4.334 | -3.403 | 0.878 |

**TAB. 4.2:** *Observable quantities as obtained from the theory of fluctuations.*

| $H_0$ km/sec/Mpc | $\tau_{Reion}$ | $\Omega_b h^2$ | $\Omega_c h^2$ | $T_{CMB}$ K |
|---|---|---|---|---|
| 71.0 | 0.09 | 0.0226 | 0.1119 | 2.725 |

**TAB. 4.3:** *Input parameters.*

Table 4.3 shows inputs from the *WMAP* seven years dataset for $\Lambda$CDM background. Table 4.4 shows the outputs as obtained from CAMB, which is in fine concord with *WMAP* seven years data. The results obtained here are for a representative value of the parameter $p = 0.6$ within its allowed range.

| $t_0$ Gyr | $z_{Reion}$ | $\Omega_M$ | $\Omega_\Lambda$ | $\Omega_k$ | $\eta_{Rec}$ Mpc | $\eta_0$ Mpc |
|---|---|---|---|---|---|---|
| 13.708 | 10.692 | 0.2669 | 0.7331 | 0.0 | 285.15 | 14347.5 |

**TAB. 4.4:** *Different physical quantities as obtained from CAMB.*

The curvature perturbation is generated due to the fluctuations in the *inflaton* and remains almost constant on the *super Hubble* scales. Long after the end of inflation it makes horizon reentry and creates matter density fluctuations through the gravitational attraction of the potential wells. These matter density fluctuations grow with time and form the structure in the Universe. So the measurement of the matter power spectrum is very crucial as it is directly related to the formation of structure. In Fig.4.1 the CAMB output for the variation of the spectrum of the matter density fluctuations with the scale for *quasi-exponential* inflation and the best fit spectrum of *WMAP*–7 for $\Lambda$CDM + TENS model [173] have been shown and it represents true behavior indeed [174].

In Fig.4.2 we confront CAMB output of CMB angular power spectrum $C_\ell^{TT}$ for *quasi-exponential* inflation with *WMAP* seven years data and the best fit spectra of *WMAP*–7 for $\Lambda$CDM + TENS. On the large angular scales, i.e., for low $\ell$, CMB anisotropy spectrum is dominated by the fluctuations in the gravitational potential leading to *Sachs-Wolfe* effect. From Fig.4.2 we see that the Sachs-Wolfe plateau obtained from *quasi-exponential* inflation is almost flat confirming a nearly scale invariant spectrum and resonating with the small spectral tilt, $1-n_S = 0.0254$, as estimated from our analysis. For larger $\ell$, CMB anisotropy spectrum is dominated by the acoustic oscillations of the baryon-photon fluid giving rise to several peaks and troughs in the spectrum. The heights of the peaks are very susceptible to the baryon fraction. Also, the peak positions are sensitive to curvature of the space and on the rate of cosmological expansion, hence on the dark energy and other forms of the matter. In `Fig.4.2` the first and the most prominent peak arises around $\ell \sim 220$ at a height of $\sim 5811\mu K^2$ followed by two equal height peaks near $\ell \sim 537$ and $\ell \sim 815$. This is in excellent agreement with *WMAP* seven years data [5] for $\Lambda$CDM background. The direct comparison of our prediction for $p = 0.60$ in `Fig.4.2` shows fine match with *WMAP* data apart from the two outliers at $\ell = 22$ and $\ell = 40$.

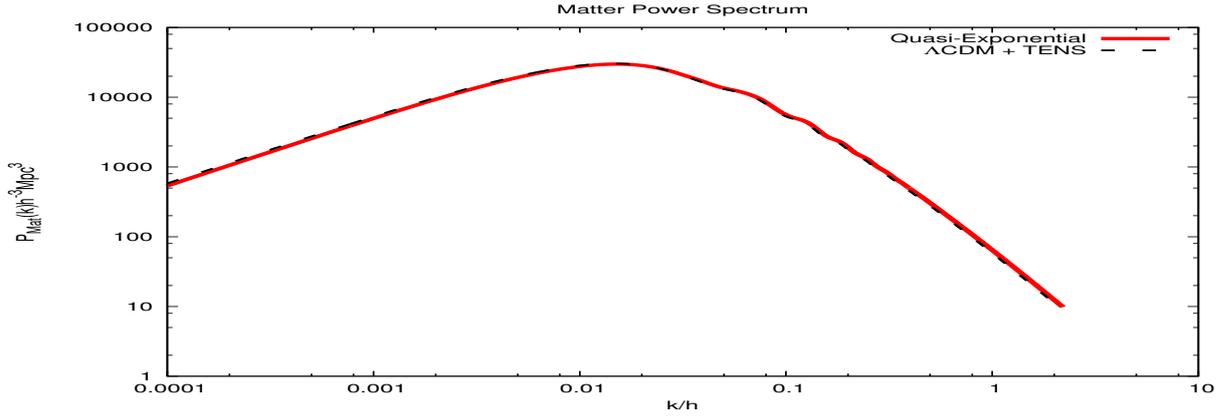

**FIG. 4.1:** *Variation of matter power spectrum $P_{Mat}(k)$ with $k/h$ in logarithmic scales for quasi-exponential inflation and the best fit spectra of WMAP–7 for $\Lambda$CDM + TENS.*

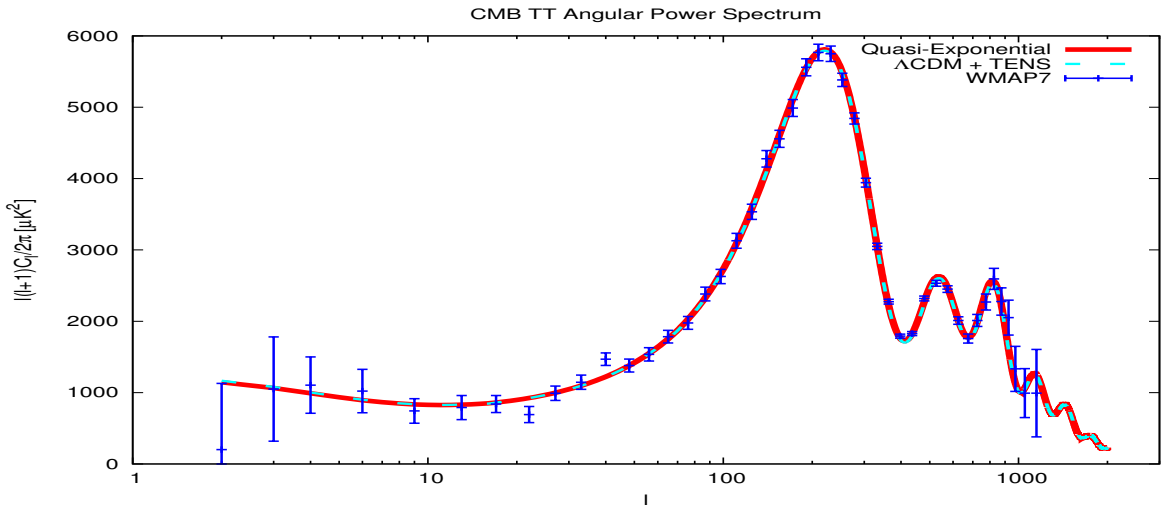

**FIG. 4.2:** *CMB angular power spectrum $C_\ell^{TT}$ for quasi-exponential inflation, the best fit spectra of WMAP7 for $\Lambda$CDM + TENS and WMAP7 data with the multipoles $\ell$.*

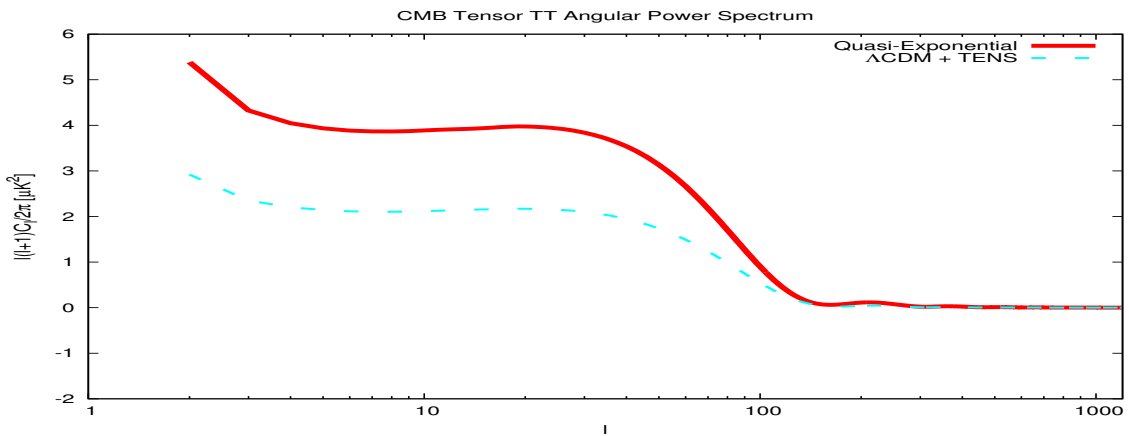

**FIG. 4.3:** *Variation of CMB tensor $C_\ell^{TT}$ angular power spectrum for quasi-exponential inflation and the predicted tensor spectra of WMAP7 for $\Lambda$CDM + TENS with the multipoles $\ell$.*

The gravitational waves generated during inflation also remain constant on *super Hubble* scales having small amplitudes. But as their wavelengths become smaller than the *horizon* the amplitudes begin to die off very rapidly. So, the small scale modes have no impact in the CMB anisotropy, only the large scale modes have little contributions which is obvious from the `Fig.4.3`. Here we see little difference between our result and *WMAP–7* prediction, this happens due to the fact that our tensor to scalar ratio is almost twice that of *WMAP–7* for $\Lambda$CDM + TENS model.

Further, in `Fig.4.4` we have plotted CMB *TE* and *EE* angular power spectra for *quasi-exponential* inflation and the best fit spectra of *WMAP–7* for $\Lambda$CDM + TENS and compared with *WMAP* seven years data. Both the plots resonate fairly well with the latest *WMAP* data [5].

Thus, from the entire analysis, it turns out that *quasi-exponential* inflation confronts extremely well with *WMAP* seven dataset. Not only that, the results are consistent with the recent data from *Planck* as well. The usual *exponential* (e.g. *de-Sitter*) inflation predicts almost *zero* tensor to scalar power ratio, but, as we have seen that for *quasi-exponential* inflation $r$ is of the order of $10^{-2}$, which may even be detected in near future. So, from the observational point of view *quasi-exponential* inflation is more attractive and at the same time consistent than the exact *exponential* inflation.

## 4.6  CHAPTER SUMMARY

In this chapter we have confronted *quasi-exponential models* of inflation with *WMAP–7* data using *Hamilton-Jacobi* formalism. We have first developed the formalism with a phenomenological *Hubble* parameter and demonstrated how and to what extent the scenario measures deviation from *de-Sitter* inflation. The deviation, incorporated through a new parameter $p$, has then been constrained by estimating the major observable parameters from the model and confronting them with *WMAP* seven year dataset.

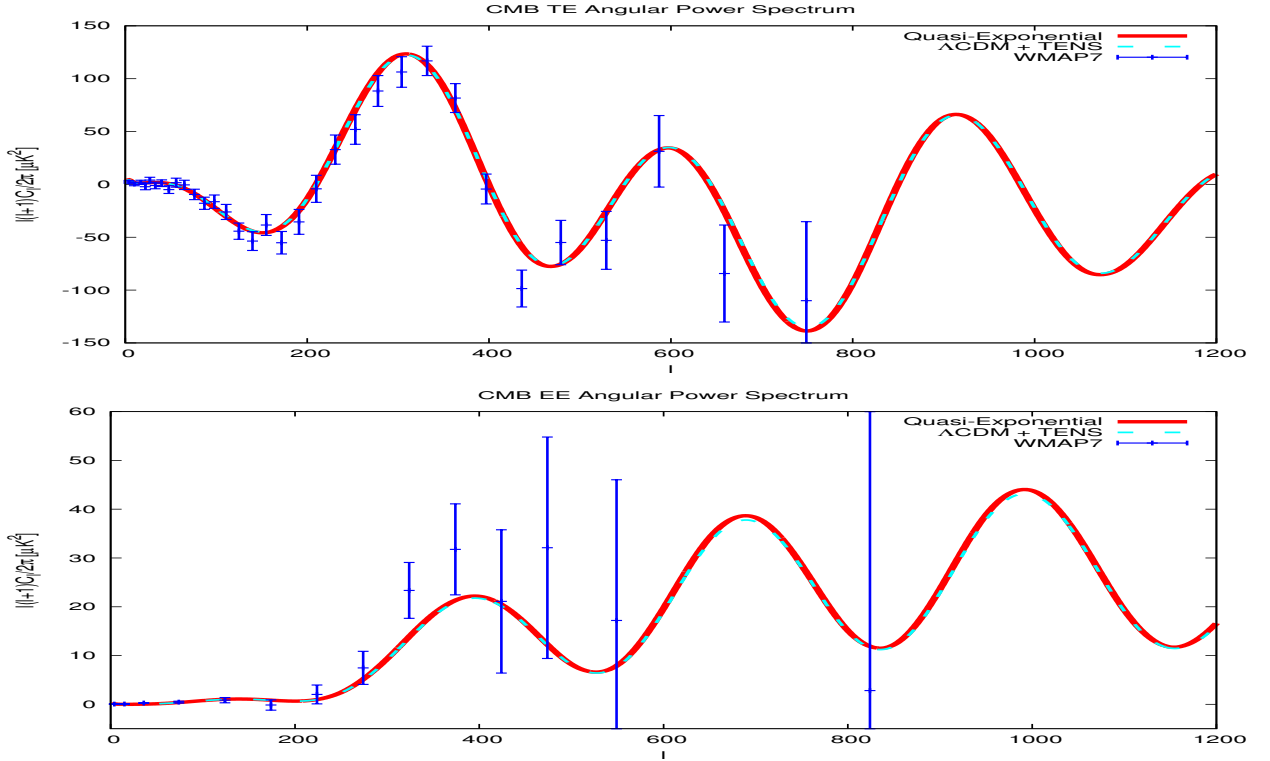

**FIG. 4.4:** *CMB angular power spectra $C_\ell^{TE}$ and $C_\ell^{EE}$ for quasi-exponential inflation and the best fit spectra of WMAP–7 for $\Lambda$CDM + TENS with the multipoles $\ell$.*

We then utilized the publicly available code CAMB [124] to compare our analyses with *WMAP–7* data [5]. CAMB outputs for an emblematic value of the parameter $p$ within its allowed range, $0.586 < p < \sqrt{2}$, are found to be in excellent agreement with the latest *WMAP* data. Values of the most significant cosmological parameters have also been calculated using CAMB and found to fair well with the observational bounds as given by *WMAP–7* data. This leads us to conclude that *quasi-exponential* inflation confronts extremely well with *WMAP* seven within a certain parameter space. The parameter $p$ could have been constrained much better, had we employed more sophisticated code, like COSMOMC [166]. But, still the results obtained through CAMB are in tune with the *WMAP–7* data.

Nevertheless, another appealing aspect of our analysis is the possibility of verifying *quasi-exponential* inflation by detecting primordial gravity waves. The current observational bound on the ratio of tensor to scalar amplitudes as given by *Planck* is $r < 0.11$ at 95% C.L. [102]. The numerical estimation reveals that the parameter $r$ is of the order of $10^{-2}$ for *quasi-exponential* inflation. Thus, *quasi-exponential* inflation can be confronted with more precise data in near future.

# Weak Gravitational Lensing of CMB

This Chapter is based on our following work:

1. Barun Kumar Pal, Hamsa Padmanabhan and Supratik Pal, `Towards reconstruction of unlensed, intrinsic CMB power spectra from lensed map`, **arXiv**: 1309.1827.

## 5.1 INTRODUCTION

The deflection of light rays by intervening structures, a phenomena referred to as *gravitational lensing*, has emerged as an extremely powerful cosmological tool, nowadays. This deflection of light directly probes the gravitational field and the matter that causes it. The gravitational field being independent of nature of the matter that creates it, lensing functions as a direct probe for the total matter in the Universe. Being a direct probe, of late, gravitational lensing has claimed special attention in observational cosmology.

Since the detection of temperature anisotropy in CMB by COBE [2, 3], being the most compelling machinery for the investigation of early universe physics, CMB has continued to stay very next to the heart of cosmologists and its prevalence is still outbreaking among different branches of physics. The latest observational probes like *Planck* [4], *WMAP* [5], ACT [165], SPT [164] have led to extremely precise data resulting in a construction of a very accurate model of our Universe. Even though the latest data from *Planck* [4] are in disagreement with the best-fit $\Lambda$CDM model for low multipoles ($\ell \lesssim 40$) at $2.5\sigma - 3\sigma$, as a whole, all the data are in excellent harmony with the $\Lambda$CDM model and *Gaussian adiabatic* initial conditions with a slightly *red-tilted* power spectrum for the primordial curvature perturbation.

All through the voyage from last scattering surface to the present day detectors, the path of CMB photon gets distorted by potential gradients along the line of sight. In the context of grav-itational lensing, CMB can be thought of as a single source plane situated at a redshift $z \sim 1100$. Being the most distant light source, CMB provides valuable information about the matter distribution on the largest scales. Though, the lensing has small contribution to CMB anisotropies, but it has non-null impact on the determination of various cosmological parameters. In addition to the temperature anisotropies, CMB is also linearly polarized, which was first detected by DASI [175]. The observed polarization field is also lensed by the potential gradients along the line of sight. Therefore, CMB radiation field provides two additional lensed observables in the form of $E$ and $B$ polarization modes.

Current observations such as LIGO [176] aim to detect primordial gravity waves which generate (along with primordial magnetic fields, if any) the large scale $B$-mode signal. CMB polarization offers a unique search for gravity waves produced during inflation, as the gravity waves from the primordial tensor perturbations produce specific pattern of polarization, which can not be generated by the scalar perturbations. The precise measurement of the primordial $B$-mode polarization, although yet to be observed (very recently SPT [164] has claimed a detection of CMB $B$- mode polarization produced by gravitational lensing [177]), is very crucial in the context of inflation as it is directly related to the inflationary energy scale [178] (assuming there are no vector modes), which may help discriminate among different classes of inflationary models. However, there is a confusion between the CMB $E$ and $B$-modes in presence of lensing [179] as the lensing of CMB $E$-mode polarization also produces non-zero $B$-mode signal [180]. Hence, the detection of a large scale $B$-mode signal in CMB polarization experiments does not ensure that we are actually observing primordial gravity waves. Also, the acoustic peaks in CMB angular power spectrum are smoothened [181] and power on the smaller scales is enhanced [182, 183] due to gravitational lensing. Accordingly, the lensing of CMB becomes more and more important on the smaller scales where there is very little intrinsic power. The lensing effect may also produce non-*Gaussian* features in the CMB maps [184]. Therefore, study of CMB lensing is very crucial to get precise handle on the true physics at the last scattering surface.

In this chapter, we start with a brief review on CMB polarization. Then we shall derive lensed CMB power spectra using correlation function technique broadly following [185, 186]. Finally we provide a new method to subtract lensing contribution from the lensed CMB spectra using *matrix inversion technique* in the simplest situation.

## 5.2 CMB Polarization

Ahead of LSS, the mean free path was small compared to the spatial scale of the perturbations and the *Thomson* scattering kept CMB radiation isotropic in the rest frame of the *electron-baryon plasma*. As protons and electrons started to *recombine* to form neutral hydrogen, the mean free path began to grow and anisotropies started to develop. The non-zero *photon quadrupole* at LSS made the CMB radiation field linearly polarized through the *Thomson* scattering process.

The CMB polarization along any line of sight, $\hat{n}$, is conveniently represented by *Stokes* parameters: $I(\hat{n})$, $Q(\hat{n})$ $U(\hat{n})$ and $V(\hat{n})$; where $I(\hat{n})$ is the total *intensity*, $Q(\hat{n})$ and $U(\hat{n})$ are linear polarizations, $V(\hat{n})$ is the *circular polarization* which will not be discussed here as it is expected to vanish for CMB. For a fixed choice of basis $(e_1, e_2)$, in the plane *orthogonal* to the direction of observation, the polarization tensor can be represented as

$$P_{ij} = \begin{pmatrix} Q & U \\ U & -Q \end{pmatrix}$$

which also defines two *Stokes* parameters $Q$ and $U$.

As the parameters $Q$ and $U$ are *basis* dependent, it is not convenient to work with them. Instead, we can define a new *spin* $-2$ quantity, $P$, which with respect to the *complex basis* $e_\pm \equiv e_1 \pm ie_2$ can be expressed as

$$P \equiv Q + iU. \tag{5.1}$$

In the flat-sky approximation $P$ can be expanded in terms of *Fourier* transform of its *electric* ($E$) and *magnetic* ($B$) components,

$$P(\bm{x}) = -\int \frac{d^2\bm{\ell}}{2\pi} \left[E(\bm{\ell}) - iB(\bm{\ell})\right] e^{-2i\phi_\ell} e^{i\bm{\ell}\cdot\bm{x}} \tag{5.2}$$

where $\phi_\ell$ is the angle made by the vector $\bm{\ell}$ with the $x$-axis. In order to calculate the correlation functions for the CMB polarizations in the flat-sky limit, we consider the basis defined by the vector $\bm{r} \equiv \bm{x} - \bm{x}'$ and the direction orthogonal to $\bm{r}$. Then in this new basis, the *spin* $-2$ polarization field $P_{\bm{r}}$ is given by

$$P_{\bm{r}}(\bm{x}) = e^{-2i\phi_r} P(\bm{x}). \tag{5.3}$$

We can now define co-ordinate independent correlation functions for CMB polarizations with respect to the newly defined basis as,

$$\xi_+(r) = \langle P_{\bm{r}}^*(\bm{x}) P_{\bm{r}}(\bm{x}')\rangle = \langle P^*(\mathbf{x}) P(\mathbf{x}')\rangle = \frac{1}{2\pi}\int \ell d\ell \left[C_\ell^{EE} + C_\ell^{BB}\right] J_0(\ell r) \tag{5.4}$$

$$\xi_-(r) = \langle P_{\bm{r}}(\bm{x}) P_{\bm{r}}(\bm{x}')\rangle = \langle e^{-4i\phi_r} P(\mathbf{x}) P(\mathbf{x}')\rangle = \frac{1}{2\pi}\int \ell d\ell \left[C_\ell^{EE} - C_\ell^{BB}\right] J_4(\ell r) \tag{5.5}$$

$$\xi_\times(r) = \langle P_{\bm{r}}(\bm{x}) T(\bm{x})\rangle = \langle e^{-2i\phi_r} P(\mathbf{x}) T(\mathbf{x})\rangle = \frac{1}{2\pi}\int \ell d\ell C_\ell^{TE} J_2(\ell r) \tag{5.6}$$

where $C_\ell^{EE}$, $C_\ell^{BB}$, $C_\ell^{TE}$ are CMB $E$-mode, $B$-mode and cross power spectra respectively. The *correlation functions* defined above, depends only on the separation, $r$, between the two points $\bm{x}$ and $\bm{x}'$. Conversely, CMB polarization and cross power spectra are given by

$$C_\ell^+ \equiv C_\ell^{EE} + C_\ell^{BB} = 2\pi \int r\, dr J_0(\ell r)\, \xi_+(r) \tag{5.7}$$

$$C_\ell^- \equiv C_\ell^{EE} - C_\ell^{BB} = 2\pi \int r\, dr J_4(\ell r)\, \xi_-(r) \tag{5.8}$$

$$C_\ell^\times \equiv C_\ell^{TE} = 2\pi \int r\, dr J_2(\ell r)\, \xi_\times(r). \tag{5.9}$$

In the spherical-sky, we expand *spin* $-2$ CMB polarization field, $P$, in the *spin-weighted spherical harmonics* as [187, 188],

$$P(\hat{\boldsymbol{n}}) = \sum_{\ell m} (E_{\ell m} - iB_{\ell m})\, {}_{-2}Y_{\ell m}(\hat{\boldsymbol{n}}) \tag{5.10}$$

where ${}_{-2}Y_{\ell m}(\hat{\boldsymbol{n}})$s are *spin-weighted spherical harmonics*. We now consider two directions, $\hat{\boldsymbol{n}}_1$ and $\hat{\boldsymbol{n}}_2$, instead of two points on the sky-surface. Now if $\hat{\boldsymbol{n}}_1$ has angular co-ordinates $(\theta_1, \phi_1)$ and $(\theta_2, \phi_2)$ be that of $\hat{\boldsymbol{n}}_2$, then the angle, $\beta$, between $\hat{\boldsymbol{n}}_1$ and $\hat{\boldsymbol{n}}_2$ is defined by

$$\mathcal{D}^{-1}(\phi_1, \theta_1, 0)\mathcal{D}(\phi_2, \theta_2, 0) = \mathcal{D}(\delta, \beta, -\gamma) \tag{5.11}$$

where $\delta$ is the angle required to rotate any vector about $\hat{\boldsymbol{n}}_1$ to bring it onto the tangent at $\hat{\boldsymbol{n}}_1$ to the *geodesic* connecting $\hat{\boldsymbol{n}}_1$ and $\hat{\boldsymbol{n}}_2$ in a *right handed sense*, $-\gamma$ is defined similarly but now at $\hat{\boldsymbol{n}}_2$ [189] and $\mathcal{D}$ is *Wigner-D* matrix. So, the *spin-weighted spherical harmonics* can be represented as

$$\mathcal{D}^\ell_{-ms}(\phi, \theta, 0) = (-1)^m \sqrt{\frac{4\pi}{2\ell+1}}\, {}_sY_{\ell m}(\hat{\boldsymbol{n}}) \tag{5.12}$$

where $(\theta, \phi)$ are angular co-ordinates of the unit vector $\hat{\boldsymbol{n}}$. We define the *geodesic-basis* such that the $x$-direction is along the *geodesic* between $\hat{\boldsymbol{n}}_1$ and $\hat{\boldsymbol{n}}_2$. We denote the quantities in this basis with an *over-bar* on them. Then, the *correlation functions* for polarizations in this *geodesic-basis* are given by [189, 190]

$$\xi_+(\beta) = \langle \bar{P}^*(\hat{\boldsymbol{n}}_1)\bar{P}(\hat{\boldsymbol{n}}_2)\rangle = \sum_\ell \frac{2\ell+1}{4\pi}\left[C_\ell^{EE} + C_\ell^{BB}\right] d^\ell_{22}(\beta) \tag{5.13}$$

$$\xi_-(\beta) = \langle \bar{P}(\hat{\boldsymbol{n}}_1)\bar{P}(\hat{\boldsymbol{n}}_2)\rangle = \sum_\ell \frac{2\ell+1}{4\pi}\left[C_\ell^{EE} - C_\ell^{BB}\right] d^\ell_{2-2}(\beta) \tag{5.14}$$

$$\xi_\times(\beta) = \langle \bar{P}(\hat{\boldsymbol{n}}_1)T(\hat{\boldsymbol{n}}_2)\rangle = \sum_\ell \frac{2\ell+1}{4\pi}C_\ell^{TE} d^\ell_{20}(\beta), \tag{5.15}$$

where $d^\ell_{mn}$s are *reduced Wigner-D matrices*. Here also we see that the correlation functions are independent of choice of *basis* and only depends upon the angle, $\beta$, between two directions $\hat{\boldsymbol{n}}_1$ and $\hat{\boldsymbol{n}}_2$. As a result, the polarization power spectra are found to be,

$$C_\ell^+ = 2\pi \int_{-1}^1 \xi_+(\beta)d^\ell_{22}(\beta)d(\cos\beta) \tag{5.16}$$

$$C_\ell^- = 2\pi \int_{-1}^1 \xi_-(\beta)d^\ell_{2-2}(\beta)d(\cos\beta) \tag{5.17}$$

$$C_\ell^{TE} = 2\pi \int_{-1}^1 \xi_\times(\beta)d^\ell_{20}(\beta)d(\cos\beta). \tag{5.18}$$

Before working out different lensed correlation functions and the corresponding lensed power spectra for CMB, we shall now define *deflection angle* and CMB *lensing potential* in the following.

## 5.3 GRAVITATIONAL LENSING IN BACKGROUND COSMOLOGY

Gravitational field distorts light paths, in other words light responds to mass. This deflection of light rays by the gravitational field is referred to as *gravitational lensing*.

### 5.3.1 DEFLECTION ANGLE

Due to lensing, an object appears to be in a deflected position. The difference between the actual and observed positions is called the *deflection angle*. Here, we shall assume that the *deflection angles* are small and only consider lensing by *density perturbations*.

We write the perturbed metric in the *conformal Newtonian gauge*, which in a spatially flat *universe* has the following form

$$ds^2 = -a^2(\eta) \left[ (1 + 2\Psi_N) d\eta^2 - (1 + 2\Phi_N) \delta_{ij} dx^i dx^j \right]. \tag{5.19}$$

where $\Psi_N$ and $\Phi_N$ are *gravitational potentials*. Since light travels along *null-geodesic*, in case of lensing, we are actually interested in the path $ds^2 = 0$. Therefore, we can use a simpler *conformally* related metric,

$$d\hat{s}^2 = -a^2(\eta) \left[ (1 + 4\Psi_W) d\eta^2 - \delta_{ij} dx^i dx^j \right] \tag{5.20}$$

where $\Psi_W \equiv \frac{1}{2}(\Psi_N - \Phi_N)$, is the *Weyl* potential. The observed deflection, $\boldsymbol{\alpha}$, of an object at a conformal distance $\chi_*$ in the direction $\hat{\mathbf{n}}$, will be sum of all deflections by the lenses between the object and the detector, so

$$\boldsymbol{\alpha}(\hat{\mathbf{n}}) = -2 \int_0^{\chi_*} d\chi \frac{\chi_* - \chi}{\chi_*} \nabla_\perp \Psi_W (\chi \hat{\mathbf{n}}; \eta_0 - \chi) \tag{5.21}$$

where $\chi(z) \equiv H_0^{-1} \int_0^z \frac{dx}{\sqrt{\Omega_\Lambda + \Omega_M (1+x)^3 + \Omega_R (1+x)^4}}$ is the conformal distance corresponding to the redshift $z$, $\nabla_\perp$ is the *covariant derivative* taken transverse to the line of sight and $\chi - \eta_0$ is the conformal time when photon was at position $\chi \hat{\mathbf{n}}$.

As, the path of CMB photon is twisted by potential gradients transverse to the line of sight, the temperature field, $T$, and the polarization field, $P$, are *remapped*. Consequently, a point $\boldsymbol{n}$ appears to be in a deflected position $\boldsymbol{n}' \equiv \boldsymbol{n} + \boldsymbol{\alpha}$, on the LSS. So, the observed lensed temperature, $\tilde{T}(\hat{\mathbf{n}})$, and polarization, $\tilde{P}(\hat{\mathbf{n}})$, in a direction $\hat{\mathbf{n}}$, actually correspond to the unlensed temperature and polarization in the direction $\hat{\mathbf{n}}' = \hat{\mathbf{n}} + \boldsymbol{\alpha}(\hat{\mathbf{n}})$, i.e.,

$$\tilde{T}(\hat{\mathbf{n}}) = T(\hat{\mathbf{n}}') = T(\hat{\mathbf{n}} + \boldsymbol{\alpha}(\hat{\mathbf{n}})), \ \tilde{P}(\hat{\mathbf{n}}) = P(\hat{\mathbf{n}}') = P(\hat{\mathbf{n}} + \boldsymbol{\alpha}(\hat{\mathbf{n}})). \tag{5.22}$$

Through out this chapter we shall work in the limit where lensing is *weak*, i.e., $|\boldsymbol{\alpha}| \ll 1$.

### 5.3.2 POWER SPECTRUM FOR LENSING POTENTIAL

The *lensing potential* $\psi$ is defined as

$$\psi(\hat{\mathbf{n}}) \equiv -2 \int_0^{\chi_*} d\chi \, \frac{\chi_* - \chi}{\chi_*} \, \Psi_W\left(\chi\hat{\mathbf{n}}; \eta_0 - \chi\right), \tag{5.23}$$

so that the deflection angle is given by $\nabla_\perp \psi(\hat{\mathbf{n}})$. To find the *power spectrum* for the lensing potential, we expand the lensing potential into *spherical harmonics* basis as,

$$\psi(\hat{\mathbf{n}}) = \sum_{\ell m} \psi_{\ell m} Y_{\ell m}(\hat{\mathbf{n}}). \tag{5.24}$$

The *angular power spectrum* for the lensing potential, $C_\ell^\psi$, turns out to be

$$\langle \psi_{lm} \psi_{l'm'} \rangle = \delta_{\ell\ell'} \delta_{mm'} C_\ell^\psi. \tag{5.25}$$

In the linear theory, $\Psi_W$ can be related to the primordial curvature perturbation, $\mathcal{R}$, generated during inflation, through the transfer function, $T(\eta; \mathbf{k})$, by the relation $\Psi_W(\eta, \mathbf{k}) = T(\eta; \mathbf{k})\mathcal{R}(\mathbf{k})$. In terms of the primordial power spectrum $\mathcal{P}_\mathcal{R}(k)$, the angular power spectrum of the lensing potential is given by

$$C_\ell^\psi = 16\pi \int \frac{dk}{k} \mathcal{P}_\mathcal{R}(k) \left[ \int_0^{z_{\text{ls}}} \frac{dz}{H(z)} j_\ell(kz) \frac{\chi(z_{\text{ls}}) - \chi(z)}{\chi(z_{\text{ls}})\chi(z)} T(z, \mathbf{k}) \right]^2 \tag{5.26}$$

where $j_\ell(kz)$ is the spherical *Bessel* function of order $\ell$ and $z_{\text{ls}}$ is the *redshift* to the surface of last scattering. Hence, given the form of the primordial power spectrum, the lensing potential power spectrum can be computed from Eqn.(5.26). This can be done, for example, using numerical code like CAMB [124].

## 5.4 THE LENSED CMB SPECTRA

Now, we shall briefly review the correlation function technique and derive the lensed CMB temperature and polarization power spectra following [181, 185, 186], first assuming the flat-sky approximation and then in the spherical-sky limit.

### 5.4.1 FLAT-SKY APPROXIMATION

In the flat-sky limit, correlation functions only depend on the separation, $r \equiv |\mathbf{x} - \mathbf{x}'|$, between two points $\mathbf{x}$ and $\mathbf{x}'$.

#### A. Lensed Temperature Correlation Function

We expand the temperature field in 2-*Dimensional Fourier* transform as,

$$T(\mathbf{x}) = \frac{1}{2\pi} \int d^2\boldsymbol{\ell} \, T(\boldsymbol{\ell}) e^{i\boldsymbol{\ell}\cdot\mathbf{x}}. \tag{5.27}$$

The correlation function and spectrum for CMB temperature anisotropy are then given by

$$\xi(r) \equiv \langle T(\mathbf{x})T(\mathbf{x}')\rangle = \frac{1}{2\pi}\int \ell d\ell\, C_\ell^{TT} J_0(\ell r) \text{ and } C_\ell^{TT} = 2\pi \int r dr\, \xi(r) J_0(\ell r). \quad (5.28)$$

Lensing remaps the temperature according to $\tilde{T}(\mathbf{x}) = T(\mathbf{x} + \boldsymbol{\alpha})$, consequently, the lensed correlation function turns out to be [181, 185],

$$\tilde{\xi}(r) \equiv \langle \tilde{T}(\mathbf{x})\tilde{T}(\mathbf{x}')\rangle = \langle T(\mathbf{x}+\boldsymbol{\alpha})T(\mathbf{x}'+\boldsymbol{\alpha}')\rangle = \int \frac{d^2\boldsymbol{\ell}}{4\pi^2} C_\ell^{TT} e^{i\boldsymbol{\ell}\cdot\mathbf{r}} \langle e^{i\boldsymbol{\ell}\cdot(\boldsymbol{\alpha}-\boldsymbol{\alpha}')}\rangle \quad (5.29)$$

where we have defined $\boldsymbol{r} \equiv \boldsymbol{x} - \boldsymbol{x}'$. The Eqn.(5.29) can be further simplified and may be written as [181, 185, 186],

$$\begin{aligned}\tilde{\xi}(r) \approx & \int \frac{d\ell'}{\ell'} \frac{\ell'^2 C_{\ell'}^{TT}}{2\pi} e^{-\ell'^2 \sigma^2(r)/2} \left[\left(1 + \tfrac{1}{16}\ell'^4 A_2(r)^2\right) J_0(\ell'r) + \tfrac{1}{2}\ell'^2 A_2(r) J_2(\ell'r) \right.\\ & \left. + \tfrac{1}{16}\ell'^4 A_2(r)^2 J_4(\ell'r)\right],\end{aligned} \quad (5.30)$$

where $A_0(r) \equiv \frac{1}{2\pi}\int \ell^3\, d\ell\, C_\ell^\psi J_0(\ell r);\ A_2(r) \equiv \frac{1}{2\pi}\int \ell^3\, d\ell\, C_\ell^\psi J_2(\ell r);\ \sigma^2(r) \equiv A_0(0) - A_0(r)$. Once the correlation function is known, we can readily define corresponding lensed temperature power spectrum, $\tilde{C}_\ell^{TT}$, by

$$\tilde{C}_\ell^{TT} = 2\pi \int r\, dr\, J_0(\ell r)\, \tilde{\xi}(r). \quad (5.31)$$

The above expansion (5.30) for the lensed temperature correlation upto the second order in $A_2(r)$ is very accurate, as the higher order terms in $A_2$ only contribute at the $O(10^{-4})$ level [185].

### B. Lensed Polarization Correlation Functions

The derivation of correlation functions for CMB polarization are almost similar to the temperature case, but with little added complicacy owing to the fact that the polarization field, $P$, has *spin*. The lensing remaps the polarization field according to $\tilde{P}(\mathbf{x}) = P(\mathbf{x} + \boldsymbol{\alpha})$, as a result, the lensed correlation functions for polarization turn out to be [180, 185, 186, 188]

$$\begin{aligned}\tilde{\xi}_+(r) \equiv & \langle P_r^*(\mathbf{x}+\boldsymbol{\alpha}) P_r(\mathbf{x}'+\boldsymbol{\alpha}')\rangle = \frac{1}{2\pi}\int \ell' d\ell' \left(C_{\ell'}^{EE} + C_{\ell'}^{BB}\right) e^{-\ell'^2\sigma^2(r)/2} \\ & \times \left[\left(1 + \tfrac{1}{16}\ell'^4 A_2(r)^2\right) J_0(\ell'r) + \tfrac{1}{2}\ell'^2 A_2(r) J_2(\ell'r) + \tfrac{1}{16}\ell'^4 A_2(r)^2 J_4(\ell'r)\right] \\ \tilde{\xi}_-(r) \equiv & \langle P_r(\mathbf{x}+\boldsymbol{\alpha}) P_r(\mathbf{x}'+\boldsymbol{\alpha}')\rangle = \frac{1}{2\pi}\int \ell' d\ell' \left(C_{\ell'}^{EE} - C_{\ell'}^{BB}\right) e^{-\ell'^2\sigma^2(r)/2} \\ & \times \left[\left(1 + \tfrac{1}{16}\ell'^4 A_2(r)^2\right) J_4(\ell'r) + \tfrac{1}{4}\ell'^2 A_2(r) [J_2(\ell'r) + J_6(\ell'r)] \right.\\ & \left. + \tfrac{1}{32}\ell'^4 A_2(r)^2 [J_0(\ell'r) + J_8(\ell'r)]\right] \\ \tilde{\xi}_\times(r) \equiv & \langle T(\mathbf{x}'+\boldsymbol{\alpha}') P_r(\mathbf{x}+\boldsymbol{\alpha})\rangle = \frac{1}{2\pi}\int \ell' d\ell'\, C_{\ell'}^{TE} e^{-\ell'^2\sigma^2(r)/2} \\ & \times \left[\left(1 + \tfrac{1}{16}\ell'^4 A_2(r)^2\right) J_2(\ell'r) + \tfrac{1}{4}\ell'^2 A_2(r)[J_0(\ell'r) + J_4(\ell'r)] \right.\\ & \left. + \tfrac{1}{32}\ell'^4 A_2(r)^2 [J_2(\ell'r) + J_6(\ell'r)]\right].\end{aligned} \quad (5.32)$$

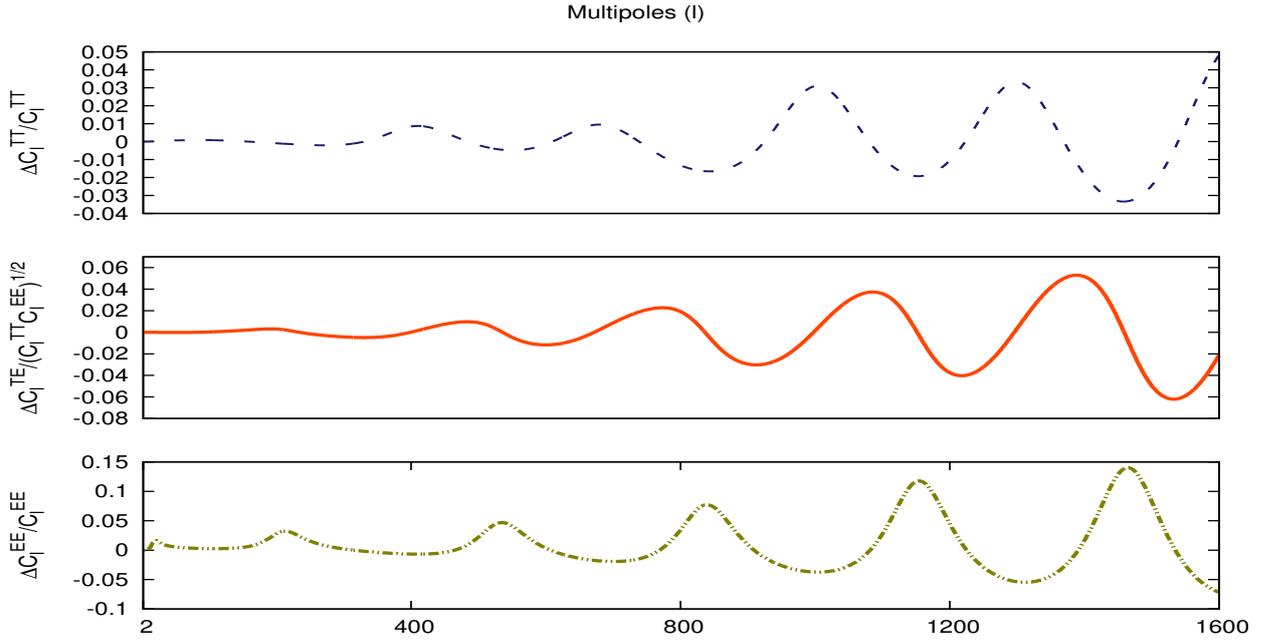

**FIG. 5.1:** *The plot shows fractional contribution to different CMB spectra from lensing.*

Consequently, we can now define lensed CMB polarization and cross power spectra as

$$\tilde{C}_\ell^+ \equiv \tilde{C}_l^{EE} + \tilde{C}_\ell^{BB} = 2\pi \int r\, dr J_0(\ell r)\, \tilde{\xi}_+(r) \tag{5.33}$$

$$\tilde{C}_\ell^- \equiv \tilde{C}_\ell^{EE} - \tilde{C}_\ell^{BB} = 2\pi \int r\, dr J_4(\ell r)\, \tilde{\xi}_-(r) \tag{5.34}$$

$$\tilde{C}_\ell^{TE} = 2\pi \int r\, dr J_2(\ell r)\, \tilde{\xi}_\times(r). \tag{5.35}$$

From the Eqns.(5.33) and (5.34), it can be seen that even if $C_\ell^{BB} = 0$, $\tilde{C}_\ell^{BB} \neq 0$. This lensed $B$ mode is purely generated by the lensing of $E$ mode signal. Consequently, detection of CMB $B$ mode signal may not necessarily imply that we are actually observing intrinsic $B$ mode signal. This is the so called CMB $E$ and $B$ mode confusion.

In Fig.5.1 fractional lensing contributions to different CMB spectra has been shown. The lensing contributes about $5\%$ to CMB $TT$ power spectrum, for CMB $TE$ spectrum it is nearly $6\%$ and $15\%$ for CMB $EE$ spectrum upto $\ell \sim 1600$, which is clear from Fig.5.1. The data for the plots are generated by CAMB [124] using *WMAP–7 best-fit parameters* for $\Lambda CDM$ + *Tens* model [5] assuming zero intrinsic CMB $B$ mode power.

### 5.4.2 SPHERICAL-SKY LIMIT

The calculation of CMB correlation functions in the spherical-sky is more involved. Here, we consider two directions $\hat{n}_1$ and $\hat{n}_2$ in the sky. Following [185, 186], we define a *spin one* deflection field by $_1\alpha \equiv \boldsymbol{\alpha} \cdot (\boldsymbol{e}_\theta + i\boldsymbol{e}_\phi)$, where $\boldsymbol{e}_\theta$ and $\boldsymbol{e}_\phi$ are the unit basis vectors in *spherical polar* *co-ordinates*. Now rotating to the basis defined by the *geodesic* connecting $\hat{n}_1$ and $\hat{n}_2$, denoted by an *over-bar*, at $\hat{n}_1$, $_1\bar{\alpha}$ has *real* and *imaginary* parts given by

$$\boldsymbol{R}[\,_1\bar{\alpha}(\hat{\boldsymbol{n}}_1)] = \alpha_1 \cos\psi_1, \ \boldsymbol{I}[\,_1\bar{\alpha}(\hat{\boldsymbol{n}}_1)] = \alpha_1 \sin\psi_1, \tag{5.36}$$

where $\alpha_1 \equiv |\boldsymbol{\alpha}(\hat{\boldsymbol{n}}_1)|$ is the length of the displacement due to lensing at $\hat{n}_1$ and $\psi_1$ is the angle it makes with the *geodesic*. Similarly at $\hat{n}_2$,

$$\boldsymbol{R}[\,_1\bar{\alpha}(\hat{\boldsymbol{n}}_2)] = \alpha_2 \cos\psi_2, \ \boldsymbol{I}[\,_1\bar{\alpha}(\hat{\boldsymbol{n}}_2)] = \alpha_2 \sin\psi_2. \tag{5.37}$$

The *covariance* of the *spin one* deflection field are found to be

$$\langle_1\bar{\alpha}(\hat{\boldsymbol{n}}_1)_1\bar{\alpha}(\hat{\boldsymbol{n}}_2)\rangle = -\frac{1}{4\pi}\sum_{\ell}\ell(2\ell+1)(\ell+1)C_{\ell}^{\psi}d_{-11}^{\ell}(\beta) \equiv -A_2(\beta) \tag{5.38}$$

$$\langle_1\bar{\alpha}^*(\hat{\boldsymbol{n}}_1)_1\bar{\alpha}(\hat{\boldsymbol{n}}_2)\rangle = \frac{1}{4\pi}\sum_{\ell}\ell(2\ell+1)(\ell+1)C_{\ell}^{\psi}d_{11}^{\ell}(\beta) \equiv A_0(\beta) \tag{5.39}$$

where $\cos\beta = \hat{\boldsymbol{n}}_1 \cdot \hat{\boldsymbol{n}}_2$.

### A. Lensed Temperature

On the full-sky, being a scalar (*spin zero*) quantity the temperature field can be expanded in usual spherical harmonics: $T(\hat{\boldsymbol{n}}) = \sum_{\ell m} T_{\ell m} Y_{\ell m}(\hat{\boldsymbol{n}})$. Therefore, the unlensed temperature correlation function and the corresponding CMB $TT$ power spectrum on the full-sky are given by

$$\xi(\beta) \equiv \langle T(\hat{\boldsymbol{n}}_1)T(\hat{\boldsymbol{n}}_2)\rangle = \sum_{\ell}\frac{2\ell+1}{4\pi}C_{\ell}^{TT}\,d_{00}^{\ell}(\beta)$$

$$C_{\ell}^{TT} = 2\pi\int_{-1}^{1}\xi(\beta)d_{00}^{\ell}(\beta)d(\cos\beta). \tag{5.40}$$

Since, the gravitational lensing of CMB remaps the temperature field in the direction $\hat{\boldsymbol{n}}$ to a new direction $\hat{\boldsymbol{n}}'$, so the lensed correlation function and the corresponding lensed temperature power spectrum turn out to be [185, 186]

$$\tilde{\xi}(\beta) \equiv \langle T(\hat{\boldsymbol{n}}_1')T(\hat{\boldsymbol{n}}_2')\rangle \approx \sum_{\ell}\frac{2\ell+1}{4\pi}C_{\ell}^{TT}\left\{X_{000}^2(\beta)d_{00}^{\ell}(\beta)\right.$$
$$\left. + \frac{8}{\ell(\ell+1)}A_2(\beta)X_{000}'^2(\beta)d_{1-1}^{\ell'}(\beta) + A_2^2(\beta)\left[X_{000}'^2d_{00}^{\ell}(\beta) + X_{220}^2d_{2-2}^{\ell}(\beta)\right]\right\} \tag{5.41}$$

$$\tilde{C}_{\ell}^{TT} = 2\pi\int_{-1}^{1}\tilde{\xi}(\beta)d_0^{\ell}(\beta)d(\cos\beta) \tag{5.42}$$

where the prime denotes derivative with respect to $\sigma^2(\beta) \equiv A_0(0) - A_0(\beta)$, and

$$X_{imn}(\beta) \equiv \int_0^{\infty}\frac{2\zeta}{\sigma^2(\beta)}\left(\frac{\zeta}{\sigma^2(\beta)}\right)^i e^{-\alpha^2/\sigma^2(\beta)}d_{mn}^{\ell}(\zeta)d\zeta. \tag{5.43}$$

### B. Lensed Polarization

On the full-sky, the three correlation functions for the polarization can be expressed upto the second order in $A_2(\beta)$, as follows [185]:

$$\tilde{\xi}_+(\beta) \equiv \langle \tilde{P}^*(\hat{\boldsymbol{n}}_1')\tilde{P}(\hat{\boldsymbol{n}}_2') \rangle \approx \sum_\ell \frac{2\ell+1}{4\pi}\left(C_\ell^{EE}+C_\ell^{BB}\right)\left\{X_{022}^2 d_{22}^\ell + 2A_2(\beta)X_{132}X_{121}d_{31}^\ell \right.$$
$$\left. + \; A_2(\beta)^2\left[X_{022}'^2 d_{22}^\ell + X_{242}X_{220}d_{40}^\ell\right]\right\} \quad (5.44)$$

$$\tilde{\xi}_-(\beta) \equiv \langle \tilde{P}(\hat{\boldsymbol{n}}_1')\tilde{P}(\hat{\boldsymbol{n}}_2') \rangle \approx \sum_\ell \frac{2\ell+1}{4\pi}\left(C_\ell^{EE}-C_\ell^{BB}\right)\left\{X_{022}^2 d_{2-2}^\ell + A_2(\beta)\times\right.$$
$$\left.\left[X_{121}^2 d_{1-1}^\ell + X_{132}^2 d_{3-3}^\ell\right] + \tfrac{1}{2}A_2(\beta)^2\left[2X_{022}'^2 d_{2-2}^\ell + X_{220}^2 d_{00}^\ell + X_{242}^2 d_{4-4}^\ell\right]\right\} \quad (5.45)$$

$$\tilde{\xi}_\times(\beta) \equiv \langle \tilde{T}(\hat{\boldsymbol{n}}_1')\tilde{P}(\hat{\boldsymbol{n}}_2') \rangle \approx \sum_\ell \frac{2\ell+1}{4\pi}C_\ell^{TE}\left\{X_{022}X_{000}d_{02}^\ell + A_2(\beta)\left[\frac{2X_{000}'}{\sqrt{\ell(\ell+1)}}\times\right.\right. \quad (5.46)$$
$$\left.\left.(X_{112}d_{11}^\ell + X_{132}d_{3-1}^\ell)\right] + \tfrac{1}{2}A_2(\beta)^2\left[(2X_{022}'X_{000}' + X_{220}^2)d_{20}^\ell + X_{220}X_{242}d_{-24}^\ell\right]\right\}$$

Again, the power spectra are related to the correlation functions through the following equations

$$\tilde{C}_\ell^{EE} + \tilde{C}_\ell^{BB} \equiv \tilde{C}_\ell^+ = 2\pi \int_{-1}^1 \tilde{\xi}_+(\beta)d_{22}^\ell(\beta)d(\cos\beta) \quad (5.47)$$

$$\tilde{C}_\ell^{EE} - \tilde{C}_\ell^{BB} \equiv \tilde{C}_\ell^- = 2\pi \int_{-1}^1 \tilde{\xi}_-(\beta)d_{2-2}^\ell(\beta)d(\cos\beta) \quad (5.48)$$

$$\tilde{C}_\ell^{TE} = 2\pi \int_{-1}^1 \tilde{\xi}_\times(\beta)d_{20}^\ell(\beta)d(\cos\beta). \quad (5.49)$$

In `Fig.5.2` we have shown fractional lensing contributions to different CMB power spectra in full-sky limit as well as in the flat-sky approximation and we see very little difference between them. The lensed CMB $B$-mode power, coming entirely from the lensing of $E$-mode polarization, has been shown in the full and flat sky limits, assuming zero intrinsic $B$-mode power. Again, the data for the plots are generated by CAMB [124] using *WMAP–7 best-fit parameters* for $\Lambda CDM +$ *Tens* model [5].

Since, the intrinsic CMB spectra are very important in the context of present day cosmology, especially as regards the primordial gravitational waves, it is important to subtract the lensing contribution. But, that is a very difficult task in practice. In what follows, I shall present a simple method to *delens* the lensed CMB power spectra under ideal conditions, i.e., without taking into account noises in measurements and errors in reconstructing the lensing potential.

## 5.5 DELENSING THE CMB POWER SPECTRA

In this section, we shall see that a simple *matrix inversion technique* (MIT) can be utilized to subtract the lensing contribution with very good accuracy. Here, we shall present both the full-sky as well as flat-sky results. For our purpose, we have used the lensing potential as given by *WMAP* seven year best fit results for $\Lambda CDM$ + *Tens* model.

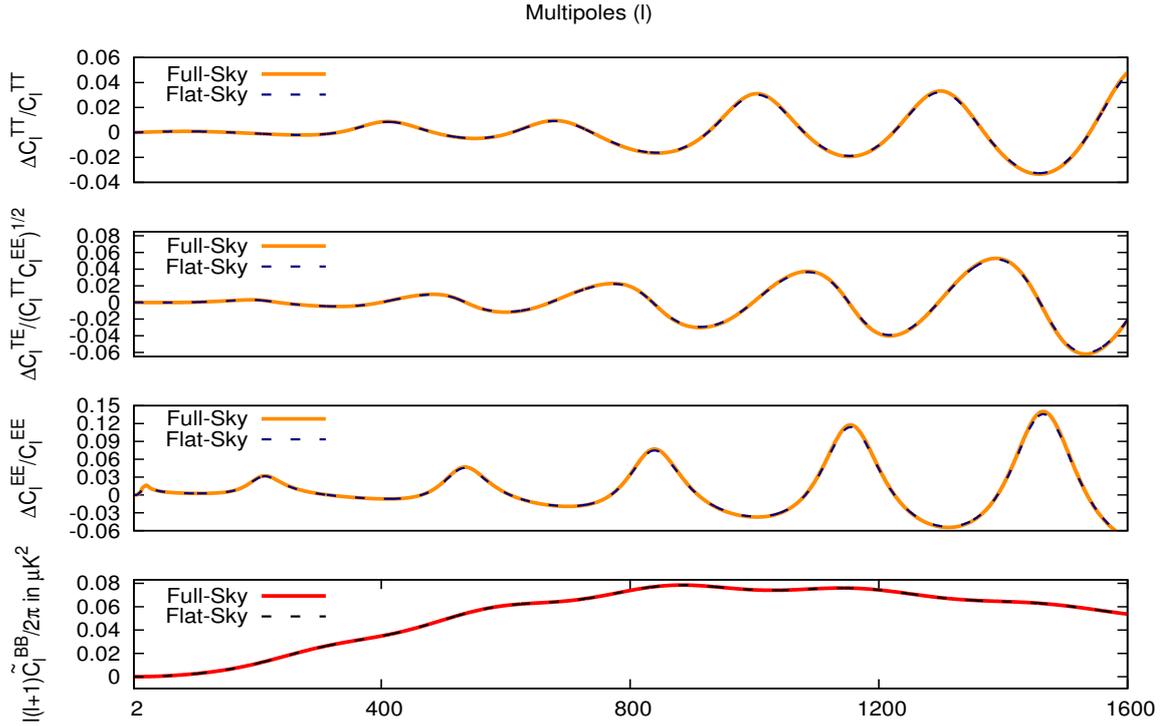

**FIG. 5.2:** *In the top three panels, we have plotted fractional differences between CMB lensed and unlensed spectra in the spherical and flat sky limits. In the bottom panel we plotted CMB B-mode power purely coming from the lensing of E-modes in the full and flat sky limits.*

### 5.5.1 FLAT-SKY ANALYSIS

The lensed CMB power spectra can be estimated directly from the observations [4, 5]. These can then be utilized to obtain the corresponding delensed power spectra provided we have the lensing potential, using MIT as follows.

We first calculate the difference between the lensed and unlensed CMB temperature anisotropy power spectra, which in the flat-sky approximation turns out to be

$$
\begin{aligned}
\tilde{C}_\ell^{TT} - C_\ell^{TT} &= 2\pi \int r\, dr J_0(\ell r) \left[\tilde{\xi}(r) - \xi(r)\right] \\
&= \int r\, dr J_0(\ell r) \int d\ell'\, \ell'\, C_{\ell'}^{TT} e^{-\ell'^2 \sigma^2(r)/2} \left[\left(1 + \tfrac{1}{16}\ell'^4 A_2(r)^2\right) J_0(\ell' r) \right. \\
&\quad + \left. \tfrac{1}{2}\ell'^2 A_2(r) J_2(\ell' r) + \tfrac{1}{16}\ell'^4 A_2^2 J_4(\ell' r)\right] - \int r\, dr J_0(\ell r) \int d\ell'\, \ell'\, C_{\ell'}^{TT} J_0(\ell' r) \\
&\equiv \int d\ell'\, C_{\ell'}^{TT}\, \delta k^T(\ell, \ell'). \qquad (5.50)
\end{aligned}
$$

In the last step, we have changed the order of integration and defined the *temperature kernel*

$\delta k^T(\ell, \ell')$, which can be read off from the above equation as

$$\begin{aligned}\delta k^T(\ell, \ell') &\equiv \ell' \int r\, dr\, J_0(lr) \left\{ e^{-\ell'^2 \sigma^2(r)/2} \left[ \left(1 + \tfrac{1}{16}\ell'^4 A_2(r)^2\right) J_0(\ell' r) \right.\right. \\ &\quad + \left.\left. \tfrac{1}{2}\ell'^2 A_2(r) J_2(\ell' r) + \tfrac{1}{16}\ell'^4 A_2(r)^2 J_4(\ell' r) \right] - J_0(\ell' r) \right\}. \end{aligned} \quad (5.51)$$

Eqn.(5.50) is the *Fredholm integral equation of the second kind*. Generally, these equations are very difficult to solve analytically as well as numerically. We now provide a numerical technique to solve this equation that works satisfactorily whenever the kernel functions are small, which is also the relevant case here, as the lensing contributions are small. To proceed further, we rewrite Eqn.(5.50) in the following form,

$$\tilde{C}_\ell^{TT} = \int d\ell'\, C_{\ell'}^{TT} \left[\delta(\ell - \ell') + \delta k^T(\ell, \ell')\right] \approx \sum_{\ell'} C_{\ell'}^{TT} \left[I_{\ell\ell'} + \delta k_{\ell\ell'}^T\right]. \quad (5.52)$$

In the last step, we have discretized the continuous expression by replacing the integral over $\ell'$ by a sum over the various $\ell'$, the delta function by its discrete counterpart, the identity matrix $I_{\ell\ell'}$ and $\delta k_{\ell\ell'}^T$ is the discrete-$\ell$ representation of $\delta k^T(\ell, \ell')$. Note that the *superscript "T"* of $\delta k_{\ell\ell'}^T$ stands for *temperature*. After defining $M_{\ell\ell'}^T \equiv I_{\ell\ell'} + \delta k_{\ell\ell'}^T$, we can rewrite the above expression for the lensed temperature power spectrum as a set of linear equations, which in the matrix notation can be expressed as

$$\tilde{\mathbf{C}}^{\mathbf{TT}} = \mathbf{M}^{\mathbf{T}} \mathbf{C}^{\mathbf{TT}}. \quad (5.53)$$

As elements of the kernel matrix $\delta k^T(\ell, \ell')$ are very small, we can solve (5.53) by expanding $(\mathbf{M}^{\mathbf{T}})^{-1}$ as a *Taylor* series in powers of $\delta \mathbf{k}^{\mathbf{T}}$. Hence, Eqn.(5.53) may be rewritten as:

$$\begin{aligned}\mathbf{C}^{\mathbf{TT}} &= (\mathbf{M}^{\mathbf{T}})^{-1} \tilde{\mathbf{C}}^{\mathbf{TT}} \\ &= \left[\mathbf{I} - \delta \mathbf{k}^{\mathbf{T}} + (\delta \mathbf{k}^{\mathbf{T}})^2 + \mathcal{O}((\delta \mathbf{k}^{\mathbf{T}})^3)\right] \tilde{\mathbf{C}}^{\mathbf{TT}}. \end{aligned} \quad (5.54)$$

For our analysis, we have retained terms upto the second order in $\delta \mathbf{k}^{\mathbf{T}}$. Since the elements of $\delta \mathbf{k}^{\mathbf{T}}$ are very small, $\mathbf{M}^{\mathbf{T}}$ is very close to identity, as a result, Eqn.(5.53) can also be solved exactly by taking the inverse of $\mathbf{M}^{\mathbf{T}}$. This enables us to extract the unlensed temperature power spectrum $\mathbf{C}^{\mathbf{TT}}$ from the lensed one in a simple manner.

In case of polarization, we can follow the same procedure to define the corresponding kernels, $\delta k^+$, $\delta k^-$ and $\delta k^\times$, in terms of which the lensed CMB polarizations and cross power spectra can be expressed as,

$$\tilde{C}_\ell^+ = \int d\ell'\, C_{\ell'}^+ \left[\delta(\ell - \ell') + \delta k^+(\ell, \ell')\right] \approx \sum_{\ell'} C_{\ell'}^+ \left[I_{\ell\ell'} + \delta k_{\ell\ell'}^+\right] \quad (5.55)$$

$$\tilde{C}_\ell^- = \int d\ell'\, C_{\ell'}^- \left[\delta(\ell - \ell') + \delta k^-(\ell, \ell')\right] \approx \sum_{\ell'} C_{\ell'}^- \left[I_{\ell\ell'} + \delta k_{\ell\ell'}^-\right] \quad (5.56)$$

$$\tilde{C}_\ell^{TE} = \int d\ell'\, C_{\ell'}^{TE} \left[\delta(\ell - \ell') + \delta k^\times(\ell, \ell')\right] \approx \sum_{\ell'} C_{\ell'}^{TE} \left[I_{\ell\ell'} + \delta k_{\ell\ell'}^\times\right] \quad (5.57)$$

where we have now defined

$$\delta k^+(\ell, \ell') \equiv \ell' \int r\, dr J_0(\ell r) \left\{ e^{-\ell'^2 \sigma^2(r)/2} \left[ \left(1 + \tfrac{1}{16}\ell'^4 A_2(r)^2\right) J_0(\ell' r) + \tfrac{1}{2}\ell'^2 A_2(r) J_2(\ell' r) \right. \right.$$
$$\left. \left. + \tfrac{1}{16}\ell'^4 A_2(r)^2 J_4(\ell' r) \right] - J_0(\ell' r) \right\} \tag{5.58}$$

$$\delta k^-(\ell, \ell') \equiv \ell' \int r\, dr J_4(lr) \left\{ e^{-\ell'^2 \sigma^2(r)/2} \left[ \left(1 + \tfrac{1}{16}\ell'^4 A_2(r)^2\right) J_4(\ell' r) + \tfrac{1}{4}\ell'^2 A_2(r) \times \right. \right.$$
$$\left. \left. [J_2(\ell' r) + J_6(\ell' r)] + \tfrac{1}{32}\ell'^4 A_2(r)^2 [J_0(\ell' r) + J_8(\ell' r)] \right] - J_4(\ell' r) \right\} \tag{5.59}$$

$$\delta k^\times(\ell, \ell') \equiv \ell' \int r\, dr J_2(lr) \left\{ e^{-\ell'^2 \sigma^2(r)/2} \left[ \left(1 + \tfrac{1}{16}\ell'^4 A_2(r)^2\right) J_2(\ell' r) + \tfrac{1}{4}\ell'^2 A_2(r) \times \right. \right.$$
$$\left. \left. [J_0(\ell' r) + J_4(\ell' r)] + \tfrac{1}{32}\ell'^4 A_2(r)^2 [J_2(\ell' r) + J_6(\ell' r)] \right] - J_2(\ell' r) \right\}. \tag{5.60}$$

Analogous to the case of temperature, we can also define the kernel matrices for the polarizations and the cross power spectra, through the relations similar to Eqn.(5.54). This can be done as follows,

$$\mathbf{C}^+ = \left[\mathbf{I} - \delta\mathbf{k}^+ + (\delta\mathbf{k}^+)^2 + \mathcal{O}((\delta\mathbf{k}^+)^3)\right] \tilde{\mathbf{C}}^+ \tag{5.61}$$
$$\mathbf{C}^- = \left[\mathbf{I} - \delta\mathbf{k}^- + (\delta\mathbf{k}^-)^2 + \mathcal{O}((\delta\mathbf{k}^-)^3)\right] \tilde{\mathbf{C}}^- \tag{5.62}$$
$$\mathbf{C}^{\mathbf{TE}} = \left[\mathbf{I} - \delta\mathbf{k}^\times + (\delta\mathbf{k}^\times)^2 + \mathcal{O}((\delta\mathbf{k}^\times)^3)\right] \tilde{\mathbf{C}}^{\mathbf{TE}}. \tag{5.63}$$

Thus, we see that with an estimate for the lensing potential, it is possible, in principle, to extract the corresponding unlensed power spectra from the lensed ones in a very simple manner. As a result, using our estimates, it may be possible to constrain the primordial $B$-mode spectrum once we have the $B$ mode signal. In the realistic situation, our formulation must be convolved with estimates for the noise in the measured spectra and uncertainties in the transfer function; the present formalism serves as a demonstration of the deconvolution of the lensing effect under ideal conditions.

### 5.5.2 FULL-SKY ANALYSIS

We now repeat the above procedure in the full-sky limit. Using Eqns.(5.42), (5.44), (5.45), (5.46), (5.47), (5.48) and (5.49), we first express the deviation of the lensed spectra from unlensed ones, for the temperature anisotropy this reads:

$$\tilde{C}_\ell^{TT} - C_\ell^{TT} = \sum_{\ell'} \frac{2\ell'+1}{2} \int_0^\pi \sin\beta\, d\beta\, d_{00}^\ell(\beta)\, C_{\ell'}^{TT} \left\{ X_{000}^2(\beta) d_{00}^{\ell'}(\beta) + \frac{8}{\ell'(\ell'+1)} \times \right.$$
$$\left. A_2(\beta) X_{000}'^2(\beta) d_{1-1}^{\ell'}(\beta) + A_2^2(\beta) \left( X_{000}'^2 d_{00}^{\ell'}(\beta) + X_{220}^2 d_{2-2}^{\ell'}(\beta) \right) \right\}$$
$$- \sum_{\ell'} \frac{2\ell'+1}{2} \int_0^\pi \sin\beta\, d\beta\, C_{\ell'}^{TT} d_{00}^\ell(\beta) d_{00}^{\ell'}(\beta)$$
$$\equiv \sum_{\ell'} C_{\ell'}^{TT} \delta K^T(\ell, \ell'). \tag{5.64}$$

In the last line, we have defined the temperature kernel in the full sky as,

$$\begin{aligned}\delta K^T(\ell,\ell') &= \frac{2\ell'+1}{2}\int_0^\pi \sin\beta\, d\beta\, d^\ell_{00}(\beta)\left\{\left[X^2_{000}-1\right]d^{\ell'}_{00}(\beta) + \tfrac{8}{\ell'(\ell'+1)}A_2(\beta)X'^2_{000}d^{\ell'}_{1-1}(\beta)\right.\\ &+ \left. A_2(\beta)^2\left(X'^2_{000}d^{\ell'}_{00}(\beta)+X^2_{220}d^{\ell'}_{2-2}(\beta)\right)\right\}.\end{aligned} \qquad (5.65)$$

Similarly, for the case of polarizations, corresponding lensed power spectra turn out to be,

$$\begin{aligned}\tilde{C}^+_\ell &= \sum_{\ell'} C^+_{\ell'}\left[I_{\ell\ell'}+\delta K^+(\ell,\ell')\right]\\ \tilde{C}^-_\ell &= \sum_{\ell'} C^-_{\ell'}\left[I_{\ell\ell'}+\delta K^-(\ell,\ell')\right]\\ \tilde{C}^{TE}_\ell &= \sum_{\ell'} C^{TE}_{\ell'}\left[I_{\ell\ell'}+\delta K^\times(\ell,\ell')\right].\end{aligned} \qquad (5.66)$$

where we have now defined

$$\begin{aligned}\delta K^+(\ell,\ell') &= \frac{2\ell'+1}{2}\int_0^\pi \sin\beta\, d\beta\, d^\ell_{22}(\beta)\left\{\left[X^2_{022}-1\right]d^{\ell'}_{22}(\beta)+2A_2(\beta)X_{132}X_{121}d^{\ell'}_{31}(\beta)\right.\\ &+ \left. A_2(\beta)^2\left[X'^2_{022}d^{\ell'}_{22}(\beta)+X_{242}X_{220}d^{\ell'}_{40}(\beta)\right]\right\}\end{aligned} \qquad (5.67)$$

$$\begin{aligned}\delta K^-(\ell,\ell') &= \frac{2\ell'+1}{2}\int_0^\pi \sin\beta\, d\beta\, d^\ell_{2-2}(\beta)\left\{\left[X^2_{022}-1\right]d^{\ell'}_{2-2}(\beta)\right.\\ &+ A_2(\beta)\left[X^2_{121}d^{\ell'}_{1-1}(\beta)+X^2_{132}d^{\ell'}_{3-3}(\beta)\right]\\ &+ \left.\tfrac{1}{2}A_2(\beta)^2\left[2X'^2_{022}d^{\ell'}_{2-2}(\beta)+X^2_{220}d^{\ell'}_{00}(\beta)+X^2_{242}d^{\ell'}_{4-4}(\beta)\right]\right\}\end{aligned} \qquad (5.68)$$

$$\begin{aligned}\delta K^\times(\ell,\ell') &= \frac{2\ell'+1}{2}\int_0^\pi \sin\beta\, d\beta\, d^\ell_{20}(\beta)\left\{\left[X_{022}X_{000}-1\right]d^{\ell'}_{20}(\beta)\right.\\ &+ A_2(\beta)\left[\tfrac{2X'_{000}}{\sqrt{\ell'(\ell'+1)}}(X_{112}d^{\ell'}_{11}(\beta)+X_{132}d^{\ell'}_{3-1}(\beta))\right]\\ &+ \left.\tfrac{1}{2}A_2(\beta)^2\left[(2X'_{022}X'_{000}+X^2_{220})d^{\ell'}_{20}(\beta)+X_{220}X_{242}d^{\ell'}_{-24}(\beta)\right]\right\}.\end{aligned} \qquad (5.69)$$

In order to delens the CMB spectra, we define kernel matrices for the temperature as well as for the polarizations and cross power spectra. Again, to the order $(\delta \mathbf{K})^2$, the expressions for the delensed spectra are given by

$$\begin{aligned}\mathbf{C^{TT}} &= \left[\mathbf{I}-\delta\mathbf{K^T}+(\delta\mathbf{K^T})^2+\mathcal{O}((\delta\mathbf{K^T})^3)\right]\tilde{\mathbf{C}}^{\mathbf{TT}} & (5.70)\\ \mathbf{C^+} &= \left[\mathbf{I}-\delta\mathbf{K^+}+(\delta\mathbf{K^+})^2+\mathcal{O}((\delta\mathbf{K^+})^3)\right]\tilde{\mathbf{C}}^+ & (5.71)\\ \mathbf{C^-} &= \left[\mathbf{I}-\delta\mathbf{K^-}+(\delta\mathbf{K^-})^2+\mathcal{O}((\delta\mathbf{K^-})^3)\right]\tilde{\mathbf{C}}^- & (5.72)\\ \mathbf{C^{TE}} &= \left[\mathbf{I}-\delta\mathbf{K}^\times+(\delta\mathbf{K}^\times)^2+\mathcal{O}((\delta\mathbf{K}^\times)^3)\right]\tilde{\mathbf{C}}^{\mathbf{TE}}. & (5.73)\end{aligned}$$

In the above expressions, the $\delta\mathbf{K}$'s are the matrices associated with the kernels in the full sky. Given the lensing power spectrum (which determines the $X_{imn}$ functions as well as $A_2(\beta)$), the kernel functions can be worked out and using these kernels, the delensed CMB power spectra are obtained by solving the Eqns.(5.70), (5.71), (5.72) and (5.73).

Hence, the unlensed CMB spectra can be extracted by subtracting the lensing artifacts using the above method. Consequently, our formulation may serve as a aid towards resolving, in principle, the confusion between the primordial $B$-mode power spectrum and that produced due to lensing of the $E$-mode if and when primordial gravity waves are detected.

### 5.5.3 NUMERICAL RESULTS

In this section, we describe a calibration of our methodology against the unlensed spectra estimated from the primordial power spectra using CAMB [124]. We also describe the numerical results obtained by applying the procedure described above to the lensed spectra taken from the best-fit *WMAP–7* [5] data for *ΛCDM+Tens model*, to calculate the corresponding delensed spectra assuming zero unlensed $B$-mode power.

#### A. Calibration

For the calibration of our technique, we start from the unlensed spectra generated by CAMB with the best-fit *WMAP–7* cosmological parameters, and use the lensing potential of CAMB to generate corresponding lensed spectra. We delens the lensed spectra thus obtained using our technique described above. Having obtained the delensed spectra, we calibrate them against the original unlensed spectra. For the comparison, we calculate different kernel matrices, first adopting the flat-sky approximation and then considering the sky to be spherical, using a FORTRAN-90 code based on the lensing module of CAMB. The code requires as input the lensing power spectrum and generates the kernel matrices. We then utilize those matrices to obtain delensed quantities from the corresponding lensed CMB spectra. For our investigation we have constructed $2000 \times 2000$ kernel matrices, and the kernels are calculated explicitly for each integer value of $\ell$, without employing any interpolation.

In `Fig.5.3`, we have plotted the fractional difference between the delensed spectra as obtained from our full-sky analysis, and the initial CAMB produced unlensed spectra started originally. It can be seen that the differences are very small, of the order of $10^{-4}$ for values of $\ell \lesssim 1000$, for each of the $TT$, $TE$ and polarization spectra. For higher multipoles, the variation increases and reaches about $0.8\%$ for $TT$ and about $0.92\%$ for $EE$, and $0.33\%$ for $TE$ spectra between $\ell \sim 1500\text{-}1600$. This serves as a calibration for the accuracy of our methodology. The spectra obtained by CAMB are calculated independently using the primordial power spectrum together with Eqn.(5.26), and we recover the spectra started with, to an accuracy of $0.33\% - 0.92\%$ around $\ell \sim 1500\text{-}1600$. Further, if it is possible to reconstruct the lensing power spectrum directly from the observational data, independent of any model, then, by following our above method, the intrinsic CMB power may also be reconstructed in a model independent manner. As an additional calibration, we have repeated the above procedure using the *WMAP–9* cosmological parameters, and we again find an identical accuracy of about $0.8\%$ for $TT$, about $0.92\%$ for $EE$, and $0.33\%$ for $TE$ spectra around $\ell \sim$ 1500-1600 with these parameters, demonstrating that the accuracy of the technique is stable to variation in the input parameters.

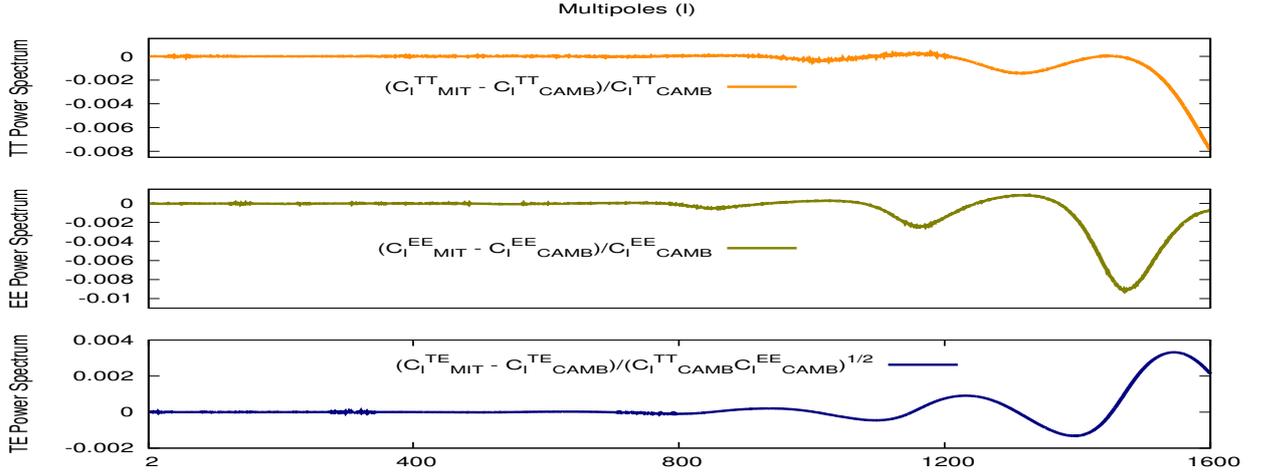

**FIG. 5.3:** *The plot shows fractional difference between the lensed spectra delensed by MIT and the unlensed CMB spectra, both the lensed and unlensed spectra generated by CAMB.*

### B. Delensed CMB Power Spectra

We now present results for the delensing of the *WMAP–7* spectra by our analysis, done in both, the flat-sky approximation as well as in the spherical sky. For this, we have utilized the lensed spectra obtained from *WMAP–7* best-fit $\Lambda CDM$ + TENS data, assuming zero unlensed $B$-mode power, and delensed the spectra using the matrix inversion technique with the kernel matrices generated using the *WMAP–7* best-fit lensing potential. In Fig.5.4, we have plotted the lensing contributions as estimated from our analysis. For the comparison, we have assumed zero intrinsic $B$-mode power. For values of $\ell \lesssim 1600$, we find that the two methods (flat-sky and full-sky) provide consistent results. The difference between the flat-sky and the full-sky results is very small in the large scale regime, and increases very slowly as we go to smaller scales, however, the fractional difference always remains less than $10^{-3}$. Fig.5.5 provides an estimate of the delensed $TT$ spectrum, as obtained from our procedure using the *WMAP–9* unbinned $TT$ spectrum and an estimate of the lensing power spectrum obtained from CAMB using the *WMAP–9* cosmological parameters. In the top panel, the delensed spectrum is over-plotted on the *WMAP–9* unbinned $TT$ spectrum. In the bottom panel, the fractional difference between the lensed and delensed spectra is plotted. It can be seen that the lensing contribution is at the level of $3\%$ for values of $\ell \lesssim 800$.

### C. Sources of Error

Finally, we furnish a brief estimate of the possible sources of error in our analysis. In the flat sky approximation, as pointed out in [185], the error in ignoring terms beyond the second order in $A_2(\beta)$ is extremely small, of the order of $10^{-4}$. The *Taylor* series expansion for evaluating the delensed spectra, Eqns. (5.54), (5.61)-(5.63) and Eqns.(5.70)-(5.73), ignores terms of the order of $(\delta K)^3$. This introduces an error of the order of $(\delta K)^3$, which we have found to be at least four orders of magnitude below the contribution from the combination of the first and second order terms, for $\ell \lesssim 500$. As we go to higher $\ell$ values, the third order contribution increases, but it still remains at most $10^{-3} - 10^{-4}$ of the first- and second-order contributions, for about $\ell \lesssim 1200$, as can be seen from Figs.5.6 and 5.7. The fractional contribution of $(\delta K)^3$ term is of the order of $10^{-3}$ for the $TT$ spectrum at large multipoles ($\ell \gtrsim 1200$), but it remains less than $10^{-4}$ for other CMB spectra. In the full-sky analysis, we have neglected the $A_0(\beta)$ terms entirely as they are very small, of the order $< 10^{-4}$ [185]. In this case also, the contributions from the $(\delta K)^3$ terms are tiny, as can be seen from Figs.5.6 and 5.7. Hence, negligence of the $(\delta K)^3$ terms do not really incorporate significant error in our analysis. Also, we did not take into account the effect of the non-linear evolution of the lensing potential which may also incorporate some additional error. However, non-linearities are increasingly important only in the very small scale regime. The integrated effect of the above errors leads to the overall accuracy of our analysis, estimated to be of the order of $0.33\% - 0.92\%$ at $\ell \sim$ 1500-1600 by the calibration technique described above. A slightly more accurate result might possibly be obtained by using the full second order expressions for the lensed correlation functions in full sky, as provided in Appendix C of Ref. [185], but would increase the computational expense.

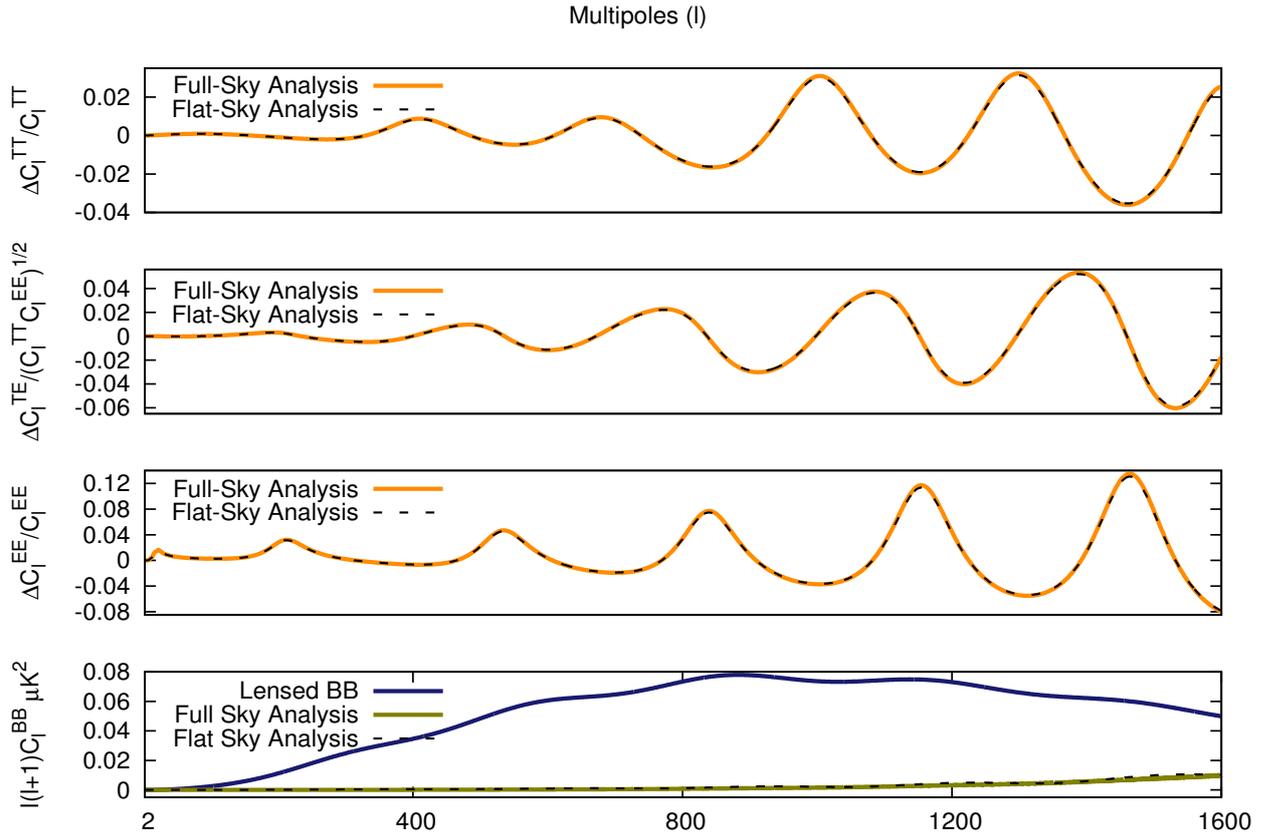

**Fig. 5.4:** *In the top three panels the fractional differences between the lensed and delensed $TT$, $TE$ and $EE$ spectra as obtained by MIT have been shown. In the bottom panel the delensed $BB$ power spectrum as obtained by MIT has been shown. The predicted lensed $BB$ spectrum from the WMAP–7 best-fit data, is also plotted for a comparison.*

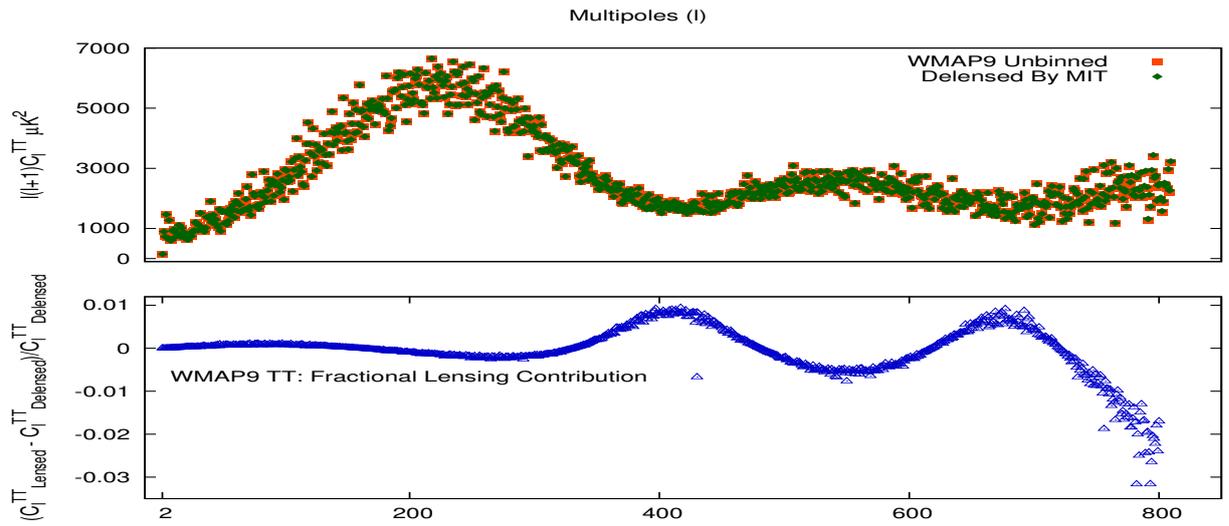

**Fig. 5.5:** *The top panel shows the delensed $TT$ power spectrum as obtained by MIT in the full-sky case over-plotted on the WMAP–9 unbinned $TT$ spectra. The bottom panel shows the fractional difference between the unbinned spectra obtained from the WMAP–9 data and the delensed $TT$ spectra as obtained from MIT.*

### 5.5.4 SIGNIFICANCE OF OUR ANALYSIS

There is a confusion between the CMB $E$- and $B$- polarization modes due to their mixing in presence of lensing. The lensing effect may even dominate the intrinsic $B$-mode power. The large scale $B$-mode signal carries valuable information about the energy scale of inflation as it may have been generated by the primordial gravity waves, i.e., tensor perturbations during inflation. However, $B$-modes are also generated due to the lensing of pure $E$-mode signal. As a result, the detection of large scale $B$-mode power does not completely constrain the power in the primordial gravity waves. But, using our matrix inversion technique, it may be possible to get hold of the intrinsic $B$-mode power by subtracting the lensing effect, and thus the confusion may be overcome, at least in principle. Of course, as we have clearly mentioned, this work is just the first step towards this reconstruction. Here, obtaining the delensed CMB spectrum and the demonstration of deconvolution of the lensing effect has been done in the ideal situation, without taking into account noise in the measured $C_\ell$'s, uncertainties in the transfer functions and in the primordial power spectrum used in the transfer functions as well as more coming from other sources as discussed earlier. In the realistic situation, the above uncertainties need to be incorporated. Further, the code requires CMB lensing potential as the only input, which is very difficult to construct in practice.

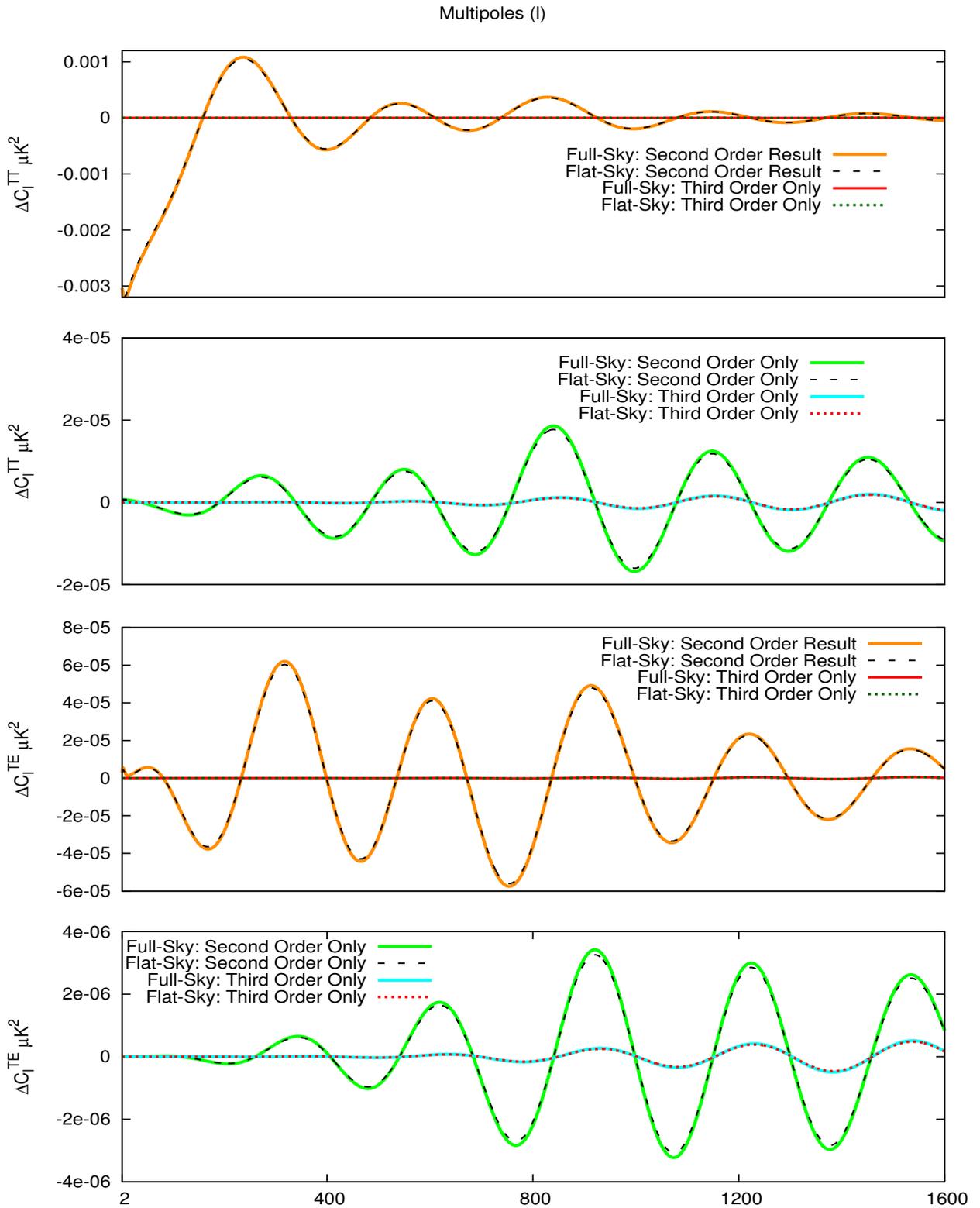

**FIG. 5.6:** *This shows the lensing contributions to the CMB $TT$ and $TE$ power spectra in units of $\mu K^2$. In the first and third panels, the lensing contributions from $(\delta K) + (\delta K)^2$ and $(\delta K)^3$ are shown. In the second and fourth panels, only the second- and third-order contributions are plotted.*

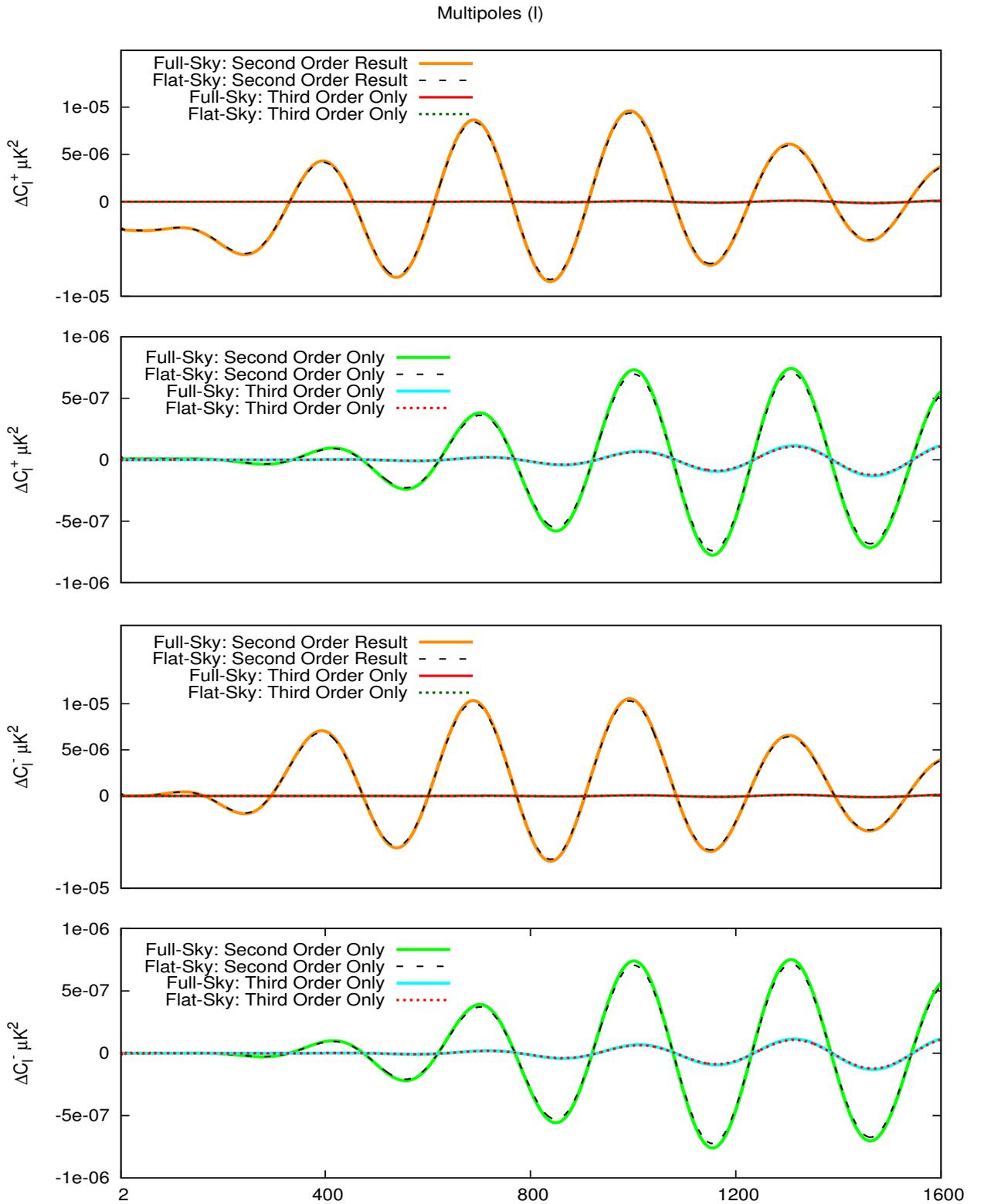

**FIG. 5.7:** *This shows the lensing contributions to the $C^+$ and $C^-$ spectra in units of $\mu K^2$. In the first and third panels, the lensing contributions from $(\delta K) + (\delta K)^2$ and $(\delta K)^3$ are shown. In the second and fourth panels, only the second- and third-order contributions are shown.*

## 5.6 CHAPTER SUMMARY

In this chapter, we have briefly reviewed *weak gravitational lensing* and *polarization*, both in the context of CMB. Following correlation function technique, we have calculated different lensed CMB power spectra. Using those results, we have presented a new method to extract the intrinsic CMB power spectra from the lensed ones by applying a matrix inversion technique. After a thorough formal development of the theoretical framework, we use a FORTRAN-90 code to compare our results with CAMB. We found that for $\ell \lesssim 1000$, both the results are almost identical with fractional deviation being of the order of $10^{-4}$ or so, but as we go beyond, the difference becomes about $0.33\% - 0.92\%$ around $\ell \sim 1600$. Thus, this new technique can very well serve as a first step towards direct reconstruction of intrinsic CMB power spectra from the lensed map.

<div style="text-align: right">**CHAPTER**
**6**</div>

# *Outcomes and Future Directions*

This dissertation can be broadly classified into two categories – (i) *theoretical aspects of inflation* and (ii) *observational aspects of inflation*. Chapters 2 and 3 mostly dealt with category (i), where we have discussed features of *mutated hilltop inflation* and derived a cosmological analogue of the *Berry phase*. In Chapters 4 and 5, some observational aspects of cosmological inflation have been discussed, where we have surveyed *quasi-exponential* models of inflation and *weak gravitational lensing* of CMB.

## 6.1 OUTCOMES OF THE THESIS

In Chapter 2 we have proposed a *variant of hilltop inflation* model. The inflationary scenario has been analyzed with a hyperbolic function for the inflaton potential which contains infinite number of terms when expanded in *Taylor Series*, that makes the *mutated hilltop inflation* more concrete and accurate at the same time. I briefly summarize major outcomes of our analysis in Chapter 2 in the following –

- The common practice in estimating the inflationary observables at the time of *horizon exit* is to adopt perfect *de-Sitter* approximation without bothering about the effect of scalar field evolution. But, if the direct effect of scalar field evolution is taken into account we end up with modified results, of course more accurate, which should be preferred by keeping in mind the accuracy level of present day observations. Apart from that, this *not so common* approach yields a brand new way of modifying the *consistency relation*. This is precisely what has been shown through the *mutated hilltop inflation* model.

- For the discrimination among different classes of inflationary models, the *running, running of the running* etc. now play crucial role. The *logarithmic scale* dependences in observable parameters for *mutated hilltop inflation*, makes it possible to get non zero *running* of any order which is one of the main intriguing feature of the *mutated hilltop inflation*. Finally, the *negative curvature* of the inflaton potential is recently preferred by the *Planck* data. This requirement is satisfied by the *mutated hilltop potential* too, making the scenario attractive from the observational point of view as well.

Chapter 3 deals with inflationary cosmological perturbations of quantum mechanical origin from a different perspective. Here, a cosmological analogue of the *Berry phase* together with the *wave-functions* for cosmological perturbations have been derived. The gist of results of Chapter 3 is given bellow –

- Quantization of the inflationary cosmological perturbations is an important area of research. But, there has been very little study in the literature which may measure the genuine quantum property of the seeds of classical cosmological perturbations. Even, the wave-functions for inflationary perturbation modes were lacking. By analytically solving the associated *Schrödinger* equation applying *Lewis-Riesenfeld* invariant operator technique we have filled (to some extent) that gap. In addition to that, the cosmological analogue of the *Berry phase*, which has been found to be a function of *spectral index*, may have the potentiality to serve as a new probe of inflationary perturbations.

- For the *Berry phase* to become a supplementary probe of inflation, it has to be measured first. A theoretical aspect may be developed in *squeezed state formalism* for the measurement. Also, the *Berry phase* has already been interpreted through the *Wigner* rotation matrices which can be represented as a measure of statistical isotropy violation in CMB temperature fluctuations. Therefore, in principle, through the measurement of statistical isotropy violation in CMB the cosmological analogue of the *Berry phase* may be estimated. Though we are far from putting any strong comment on this.

In Chapter 4 we have confronted *quasi-exponential models of inflation* with *WMAP*–7 data using publicly available code CAMB. Here, we have applied the *Hamilton-Jacobi* formulation to model the *quasi-exponential inflation*. The essence of Chapter 4 is as follows –

- The *Hamilton-Jacobi* formulation allows us to survey inflationary scenario with or without *slow-roll approximation*. Also, the exact end point of inflation is provided by this formalism making it superior choice over the potential based *slow-roll* approximation. The *exact exponential* inflation was very appealing from theoretical point of view as it yields analytical expressions for the involved quantities. But, keeping in mind non-zero values for the *spectral tilt, running of the spectral index, running of the running* etc., certain deviation from exact exponential behavior was necessary without deviating too far which is also observationally forbidden. The *quasi-exponential* inflation just does that trick.

- In recent era of precision cosmology analytical results are no more sufficient, some numerical codes have to be employed to test any theoretical prediction with the observations. CAMB being the simplest, has been utilized in order to confront the predictions from *quasi-exponential inflation* with the *WMAP*–7 data. The analysis also yields tensor to scalar ratio of the order of $\sim 10^{-2}$ much higher than what exact exponential inflation predicts, which is within the recent observational bound as well, even more importantly it is in the range of near future detection. The predictions from *quasi-exponential inflation* are found to be in very good agreement with *WMAP*–7 data.

Finally, in Chapter 5 we have studied effect of gravitational lensing in CMB. Latest observations are now endowing with highly precise data. In order to increase the precision level further, the gravitational lensing effect should now be properly taken into account. The outcomes of this Chapter 5 are given bellow –

- The recent technology is so precise, that the contribution from the gravitational lensing for the determination of different cosmological parameters plays vital role. But, as we have only one sky to look into, the subtraction of lensing contribution is not so trivial. So, matrix inversion technique may serve as an aid to that problem. Though, this algorithm has been developed for idealistic situation, but as a first step towards direct reconstruction of intrinsic CMB spectra it may be very useful.

- The code that has been developed for matrix inversion entirely relies on the lensing potential, which is extremely difficult to construct. But, still this lensing reconstruction may be possible when complete CMB polarization data are in hand. Due to lensing the CMB $E$ and $B$ polarization modes are mixed, which ends up in a confusion between those two modes. But, the large scale unlensed $B$ modes are very important in order to get hold of typical energy scale of inflation (provided inflationary tensor perturbations are the only source of $B$ mode signal). Our methodology provides a simple algorithm towards the reconstruction of *intrinsic* $B$ mode signal, which may be applied to remove that confusion.

Here, the model parameters as well as the prediction for the observational parameters from a typical model of inflation could have been constrained much better had we employed sophisticated code like COSMOMC which is hoped to be addressed in near future.

## 6.2   FUTURE DIRECTIONS

There are many topics that are very much related to this thesis, *viz. non-Gaussianity, Primordial Gravity Waves, Integrated Sachs-Wolfe Effect, Dark Energy* etc., but not discussed here. In the following, I briefly summarize the research work that can be carried out based on the thesis presented here.

The future research can be classified into three broad categories –

1. *Physics of the Cosmic Microwave Background*
2. *Primordial Gravity Waves*
3. *Search for New Probes of Inflationary Perturbations*.

## 6.2.1 PHYSICS OF THE COSMIC MICROWAVE BACKGROUND

CMB has remained the most powerful and engrossing source that carries the best information about the early Universe. Inflation – so far the best physically motivated paradigm for the early Universe is in vogue for more than three decades. But, the large observational window is still allowing a number of inflationary models. To zero in on specific inflationary models as well as cosmological parameters, we need more precise CMB measurements where secondary effects like *weak gravitational lensing, non-Gaussianity* etc. come into play.

### A.  Weak Lensing of CMB

The gravitational lensing remaps the CMB temperature and polarization fields. The remapping of power of the CMB anisotropy spectra by the effect of lensing is a very small effect, but it is important in the context of present era of precision cosmology. Not only that, lensing also introduces *non-Gaussianity* in the CMB maps. So, lensing effects should be accurately taken into account.

As we have only one sky to look into, it is very difficult to delens the observed CMB spectra. But, if we can reconstruct the lensing potential, then it is possible to delens the CMB spectra which will definitely enrich our knowledge about various cosmological parameters. So, lensing reconstruction will be an important prospect. This may also get rid of the confusion between CMB $E$ and $B$ polarization modes in presence of lensing.

### B.  Non-Gaussianity

*Non-Gaussianity* has now become very interesting area of research with *Planck*'s recent claim that *non-Gaussianity* may not be small. Though in general single field models of inflation are expected to produce small local type of *non-Gaussianity* in the *squeezed limit* according to *Maldecena* theorem but, it may be possible to produce large (within current observational bound) local type of *non-Gaussianity* even from single field inflationary models, which will be very interesting subject matter to look into.

In case where variable sound speed for the perturbations are considered, it is difficult to get higher *tensor to scalar ratio* and large *non-Gaussianity* simultaneously. So, simultaneous production of large $r$ and large *non-Gaussianity* will be an interesting business.

### C.  Integrated Sachs-Wolfe Effect

The *Integrated Sachs-Wolfe* (ISW) effect, i.e., the early rise in the CMB temperature anisotropy map, provides crucial information about the nature of *dark energy*. So, precise measurement of ISW effect help us discriminate among different dark energy models, to mention the least, it will help us infer whether dark energy is made of a dynamical field or cosmological constant. The accuracy level may be increased by combining the Lensing and SNIa data with CMB, which will presumably increase our knowledge about ISW effect.

### 6.2.2  PRIMORDIAL GRAVITY WAVES

The most recent trend in cosmology is the search for primordial *Gravity-Waves* through the large scale $B$-mode signal. Gravity waves are expected from the *Big-Bang* and in course of inflation their amplitudes will be amplified. This will in turn provide the best information along with CMB about the era of cosmic inflation. But the gravity waves are yet to be detected (all we have is the indirect detection via change in period of binary stars) and even if it is detected, we may not be able to make conclusive comments about the primordial gravity waves, since, the $B$-mode like signals that we are expecting to detect for gravity waves are also generated from lensing of purely $E$-mode signals. So, the separation of pure primordial gravity waves from observed data will be an important aspect. At present this will be very interesting in the light of SPT's recent claim about the detection of CMB $B$-mode polarization produced by gravitational lensing.

### 6.2.3  SEARCH FOR NEW PROBES OF INFLATIONARY PERTURBATIONS

A cosmological analogue of *Berry phase* in the context of inflationary cosmological perturbations has already been derived. The interesting point about the *Berry phase* is that, it has been interpreted in terms of the *Wigner rotation matrices*. Also, the *Wigner* rotation matrix can be represented as a measure of *statistical isotropy violation* in CMB. So, at least in principle, this *statistical isotropy violation* in CMB can be related with the cosmological analogue of the *Berry phase* and vice versa. The exploration of this theoretical relation, may end up with a signature of the *Berry phase* in CMB.

Last but not the least, the combined data from *Planck, WMAP, SNIa, SDSS* etc., will definitely increase the accuracy level of different cosmological parameters and we can put stringent constraints on various cosmological models. So, the combination of individual datasets of CMB, Gravitational Lensing, SNIa, Baryonic Acoustic Oscillations etc. and their cross-correlations will make the observational window smaller than what can be achieved by individual datasets alone. This will in turn provide more precise information about our Universe.